\documentclass[12pt]{iopart}
\usepackage{setstack}
\usepackage{amssymb}
\usepackage{bm}
\usepackage{graphicx}

\begin{document}
\rightline{{\bf New J.\ Phys.\ 9 (2007) 377}, Focus Issue on Dark Energy}
\font\bm=cmmib10 \def\B#1{\hbox{\bm#1}} \def\Bom{\B{\char33}}
\def\dd{{\rm d}} \def\ds{\dd s} \def\e{{\rm e}} \def\etal{{\em et al}.}
\def\al{\alpha}\def\be{\beta}\def\ga{\gamma}\def\de{\delta}\def\ep{\epsilon}
\def\et{\eta}\def\th{\theta}\def\ph{\phi}\def\rh{\rho}\def\si{\sigma}

\def\mean#1{{\vphantom{\tilde#1}\bar#1}}
\def\bH{\mean H}\def\OM{\mean\Omega}\def\ab{\mean a} \def\bz{\mean z}
\def\rhb{\mean\rh}\def\bq{\mean q}\def\bT{\mean T}\def\bn{\mean n}
\def\bD{\mean D}\def\mx{\mean x} \def\rb{\mean r}
\def\OmB{\Omega\Z B}\def\bOmB{\mean\OmB}\def\bnB{\bn\Z B}
\def\gb{\mean\ga}\def\gc{\gb\Z0}\def\etb{\mean\eta}
\def\OmBw{\Omega\Z{B\hbox{\sevenrm w}}}
\def\zdec{z\ns{dec}} \def\OmBn{\Omega\Z{B0}}\def\bOmBn{\mean\Omega\Z{B0}}
\def\kB{k\Z B} \def\Ns#1{\Z{\hbox{\sevenrm #1}}}\def\ze{\,\zeta(3)\,}
\def\dOMav{\dd\Omega\ns{av}^2} \def\DE{\Delta} \def\la{\lambda}
\def\w#1{\,\hbox{#1}} \def\Deriv#1#2#3{{#1#3\over#1#2}}
\def\Der#1#2{{#1\hphantom{#2}\over#1#2}} \def\br{\hfill\break}
\def\ts{t} \def\tc{\tau} \def\Dtc{\mathop{\hbox{$\Der\dd\tc$}}}
\def\ns#1{_{\hbox{\sevenrm #1}}} \def\dOM{\dd\Omega^2}
\def\goesas{\mathop{\sim}\limits}
\def\zb{{\bar z}} \def\bxi{{\mathbf\xi}}
\def\Y#1{^{\raise2pt\hbox{$\scriptstyle#1$}}}
\def\Z#1{_{\lower2pt\hbox{$\scriptstyle#1$}}}
\def\X#1{_{\lower2pt\hbox{$\scriptscriptstyle#1$}}}
\def\av{{a\ns{v}\hskip-2pt}} \def\aw{{a\ns{w}\hskip-2.4pt}}
\def\ac{a} \def\an{\ac\Z0} \def\QQ{{\cal Q}} \def\qh{q}
\def\tv{\tc\ns{v}} \def\tw{\tc\ns{w}} \def\kv{k\ns v} \def\kw{k\ns w}
\def\Om{\Omega\Z M} \def\gw{\gb\ns w}
\def\fw{{f\ns w}} \def\fvi{{f\ns{vi}}} \def\fwi{{f\ns{wi}}}
\def\rhw{{\rh\ns w}} \def\rhv{\rh\ns v}
\def\etw{\eta\ns w} \def\etv{\eta\ns v} \def\rw{r\ns w}
\def\gv{\gb\ns v} \def\fv{{f\ns v}} \def\Hv{H\ns v} \def\Hw{H\ns w}
\def\OmD{\Omega^{\cal D}} \def\OmDm{\OmD\Z M} \def\OmDR{\OmD_k}
\def\OmDQ{\OmD\Z{\cal Q}} \def\FF{{\cal F}} \def\FI{\FF\Z I} \def\Fi{\FF\X I}
\def\OMM{\OM\Z M}\def\OMk{\OM_k}\def\OMQ{\OM\Z{\QQ}}\def\OMR{\OM\Z R}
\def\OmM{\Omega\Z M} \def\OmR{\Omega\Z R} \def\OMMn{\OM\Z{M0}}
\def\OmMw{\Omega\Z{M\hbox{\sevenrm w}}} \def\OmMn{\Omega\Z{M0}}
\def\gwf{\left(1-\gw^{-1}\right)} \def\gbff{\Bigl(1-{1\over\gb}\Bigr)}
\def\fvf{\left(1-\fv\right)} \def\Hb{\bH\Z0} \def\Hh{H} \def\Hm{H\Z0}
\font\sevenrm=cmr7 \def\ws#1{_{\hbox{\sevenrm #1}}} \def\rs{r\ns s}
\def\ave#1{\langle{#1}\rangle} \def\Rav{\ave{\cal R}} \def\hr{h_r}
\def\gd{{{}^3\!g}} \def\pt{\partial} \def\noi{\noindent}
\def\half{\frn12} \def\DD{{\cal D}} \def\bx{{\mathbf x}} \def\Vav{{\cal V}}
\def\frn#1#2{{\textstyle{#1\over#2}}} \def\LL{${\cal L}$}
\def\SS{${\cal S}$} \def\BB{${\cal B}$} \def\TT{${\cal T}$} \def\VV{${\cal V}$}
\def\lsim{\mathop{\hbox{${\lower3.8pt\hbox{$<$}}\atop{\raise0.2pt\hbox{$\sim$}}
$}}} \def\dL{d\Z L} \def\dA{D\Z A} \def\rhcr{\rh\ws{cr}}
\def\dAdec{d\Z{A\,\hbox{\sevenrm dec}}} \def\const{\hbox{const}}
\def\kmsMpc{\w{km}\;\w{sec}^{-1}\w{Mpc}^{-1}}
\def\etBg{\et\Z{B\ga}} \def\ab{{\bar a}} \def\LCDM{$\Lambda$CDM}
\def\epi{\epsilon_i}\def\gbi{\gb_i}\def\Omi{\OM_i}
\def\lcdm{(\Lambda\hbox{\sevenrm CDM})}
\def\beq{\begin{equation}} \def\eeq{\end{equation}}
\def\bea{\begin{eqnarray}} \def\eea{\end{eqnarray}}
\def\PRL#1{Phys.\ Rev.\ Lett.\ {\bf#1}} \def\PR#1{Phys.\ Rev.\ {\bf#1}}
\def\ApJ#1{Astrophys.\ J.\ {\bf#1}} \def\PL#1{Phys.\ Lett.\ {\bf#1}}
\def\AsJ#1{Astron.\ J.\ {\bf#1}}
\def\MNRAS#1{Mon.\ Not.\ R.\ Astr.\ Soc.\ {\bf#1}}
\def\CQG#1{Class.\ Quantum Grav.\ {\bf#1}}
\def\GRG#1{Gen.\ Relativ.\ Grav.\ {\bf#1}}
\def\AA#1{Astron.\ Astrophys.\ {\bf#1}}
\def\figfi{\centerline{\scalebox{0.75}{\includegraphics{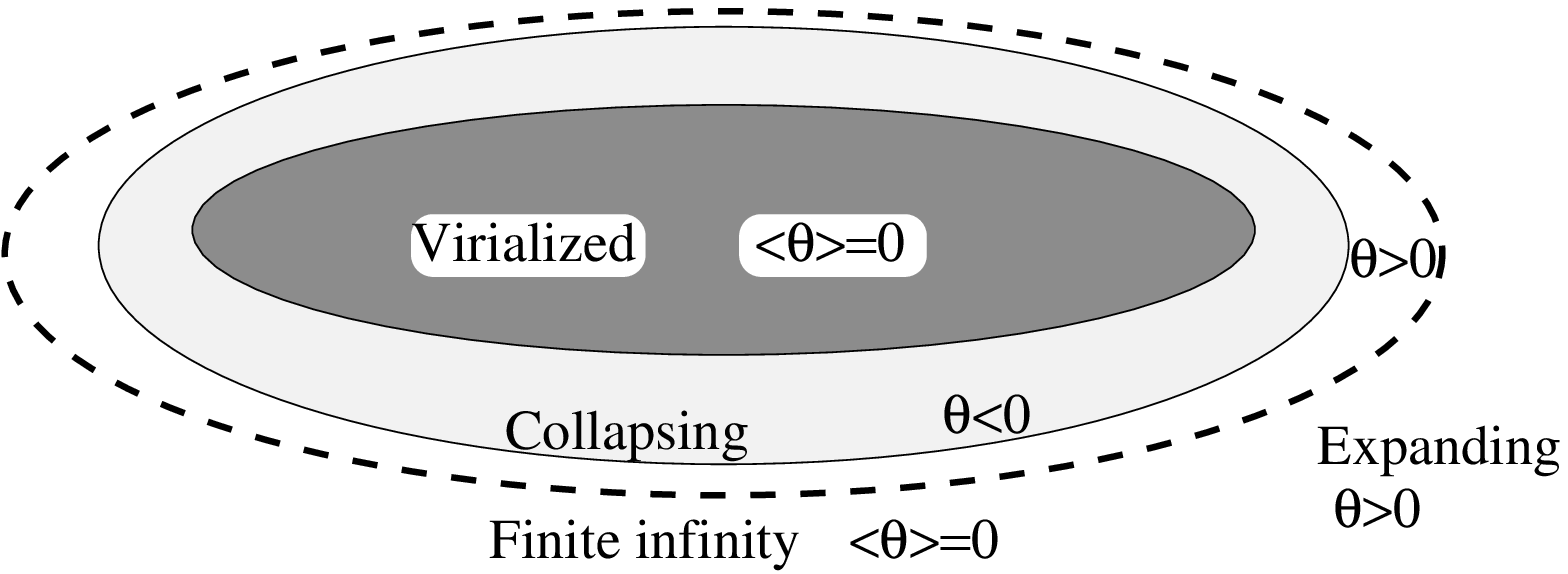}}}}
\def\figld{\centerline{\scalebox{0.82}{\includegraphics{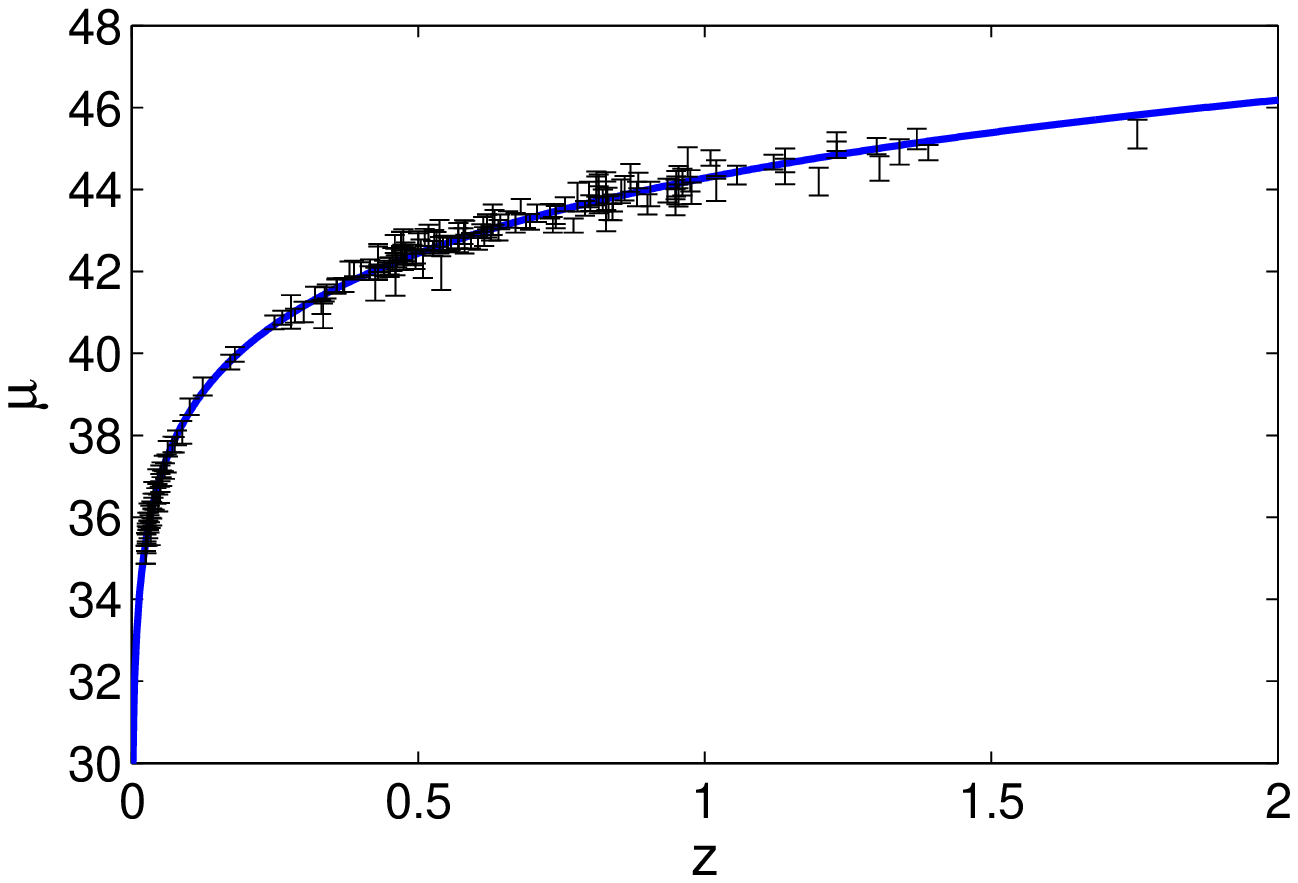}}}}
\def\figfvgb{\centerline{\scalebox{0.75}{\includegraphics{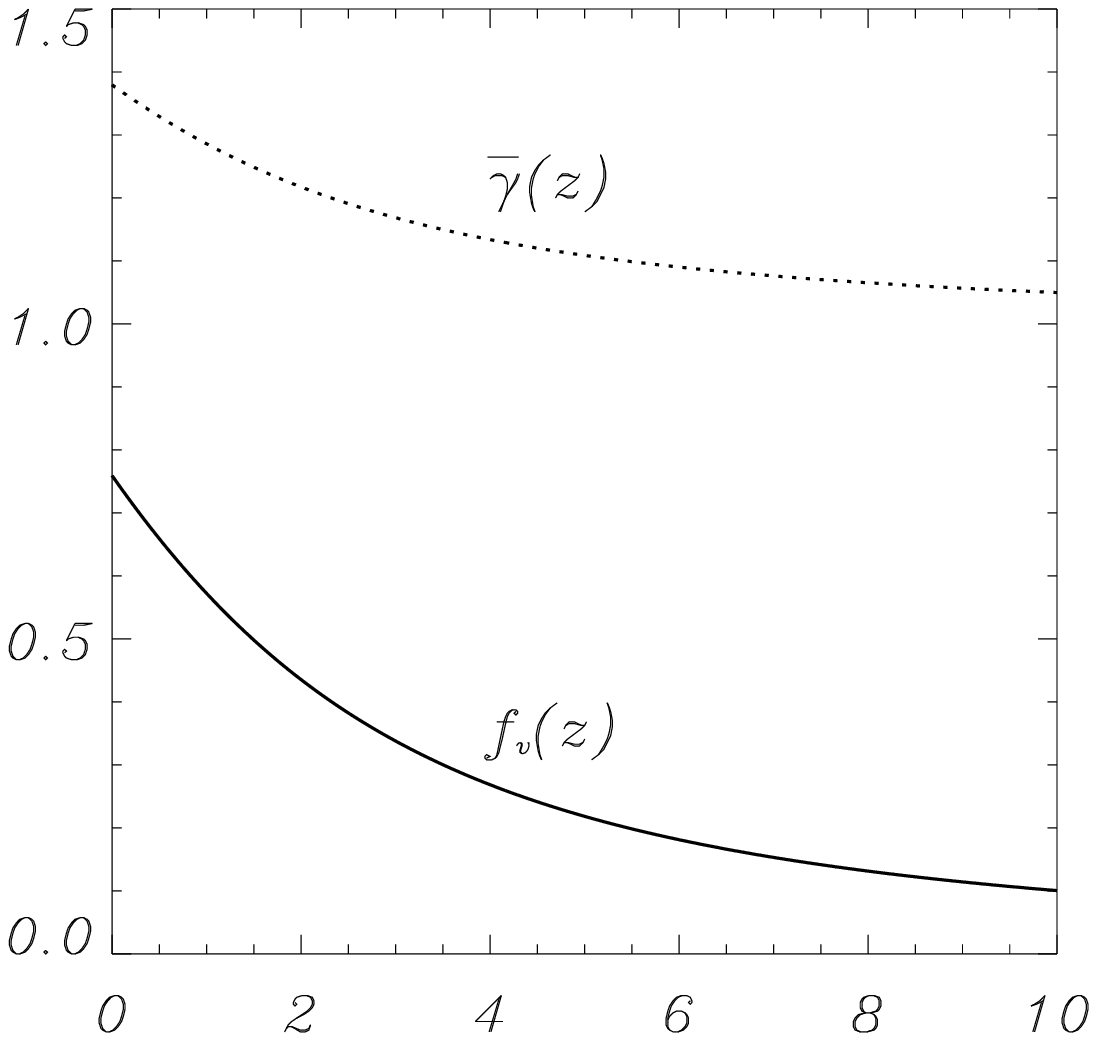}}}}
\def\figdelmu{\centerline{\scalebox{0.9}{\includegraphics{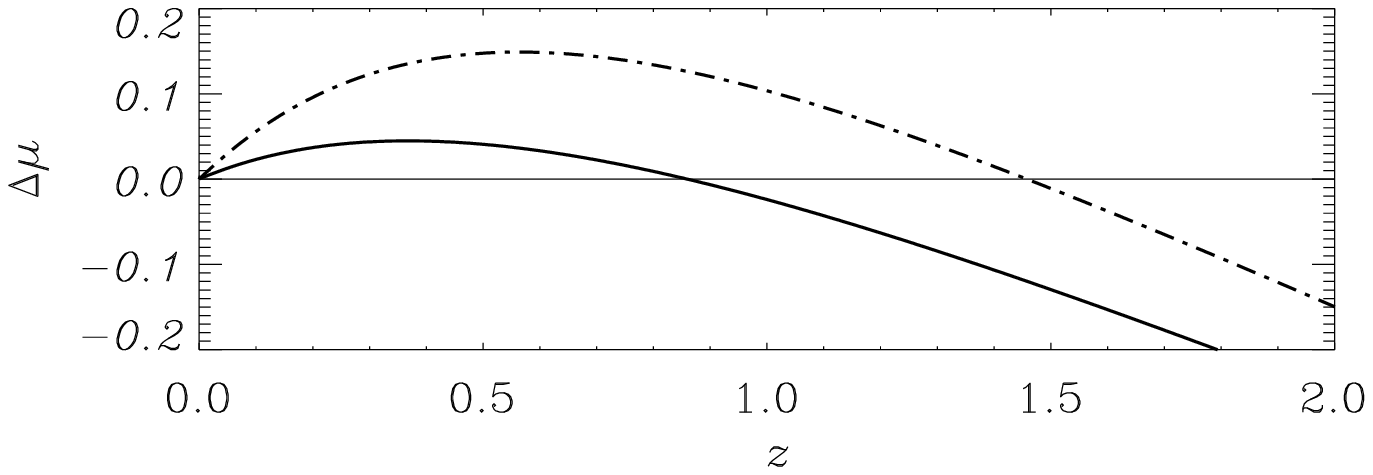}}}}
\def\figage{\centerline{\scalebox{0.8}{\includegraphics{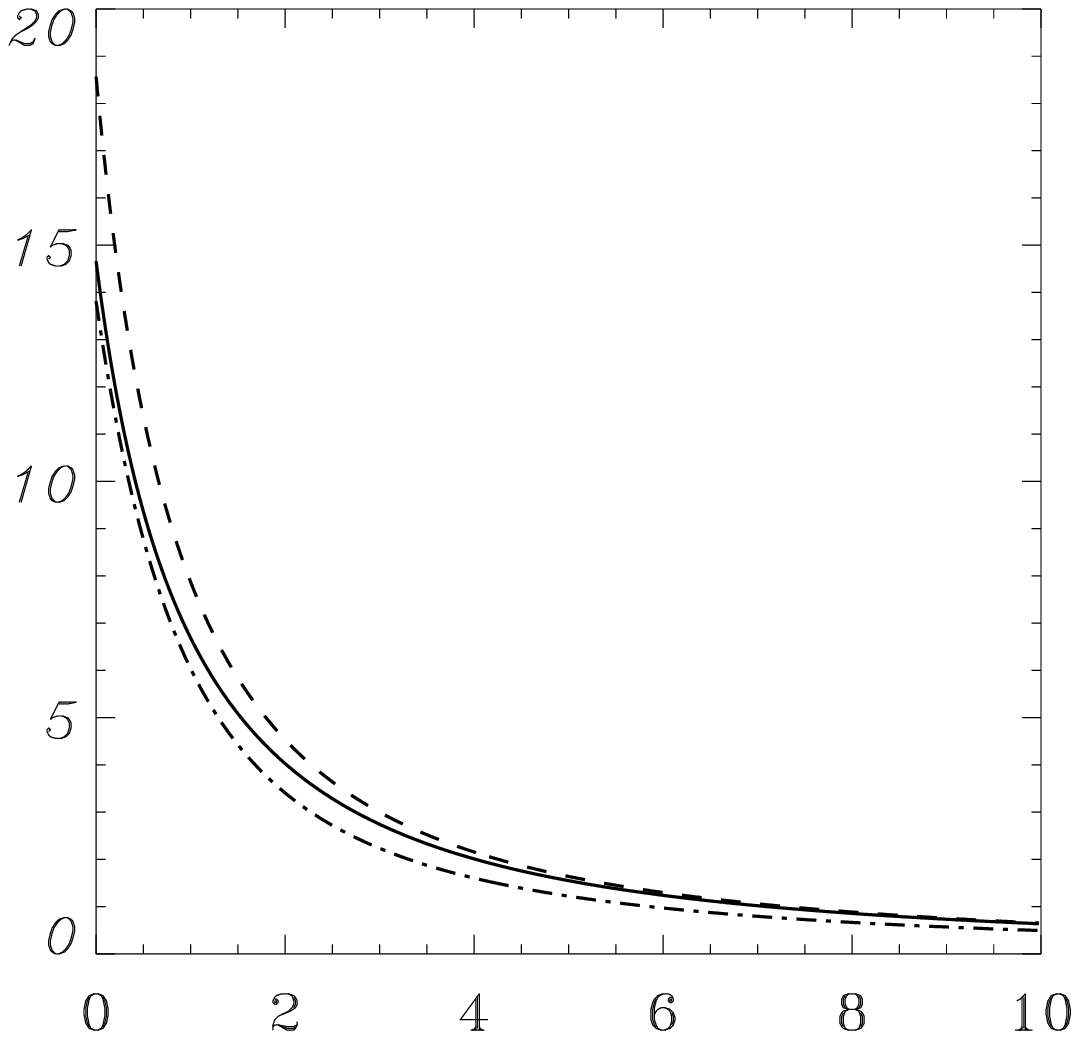}}}}
\title{Cosmic clocks, cosmic variance and cosmic averages}
\author{David L. Wiltshire}
\address{Department of Physics and Astronomy, University of Canterbury,
Private Bag 4800, Christchurch 8140, New Zealand}
\eads{\mailto{David.Wiltshire@canterbury.ac.nz}\\
http://www2.phys.canterbury.ac.nz/$\goesas$dlw24/}

\begin{abstract}
Cosmic acceleration is explained quantitatively, purely in
general relativity with matter obeying the strong energy condition, as an
{\em apparent} effect due to quasilocal gravitational energy differences
that arise in the decoupling of bound systems from the global expansion of
the universe. ``Dark energy'' is recognised as a misidentification of those
aspects of gravitational energy which by virtue of the equivalence principle
cannot be localised. Matter is modelled as an
inhomogeneous distribution of clusters of galaxies in bubble walls
surrounding voids, as we observe. Gravitational energy differences between
observers in bound systems, such as galaxies, and volume--averaged comoving
locations in freely expanding space can be so large that the time dilation
between the two significantly affects the parameters of any effective
homogeneous isotropic model one fits to the universe. A new approach to
cosmological averaging is presented, which implicitly solves the Sandage-de
Vaucouleurs paradox. Comoving
test particles in freely expanding space, which observe an isotropic cosmic
microwave background (CMB), possess a quasilocal ``rest'' energy
$E=\ave{\ga(\tc,\bx)} m c^2$ on the spatial hypersurfaces of
homogeneity. Here $1\le\ga<\frac32$: the lower bound refers to fiducial
reference observers at ``finite infinity'', which is defined technically in
relation to the demarcation scale between bound systems and expanding space.
Within voids $\ga>1$, representing the quasilocal gravitational energy of
expansion and spatial curvature variations. Since all our cosmological
measurements apart from the CMB involve photons exchanged between objects
in bound systems, and since clocks in bound systems are largely unaffected,
this is entirely consistent with observation. When combined with a non--linear
scheme for cosmological evolution with back--reaction via the
Buchert equations, a new observationally viable model of the universe
is obtained, without ``dark energy''. A quantitative scheme is presented for
the recalibration of average cosmological parameters. It uses boundary
conditions at the time of last scattering consistent with primordial
inflation. The expansion age is increased, allowing more time for structure
formation. The baryon density fraction obtained from primordial
nucleosynthesis bounds can be significantly larger, yet consistent with
primordial lithium abundance measurements. The angular scale of the first
Doppler peak in the CMB anisotropy spectrum fits the new model despite an
average negative spatial curvature at late epochs, resolving the anomaly
associated with ellipticity in the CMB anisotropies. Non--baryonic dark
matter to baryonic matter ratios of about 3:1 are typically favoured by
observational tests. A number of
other testable consequences are discussed, with the potential to profoundly
change the whole of theoretical and observational cosmology.
\end{abstract}
\maketitle

\section{Introduction}

It is a cornerstone of cosmology that the observed near exact
isotropy of the cosmic microwave background radiation (CMBR),
together with the assumption that our spatial location is not
special -- the {\em Copernican} or {\em Cosmological Principle} --
leads to the conclusion that, to a reliable degree of approximation,
we live in a homogeneous isotropic universe characterised by a
Friedmann--Lema\^{\i}tre--Robertson--Walker (FLRW) geometry.
However, although the matter distribution was certainly very
homogeneous at the epoch of last--scattering when the CMBR was
laid down, in the intervening aeons the matter distribution has become
very inhomogeneous through the growth of structure, and the problem of
fitting a smooth geometry to a universe with a lumpy matter distribution
\cite{fit1,fit2} is central to relating observations to the numerical values
of the averaged parameters which describe the Universe and its evolution
as a whole.

Our conventional interpretation of observations within the FLRW models
has yielded a standard model of cosmology in broad agreement with
observations. However, this model requires that most of the matter in
the universe is in forms of clumped non--baryonic dark matter and
smooth dark energy, the nature of which has been described by many
commentators as the greatest challenge to science. Furthermore,
even if a mysterious dark energy is accepted, the parameters which provide
the best fit to
observation also give a model with a number of puzzling anomalies. These
include: the apparent very early formation of galaxies \cite{red}; the low
power of the quadrupole in the spectrum of the CMB radiation anisotropies
and the unexplained alignments of low multipoles which theory says should be
random \cite{axis}; ellipticity in the CMB anisotropies consistent with
the geodesic mixing expected from average negative spatial curvature
\cite{elliptic2}; and the fact that the baryon to photon ratio
which best fits CMBR anisotropy data \cite{wmap} gives, via big bang
nucleosynthesis, a predicted lithium abundance at variance with what is
observed \cite{lithium}. Recent astronomical observations of one globular
cluster suggest a possible resolution of the lithium anomaly in terms of
stellar astrophysics \cite{lithium2}, but this remains to be confirmed by
modelling.

Another puzzle, serious enough to be described as a ``crisis'' \cite{P_void},
is that our present universe is dominated by
voids which are far emptier than predicted by models of structure formation.
The present clumped matter distribution is inhomogeneous, with an observed
hierarchical structure, the largest structures being clusters and
superclusters of galaxies bound in walls and filaments surrounding voids.
Some 40--50\% of the volume \cite{HV} of the universe at the present epoch
is in voids of order $30h^{-1}$Mpc in diameter, $h$ being the dimensionless
Hubble parameter, $\Hm=100h\kmsMpc$, and there is much evidence for voids
of 3--5 times this size \cite{Tom1}, as well as local voids on smaller scales
\cite{minivoids}. When
considering the volume fraction of voids on all scales, it appears that our
observable universe is ``void--dominated'' at the present epoch.

In this paper I propose that the resolution of various cosmological
anomalies is related to understanding why our universe is dominated by
voids, and to taking the correct average of such an inhomogeneous matter
distribution to obtain the smoothed Hubble flow. This involves careful
examination of the operational understanding of the relationship between
our measurements and the average FLRW geometry. I will do so by addressing
central issues which are ambiguous in general relativity and often ignored
in cosmology: the definition of gravitational energy, and the question of
what is meant by the expansion of space and its influence on bound systems.

While the definition of gravitational energy in general is difficult and
possibly unresolvable in general relativity on a completely arbitrary
background, I propose that
in model universes which are inhomogeneous but began from very close to an
homogeneous state with scale--invariant density perturbations, as ours did,
there is a clear physical answer operationally. The answer put forward here
has the consequence that all average cosmological parameters must be
recalibrated, as we have systematically ignored the variation in quasilocal
gravitational energy in the universe at late epochs in so far as it
affects the rest--energy of ideal comoving observers, and the synchronisation
of their clocks with respect to ours. Understanding this point may make it
possible to obtain a viable model of the universe, without a substantial
fraction of dark energy at the present epoch. Furthermore, the new
paradigm suggests a framework in which other cosmological puzzles could
conceivably be resolved, as shall be discussed.

There are two key elements to the new solution. The first is the fact that
in an inhomogeneous universe the appropriate averaged Einstein equations
which describe the dynamical evolution of the universe are not the Friedmann
equation, but modified equations obtained by a suitable non--linear averaging
scheme, for which a number of possible alternatives exist \cite{buch1,Zal1}.
Such schemes cannot succeed, however, unless the average quantities
are understood operationally rather than by convention. In this paper
I argue that the second key element to obtaining a viable cosmology
involves a careful examination of the physical meaning of the time
parameter of the averaging scheme and its operational relationship to the
proper time of observers in galaxies. Since time parameterisations in an
arbitrary inhomogeneous universe are also inhomogeneous, this is not a
trivial issue.

In this paper I will propose a new relationship between
our measurements and those of volume--averaged observers.
Since the proposal overturns a standard simplifying assumption that
has been made for 80 years, but is not demanded by general relativity or
observation, its intuition -- already briefly outlined in a limiting case
in a previous paper \cite{paper0} -- may not be obvious. The purpose of
this paper therefore is to provide an expository clarification of the
physical basis of the new solution to the fitting problem
at a conceptual level. Sufficient quantitative details and numerical
examples will be
provided to demonstrate that it is likely that a new concordance
cosmology can be found purely with matter obeying the strong energy
condition. However, the numerical derivation of best--fit
cosmological parameters is left to other papers \cite{paper1,paper2}.

Firstly let me briefly outline what is already well--known about cosmic
averages in an inhomogeneous universe. Conventionally we assume that the
universe is homogeneous and isotropic on a suitably large scale, take
smoothed averages of the curvature and energy--momentum and substitute
these in the Einstein equations, giving the standard Friedmann and
Raychaudhuri equations. However, since the Einstein equations are local
and non--linear, strictly speaking we should start with an inhomogeneous
curvature and energy--momentum, evolve these via the Einstein equations
and then take the average. This yields equations corrected by a
{\em back--reaction}.

There are many alternatives to averaging, depending on whether one chooses
to begin by averaging tensor quantities \cite{Zal1}, or scalars \cite{buch1},
for example, and whether one chooses spacelike volume averages or null cone
averages. As one example,
for irrotational dust cosmologies, characterised by an energy density,
$\rh(t,\bx)$, expansion, $\th(t,\bx)$, and shear, $\si(t,\bx)$, on a compact
domain, $\DD$, of a suitably defined spatial hypersurface of constant average
time, $t$, and spatial 3--metric, $\gd_{ij}(t,\bx)$, average cosmic evolution
in Buchert's scheme \cite{buch1} is described by the exact equations
\bea
3{\dot\ab^2\over\ab^2}&=&8\pi G\ave\rh-\half\Rav-\half\QQ,\label{buche1}\\
3{\ddot\ab\over\ab}&=&-4\pi G\ave\rh+\QQ,\label{buche2}\\
\pt_t\ave\rh&+&3{\dot\ab\over\ab}\ave\rh=0,
\label{buche3}\eea
where $\ab(t)\equiv\left[\Vav(t)/\Vav(t\Z0)\right]^{1/3}$ with
$\Vav(t)\equiv\int_\DD\dd^3x\sqrt{\det\gd}$,
overdot denotes a $t$--derivative,
\bea
\QQ&\equiv&\frac23\left\langle\left(\th-\langle\th\rangle\right)^2\right\rangle
-2\langle\si\rangle^2\nonumber\\
&=&\frac23\left(\langle\th^2\rangle-\langle\th\rangle^2\right)-
2\langle\si\rangle^2\,,
\label{backr}\eea
and angle brackets denote the spatial volume average of a quantity, so that
$\Rav\equiv\left(\int_\DD\dd^3x\sqrt{\det\gd}\,{\cal R}(t,\bx)\right)/\Vav(t)$
is the average spatial curvature, for example. The following integrability
condition follows from (\ref{buche1})--(\ref{backr}):
\beq \pt_t\left(\ab^6\QQ\right)+\ab^4\pt_t\left(\ab^2\Rav\right)=0.
\label{intQ}\eeq
The extent to which the back--reaction, $\QQ$, can lead to apparent cosmic
acceleration or not has been the subject of much debate
\cite{buch2}--\cite{IW}.

Despite the common misrepresentation in the literature that ``we
measure acceleration'' since we actually measure apparent magnitudes of
type Ia supernovae (SneIa), we are in reality determining luminosity {\em
distances} as a function of redshift. The deduction of acceleration requires
two time derivatives. The issue of how any time parameter, $t$, such as
that used in (\ref{buche1})--(\ref{intQ}), is related to our own clocks
must therefore be central to the debate about whether cosmic acceleration
can be obtained from back--reaction, and if so what its magnitude is.
While the relationship between any two average cosmic time parameters
can be expected to be monotonic on physical grounds, in the context of
inhomogeneous cosmology it cannot simply be assumed that any time parameter
we write down is necessarily the time on our own clocks. The operational
meaning of average cosmic time parameters is central and cannot be ignored.

The subject of inhomogeneous cosmology is a vast one \cite{Krasinski},
and in examining the question of back--reaction I shall make particular
choices. Firstly, while cosmological perturbation theory is highly
important near the epoch of last--scattering, I do not believe that
issues concerning back--reaction at the present epoch can be resolved
in the context of perturbation theory. While perturbative approaches have
naturally led to realisation of the significance of back--reaction
\cite{kolb2}, to account for ``74\% dark energy'' the effect of
back--reaction on the background of the universe would be so great that
a viable quantitative model is beyond the domain of applicability of
perturbation theory, and so these approaches will not be considered further
in this paper. Secondly, while exact inhomogeneous models such as the
spherically symmetric Lema\^{\i}tre--Tolman--Bondi (LTB) models
\cite{LT,Bondi} are immensely useful, both as exact models for isolated
systems in an expanding universe, or as toy models for understanding
fundamental concepts, they could only be applied to the universe as a whole
if one abandoned the Copernican Principle. I am interested in the problem
of obtaining the correct homogeneous average from the inhomogeneous
geometry within our present particle horizon volume, and I shall retain the
Copernican Principle. Thus I shall not further consider approaches based on
the exact LTB models \cite{LT,Bondi} or the exact Szekeres models
\cite{Szek}.

The outline of this paper is as follows: an outline of broad conceptual
issues and some important preliminary definitions, in particular the notions
of finite infinity and the true critical density are first given in
\S\ref{Cop} and \S\ref{infty}. The key physical ideas are then laid out.
In \S\ref{surfs} an implicit solution of the Sandage--de Vaucouleurs paradox
is given. In \S\ref{Have} and \S\ref{obs} a mathematical model is
presented, which may offer a simple viable model of the universe
that successfully describes observations from the time of last
scattering to the present epoch, without dark energy. The proof of
principle is demonstrated by a numerical example. In \S\ref{cmb}
specific quantities associated with the CMB and the early universe are
recalibrated, including the baryon--to--photon ratio, the sound horizon
and the angular scale associated with the first Doppler peak. It is
demonstrated that the broad features of the CMB anisotropy spectrum
may be fitted while simultaneously resolving a number of observational
anomalies. The scale of homogeneity is physically identified with the
``comoving'' baryon acoustic oscillation scale and consequences for variance of
the Hubble flow are discussed. In \S\ref{Newton}
the physical implications of the model on relatively small cosmological
scales are considered, and the potential for further cosmological tests
discussed. In \S\ref{inflation} arguments are presented as to why the
new model universe is a natural consequence of primordial inflation,
and broader questions concerning the
primordial perturbation spectrum are discussed. A concluding discussion
is presented in \S\ref{dis}, to which a reader without much time is referred
for a summary of the main results.

\section{The Copernican Principle\label{Cop}}

Much of the debate about inhomogeneous back--reaction has centred on
perturbation theory, and has overlooked the key point that in taking
averages we must be certain that we are correctly relating
observations to parameters of the average geometry. In particular,
the assumption that the background geometry is very close to an FLRW
geometry, with the metric
\beq\label{FLRW}
\dd\bar s^2 = - \dd\ts^2 +\ab^2(\ts)\dd\OM^2_{k}
\eeq
where $\dd\OM^2_{k}$ is the 3--metric of a space of constant curvature,
with $k=-1,0,+1$, is well justified by the Copernican Principle, almost
exact isotropy of the CMBR and the average isotropic Hubble flow.

In interpreting (\ref{FLRW}), it has been assumed since the early days of
relativistic cosmology that the time parameter $\ts$, is to a good
approximation the time on our own clocks. This simplifying assumption, made
at an early point in most cosmology texts, is not required by either theory,
principle or observation, as I shall now argue. Furthermore, it is not the
natural choice in a far--from--equilibrium universe with the inhomogeneous
structure that grows from an initially almost scale--free spectrum of
density perturbations, as one would expect from primordial inflation.

By the Copernican Principle if we reside at an ``average point'' on
a spatial hypersurface then observers at other ``average points''
also measure a near to isotropic CMBR, apart from a dipole anisotropy
of order $10^{-3}$ of the mean CMBR temperature due to small peculiar
velocities. However, taking our
clocks rates to be very nearly the same as the clock rates at other
average points on an appropriate spatial hypersurface,
involves an implicit assumption over--and--above that implied by
the Copernican Principle. In particular, the na\"{\i}ve identification
of surfaces of homogeneity with surfaces of synchronicity implies that there
is a single class of average ``comoving'' observers who measure an almost
identical mean CMBR temperature. This need not be the case.

In a universe of voids and galaxies confined to bubble walls there are in
fact two classes of average points to consider: (1) the mass--averaged
observers residing in bound systems, typically in galaxies, where space is
{\em not} expanding; and (2) the volume--averaged observers in freely
expanding space, typically in voids which occupy the largest volume of space.
Both classes of observers can measure an isotropic CMB in accord with the
Copernican Principle, while measuring a {\em different mean temperature}
and a {\em different angular anisotropy scale}.

Imagine a static geometry such as that of a system of an infinite number of
equidistant black holes of equal electric charges and masses held in
equilibrium by mutual gravitational attraction and Coulomb repulsion. On a
large scale, this geometry is effectively a homogeneous landscape, whether
viewed from deep in a gravitational well close to a mass source or in an
almost asymptotic region equidistant from the closest sources. By
gravitational time dilation, however, clock rates at the two vantage points
differ, despite the overall homogeneity. In an expanding universe, the
situation is more subtle, but actually quite similar. In particular, total
gravitational energy is very important in inhomogeneous universes -- a point
which was realised long ago by Bondi \cite{Bondi} in his study of the
spherically symmetric LTB models \cite{LT}.
In freely expanding situations, {\em quasilocal}
gravitational energy differences can also be significant.

We reside in a gravitationally bound system: locally space has not
been expanding for over 10 billion years. This has the consequence that
measurements by our local clocks relate parametrically to solutions of the
geodesic equations in the Schwarzschild geometry centred on our sun, with
a time parameter related to what is effectively almost a Killing vector,
$\pt/\pt\tc$. To match our clock rate, $\tc$, which has been effectively
frozen in for billions of years to the clock rate of the idealised
comoving observers at {\em volume--averaged points} of the average geometry
(\ref{FLRW}) involves matching geometry from the scale of stars, to galaxies,
to galaxy clusters, to bubble walls, and ultimately a homogeneous scale.
Until one solves this fitting problem \cite{fit1,fit2} the assumption that our
clock rate, $\tc$, closely matches the parameter, $t$, of (\ref{FLRW}) is
an ansatz.

The typical justification for assuming the conventional clock ansatz to be
reliable comes from considerations such as that of particle motion in the
Schwarzschild and Kerr geometries. In the Schwarzschild geometry, a
radially moving clock of rest mass, $m$, with locally measured velocity
$\be c$ on a geodesic possesses a conserved energy of
\beq E=(1-\rs/r)\gamma mc^2,\qquad\gamma=(1-\be^2)^{-1/2},\label{bind}\eeq
as follows directly from Killing's
equations. Given values of the ratio of the Schwarzschild radius, $\rs$, of
galaxies to their radius, $r$, and similar calculations for objects in
closed bound orbits, the effect of gravitational binding energy appears to be
negligible\footnote{More rigorous discussions of gravitational binding
energy for a perfect fluid source in a quasilocal framework in stationary
spacetimes have been given by Katz, Lynden-Bell and Bi\v{c}\'ak \cite{KLB}.},
and is only significant in the vicinity of highly compact objects such as
neutron stars and black holes. However, considerations with regard to
gravitational binding energy do not translate directly to the quasilocal
kinetic energy of expansion and gravitational energy associated with spatial
curvature variations, as I shall argue further in the next section.

The conventional clock ansatz was verified as being consistent by Einstein
and Straus \cite{gruyere} in the Swiss cheese model in which spheres are
excised from a dust FLRW model and replaced by point sources of mass equal
to that of the excised dust in the holes\footnote{The clock rate for an
observer in the Einstein-Straus vacuole differs from that of the comoving
clock \cite{schu}. However, the differences turn out to be observationally
negligible, as has been discussed in detail by Harwit \cite{H}. I thank a
referee for bringing this work to my attention.}.
Such a model would be accurate if the universe did actually consist of
isolated galaxies moving uniformly in the ``coins on the balloon'' analogy
described in undergraduate texts. However,
the observed void / bubble wall structure indicates that the actual universe is
quite different from this hypothetical situation. In fact, the difference is
important enough to be a phenomenological puzzle. This puzzle, the
{\em Sandage--de Vaucouleurs} paradox\footnote{My nomenclature derives from
the fact that this problem was originally raised by Sandage and collaborators
\cite{STH} in objection to de Vaucouleurs' hierarchical cosmology \cite{deVau}
before the evidence for the void structure of the universe was as good as it
is now. In the literature it sometimes called the ``Hubble--de Vaucouleurs
paradox'' \cite{paradox1,bary} and sometimes the ``Hubble--Sandage paradox''
\cite{paradox2}.}, arises since we expect that the statistical scatter in
peculiar velocities of galaxies as a fraction of their ``recession velocity''
should be large until the scale of homogeneity is approached. In fact,
on the scale of 20 Mpc -- of order 10\% of the scale of homogeneity -- the
scatter ought to be so large that no linear Hubble flow should be derivable,
statistically speaking. Yet, 20 Mpc is the local scale over which Hubble
originally obtained his famous linear law. By conventional understanding this
does not make sense.

\section{Where is infinity?\label{infty}}

It is well--known that the definition of gravitational energy in general
relativity is difficult, since space itself carries energy and momentum.
The dynamical nature of space in general relativity is well illustrated by
the phenomenon of frame--dragging in the Kerr geometry, whereby an infalling
test particle initially on a radial geodesic will rotate more and more with
respect to spatial infinity, the closer it gets to the source of the
geometry. The best attempts at quasilocal definitions of energy in general
relativity involve integrals over 2--surfaces \cite{quasi_rev}--\cite{quasi2},
but have not been widely applied in cosmology.

It is intrinsic to the physical assumptions of general relativity that an
expanding universe must possess some sort of gravitational {\em kinetic
energy of expansion}; yet this is an issue which is not considered much beyond
loosely identifying the l.h.s.\ of eq.~(\ref{buche1}) with such a term when
we consider the Friedmann equation (with $\QQ=0$). Our understanding of such
a kinetic energy is based largely on Newtonian thinking, and indeed in the
ad hoc Newtonian derivation of the Friedmann equation, the correspondence
is exact, while the total energy is associated with the spatial curvature
term up to an overall sign. To make progress, we must extend these concepts
beyond the Newtonian limit.

Our best understanding of gravitational energy is for exact solutions where
space is assumed to be asymptotically flat or (anti)-de Sitter. Since the
universe is not asymptotically flat, we need an understanding of
gravitational energy that extends beyond bound regions. In his pioneering work
on the fitting problem \cite{fit1}, Ellis introduced the notion of
{\em finite infinity}, ``{\em fi}'' ,
as being a timelike surface within which the dynamics of an isolated
system such as the solar system can be treated without reference to the
rest of the universe. No system is truly isolated since we continually
receive electromagnetic and gravitational radiation from distant parts
of the universe. However, this radiation is so small that the solution
to the problem of geodesic motion within the isolated system can be
considered without reference to the rest of the universe. Ellis' suggestion
provides a notional answer to the question of: ``Where is infinity?''.
Within finite infinity a solution might be considered to be almost
asymptotically flat, and governed by ``almost'' Killing vectors.

I now propose to modify Ellis' suggestion slightly and will identify finite
infinity in terms of average expansion and its relationship to the
{\em true critical density}.

\subsection{The true critical density}

It is a direct consequence of averaging in an inhomogeneous universe
that the average internal energy density of a dust universe that we would
measure at the present epoch
is not the same as the time evolution of the average density that we would
have measured at some time in the past. In terms of Buchert's scheme,
the non--commutativity of averaging and time evolution is described by
the exact relation \cite{buch1}
\beq\Deriv{\dd}t{}\ave\Psi-\ave{\Deriv{\dd}t\Psi}=\ave{\Psi\th}
-\ave\th\ave\Psi\label{comm}\eeq
for any scalar, $\Psi$, such as the internal energy density, $\rh$.

In Buchert's scheme it is natural to rewrite (\ref{buche1}) as
\beq
\OmDm+\OmDR+\OmDQ=1,\label{eqn0}
\eeq
where
\beq
\OmDm\equiv{8\pi G\ave\rh\over3\bH^2},\qquad
\OmDR\equiv-{\ave{\cal R}\over6\bH^2},\qquad
\OmDQ\equiv-{{\cal Q}\over6\bH^2},
\label{undress1}\eeq
and $\bH\equiv\dot\ab/\ab$. It should be noted that density parameters
are considered as fractions of the region--dependent quantity
$3\bH^2/(8\pi G)$, which does not, however, play the role of a critical
density in delineating the critical case between a ``closed'' and an ``open''
universe.

If we fit an FLRW model to the averaged geometry then there are further
issues associated with matching volumes of a homogeneous isotropic model,
which Buchert and Carfora have described by the slogan \cite{BC}:
``Cosmological parameters are dressed''. Quantities such as $\OmDm$ must still
be corrected by volume factors to get the equivalent ``dressed'' parameters
of the equivalent FLRW model, which according to an earlier estimate of
Hellaby \cite{Hellaby} can give corrections of 10--30\%. Once again, however,
the dressed parameters are still region dependent quantities, and one does
not have the notion of a critical density. So is there a notion of critical
density, and if so where is it to be found?

I will now make the following crucial physical observation. By the evidence
of the CMB, the universe at last scattering was very close to being truly
homogeneous and isotropic. Therefore
an operational definition of critical density does exist, provided we assume
the Copernican Principle and accept that the universe was globally smooth
at that epoch and not just in our present past horizon volume. This might
also be viewed as a direct consequence of primordial inflation, as is
further discussed in \S\ref{inflation}.

At the epoch of last scattering, $t_i$, the Hubble expansion was uniform, as
the local velocity perturbations were tiny. Given a uniform initial expansion
rate there must have existed a uniform critical density of matter required
for gravity to be able to eventually bring that expansion to zero. This
critical density, $\rhcr(t_i)$, therefore sets {\em a universal
scale} which delineates the boundary between density perturbations which
will become bound, as opposed to density perturbations which are unbound.

This may seem a trivial point. However, when one considers
an inhomogeneous evolution it is clear that the average Hubble parameter
on a given domain does not correspond to the time evolved critical density,
whether in a dressed or undressed form. The na\"{\i}ve use of the Friedmann
equation in cosmology to date means that we could well be making a gross
error in choice of background in structure formation studies. In
particular, we estimate the critical density by extrapolating back in time
using our present Hubble parameter, $\Hm$, assuming the evolution of the
universe is smooth and that $3\Hm^2/(8\pi G)$ is the critical density at the
present epoch, when it is not.
If the total density of the universe is very close to one, this has the
effect that we can {\em mis-estimate} the background density of the universe
at early epochs where structure formation boundary conditions are set.
We are in effect perturbing about the wrong background by implicitly
assuming that the average density at the present epoch, as determined by
the recent past within our past light cone, is identical to the universe
as a whole. By cosmic variance there is no reason to expect this to be
the case.

\subsection{Location of finite infinity}

We will define {\em finite infinity} as the timelike boundaries which
demarcate the boundary between bound -- or more strictly {\em potentially}
bound -- systems and unbound systems, assuming that the observable universe is
at present void--dominated. At the technical level we will
propose the following working definition:
\begin{itemize}
\item[] With respect to a foliation of spacetime by spacelike
hypersurfaces, finite infinity is identified with the set of {\em timelike
boundaries} of (disjoint) compact domains, $\FI$, within which the
{\em average} expansion vanishes, while being positive outside:
\beq\vbox{\halign{#\hfil\ &#\hfil\cr(i)&$\ave{\th(p)}\Z{\Fi}=0$;\cr
(ii)&$\exists$ $\DD\Z I$ such that $\FI\subset\DD\Z I$ and
$\th(p)>0\quad\forall$ $p\in\DD\Z I\!\smallsetminus\FI$\cr}}\label{ddi}\eeq
\end{itemize}
\noi The index, $I$, is taken to run over the disjoint domains. Thus
finite infinity, {\em fi} $\equiv\cup\Z I\pt\FI$. Finite infinity becomes
operationally defined only once collapsing regions form, and each $\FI$ is
centred on a region which is initially collapsing. Furthermore, the average
is a volume average over the {\em smallest} scales on which (\ref{ddi})
applies. Each $\DD\Z I$ must be contained within the particle horizon volume
at any epoch.
\begin{figure}[htb]
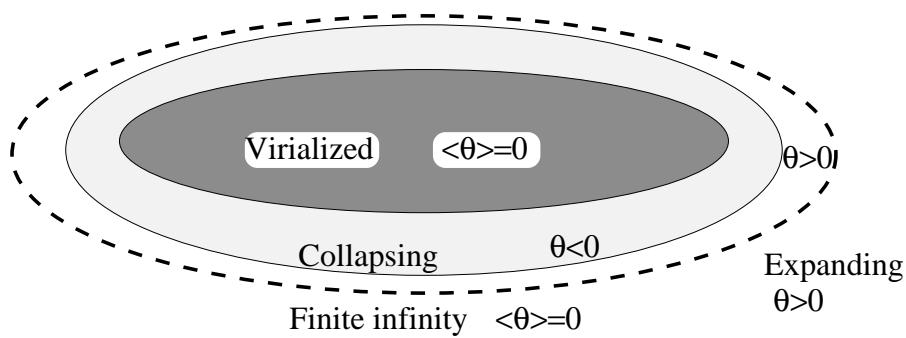

\vbox{\figfi
\caption{\label{fig_fi}%
{\sl A schematic illustration of the notion of finite infinity, {\em fi}:
the boundary (dashed line) to a region with average zero expansion inside,
and positive expansion outside. It may or may not contain collapsing
regions.}}}
\end{figure}

Some comments are in order. At first sight, it may seem more
natural to identify finite infinity with the set of timelike
boundaries of (disjoint) compact domains, $\FI$, outside which the
local expansion is positive, rather than using an average.
However, we must bear in mind that pressure,
shear and vorticity are neglected here. Vorticity certainly becomes
important when regions collapse, and our present study is within the
context of an averaging scheme in which the vorticity is neglected.
Therefore it seems that we can only talk about {\em averaged} expansion.
The specification of the spatial hypersurfaces on which domains are
averaged is deferred until the next section.

We also note that while each $\FI$ is centred on an initially collapsing
region, to obtain an averaged zero expansion one must extend $\FI$
outwards to connected regions with positive expansion which will ultimately
cease expanding. There is an intrinsic ambiguity in assigning the shape
to the boundary; though there are probably ways to do this in terms of
a surface of minimal proper distance to regions with $\th(t,\bx)\le0$
on the hypersurfaces in question. Thus the {\em zero expansion boundaries},
at which $\th(t,\bx\Z0)=0$, lie within the finite infinity regions. The
zero expansion boundary may be thought of as the instantaneous
``tipping point'' which separates regions where space is not expanding
from regions where it is expanding. Finite infinity represents a boundary
to the region within which space would {\em ultimately} stop expanding if its
entire future evolution were determined from the local dynamics of
the $\FI$ region alone; a local, rather than a global statement.

The idea behind our definition -- which is the crucial property we wish
to preserve should the technical definition (\ref{ddi}) require further
refinement -- is that the true critical density may now be {\em defined} by
\beq
\rhcr(\tc)=\ave{\rh(\tc,\bx)}\Z{\Fi}
\label{truecr}\eeq
for each $\FI$ defined by (\ref{ddi}). Our definition encompasses more
situations than envisaged in the original suggestion of Ellis \cite{fit1},
since it is our aim that it should apply as soon as regions start collapsing.
Since galaxies and clusters of galaxies are still growing by matter infall
at the present epoch, it would not be reasonable to try to define finite
infinity for groups of galaxies as a static boundary at a fixed proper
distance from the barycentre of a mass concentration. Thus while Ellis
may have had something closer to the zero expansion surfaces in mind,
our version has greater utility as it will correspond to a surface where
space is locally expanding as long as the observed universe is, and the
true critically density will be directly related to the local expansion,
$\th\Z{\Fi}=3\bH$, at finite infinity. Here local expansion refers to a
small region straddling the finite infinity boundary in Fig.~\ref{fig_fi},
within which $\th>0$.

Physically finite infinity would have to be located well outside the
concentrated visible mass and invisible halo mass of galaxies, and clusters
of galaxies. Essentially one needs to include enough low density regions
surrounding the dense cores of bound structures in order to get
an average density which corresponds to the present true critical density.
As long as bound structures are growing the proper distance from the
barycentre of some $\FI$ to its finite infinity boundary will increase.
The rate of increase of this distance will be commensurate with the locally
measured Hubble flow, $\bH$, when the growth of density contrasts is in the
linear regime but will differ in general, particularly at late epochs.

By our definition the regions $\FI$ should be seen as analogues of
the spheres cut out in the Einstein--Straus solution, which also
possess a non--static boundary. The principal
differences are: (i) finite infinity contains an average rather than exact
geometry; (ii) no assumptions about homogeneity are made beyond
finite infinity. Beyond finite infinity we find regions below the
true critical density, which are non--uniformly distributed. In particular,
we expect a non--uniform situation in which the expansion within the
filamentary structures of the bubble walls is different to the expansion
in the voids, if referred to a single set of clocks.

\section{Gravitational energy and the definition of homogeneity\label{surfs}}

Since a broadly isotropic Hubble flow is observed, it is clear that an
average sense of homogeneity must exist despite the observed large
inhomogeneities in the matter distribution at the present epoch. Reconciling
these two facts -- which find their quantitative expression in the
Sandage--de Vaucouleurs paradox -- one must face what critics of inhomogeneous
back--reaction have described as the ``seemingly impossible burden of
explaining why the universe appears to be so well described by a model
that has only very small departures from a FLRW metric'' \cite{IW}.

This section contains the essential conceptual arguments of my proposal.
The ``seemingly impossible burden'' can be resolved not by departing from an
average FLRW geometry, albeit one that evolves on average with the inclusion of
back--reaction, but by a careful reinterpretation of the relationship between
the parameters of the average homogeneous geometry and our own observations.
My proposal will simultaneously clarify the role of gravitational energy in
the averaged cosmology, and resolve the Sandage--de Vaucouleurs paradox,
issues which are fundamental but otherwise unanswered.

\subsection{The locally synchronous gauge}
To set the scene, let us carefully examine the gauge choices that have
been made in Buchert's averaging scheme, and the interpretation of the
parameters. For a globally hyperbolic manifold, we can make a standard
$3+1$-split
\beq\ds^2=-\Bom^0\otimes\Bom^0+g_{ij}(\tc,\bx)\,\Bom^i\otimes\Bom^j,
\label{split}\eeq
where
\bea\Bom^0&=&\ga(\tc,\bx)\,\dd\tc,\nonumber\\
\Bom^i&=&\dd x^i+\be^i(\tc,\bx)\,\dd\tc.\label{shift}
\eea
Here $\ga(\tc,x^k)$ is the {\em lapse function}, which measures the difference
between coordinate time, $\tc$, and proper time, $t$, on curves normal to
hypersurfaces $\Sigma_\tc$, the unit normal being $n_\al=(-\ga,0,0,0)$, or
$n^\al=\ga^{-1}(1,\be^i)$. Also,
$\be^i(\tc,x^k)$ is the {\em shift vector}, which measures the difference
between a worldline with fixed spatial coordinate, and the point reached by
following the worldline along the normal $\mathbf n$ from one hypersurface
to the next.

On account of the Bianchi identities, the Einstein equations include
the Hamiltonian and momentum constraints. As a consequence, four out of the
ten algebraically independent components of the metric
(\ref{split})--(\ref{shift}) can always be removed locally by gauge choices
at the dynamical level. The choice made by Buchert is to assume that
\begin{itemize}
\item[(i)] $\be^i(\tc,\bx)=0$, which is to say that the coordinates are
{\em comoving};
\item[(ii)] $\ga(\tc,\bx)=1$, which is to say that if the unit normal, $\B n$,
is assumed to be the 4--velocity then $\tc$ is the local proper
time\footnote{The choice $\ga=1$ is usually part and parcel of what are
termed ``comoving coordinates''. However, strictly speaking $\ga=1$ is an
independent gauge choice from $\be^i=0$.}.
\end{itemize}

Buchert shows that these choices can always be made in irrotational dust
cosmologies \cite{buch1}. In the presence of a perfect fluid with non--zero
pressure, the choice $n^\al n_\al=-1$ does not generally lead to
$\ga(\tc,\bx)=1$ \cite{buch3}. However, our concern here is primarily with
dust cosmologies since the universe can be assumed homogeneous and isotropic
before last scattering at the level of accuracy we require. We will be
averaging over scales larger than the domains bounded by finite infinity,
and it will be assumed that at those scales pressures, shear and vorticity
can be neglected.

The physical implications of Buchert's gauge choices need to be examined.
The choice that $\be^i(\tc,\bx)=0$ can be understood as the statement that
the momentum flux within space is neglected at the level of averaging.
In particular, spatial variations in vorticity, i.e., the angular momentum
content of space, are neglected, as are non--expanding regions as compared to
expanding regions. The choice that $\ga(\tc,\bx)=1$ can be understood as
the statement that variations in the gravitational energy of space can
also be neglected at the level of averaging.

It is clear that neither of these gauge choices is appropriate if one actually
wishes to consider the worldlines in collapsing or virialised regions as
physically distinct from worldlines in expanding regions. Since timelike
geodesics converge in collapsing regions comoving coordinates cannot be
adopted globally, but only in an averaged sense. With respect
to comoving observers in expanding regions, collapsing regions should
be physically distinguishable in terms of a momentum flux which differs
from the average expansion within the
spatial hypersurfaces. Nonetheless, provided the averaging scale is
much larger than collapsing or virialised regions then the choice $\be^i=0$ is
a reasonable one provided all measurements are referred to observers
in expanding regions. Furthermore, since angular momentum perturbations decay
in the linear regime, neglecting vorticity is reasonable. It is only
after overdense regions break away from the Hubble flow that rotation cannot
be neglected. As long as we are averaging over suitably large scales, the
effect of vorticity will be a correction, but a small one. The magnitude of
its correction might be judged from any net circulation of galaxy clusters
around voids.

The choice that $\ga(\tc,\bx)=1$ is less reasonable as a physical assumption,
since density contrasts grow even in the linear regime when one is very
close to matter homogeneity, suggesting possible large variations in
gravitational energy. However, provided all measurements are referred to the
same single class of average observers in freely expanding space, then of
course Buchert's choice $\ga(\tc,\bx)=1$ is consistent.

The crucial issue, however, is that all observers in bound systems --
such as ourselves and all other galaxies -- do not reside in regions of
locally expanding space. Therefore they do not follow {\em average} geodesics
tangent to the average fluid 4--velocity, $\bn^\mu$. While Buchert's scheme
may be consistent if applied to volume--averaged clocks, variations in
gravitational energy and its effect on physical clocks {\em cannot} be
neglected once back--reaction becomes completely non--linear. The average
surfaces of homogeneity are not necessarily surfaces of synchronicity
of local proper time.

To make this point more transparent we must recall that particle interactions
in relativity involve conservation of 4--momentum, not 4--velocity. Similarly,
Einstein's equations involve the energy--momentum tensor.
Although a dust cosmology has an energy--momentum tensor
\beq T^{\mu\nu}=\rh\, \bn^\mu\bn^\nu\label{stress}\eeq
this merely represents the {\em internal energy} of the fluid. In cosmology
the kinetic energy of the expansion of space is contained in the l.h.s.\ of
Einstein's equations. This is an inevitable consequence of the fact that on
account of the equivalence principle there is no local definition of
gravitational energy, but at best quasilocal definitions (see, e.g.,
\cite{quasi_rev}--\cite{quasi2}). The kinetic energy content of an expanding
space involves motion and cannot be localised, and so cannot be expressed
in the internal form appropriate to an energy--momentum tensor; instead it
is contained in the Einstein tensor. Similarly spatial curvature cannot
be measured at a point but only by geodesic deviation. Thus variations in
spatial curvature between galaxies and voids are encoded in variations
of the total quasilocal gravitational energy within the Einstein tensor.
Since the walls containing galaxies are assumed to be near
to spatially flat, with large negative spatial curvatures only pertaining to
voids, the quasilocal curvature variations can well be the dominant
contribution to gravitational energy differences.

By the principles of general relativity, the physical effect of gravitational
energy on local clocks must nonetheless be real, even if it cannot be
localised, except in an averaged sense over a region. Furthermore, if
local spatial curvature differs systematically between bound systems and
the volume average, then we cannot na\"{\i}vely assume that our measured
angular positions of the Doppler peaks in the spectrum of CMB anisotropies
will be the same as those at the volume average. As soon as one
is dealing with a genuinely inhomogeneous cosmology then the operational
understanding of our measurements in relation to those of the average
geometry of the universe is a nontrivial question which must be confronted.

\subsection{The quasilocally uniformly expanding gauge\label{quasiexp}}

I propose that the ultimate rigorous understanding of the relationship
between local measurements in bound systems and measurements at an average
location in freely expanding space, will be through an extension of quasilocal
formulations \cite{KLB},\cite{quasi_rev}--\cite{quasi2} to cosmology. Such an
approach has yet to be developed; what is outlined here is an attempt
to come to grips with its essential physical principles. As such the
proposal is not yet complete. I will write down a quantitatively viable model
universe in \S\ref{Have}, while leaving numerous issues to be further
refined.

One general feature of many quasilocal approaches is the subtraction of an
energy--momentum integral of a fiducial geometry from an energy--momentum
integral of some more general system -- usually a bound system. Typically
spatial infinity in an asymptotically flat spacetime provides the fiducial
geometry. There one has a notion of static or
zero--angular momentum observers, with tangent
vectors appropriately defined in terms of timelike Killing vectors. This leads
to an appropriate notion of gravitational binding energy \cite{KLB}.
The idea here, is that in the actual universe similar observers are still
the relevant class for defining the reference
point with respect to which we must quantify the gravitational energy
of expansion and global spatial curvature variations relative to the bound
systems within which we live. We no longer have exact Killing vectors, but
approximate Killing vectors within virialised regions, and spatial
infinity in asymptotically flat geometries is replaced by finite infinity,
defined in terms of the scale set by the true critical density.

It is usually assumed implicitly that stars and black holes in distant galaxies
are described in terms of asymptotically flat geometries with a stationary
Killing vector, $\xi^\mu$, with clock rates synchronised to our own
despite the fact that the intervening space is expanding and
not described by timelike Killing vectors. The issue of how this is to be
achieved mathematically is generally not addressed beyond the
approximations used in the Swiss cheese model. I will begin from the premise
that the specification of finite infinity -- a direct physical consequence
of the initial uniform expansion afforded by primordial inflation -- does
allow a means of ``asymptotically'' synchronising the clocks corresponding
to approximate Killing vectors in disjoint bound systems, even when
clocks in the freely expanding space between such regions are not
synchronised but vary in a way which reflects the underlying gravitational
energy variations.

The global vector field, $\xi^\mu=\left(\pt\over\pt\tc\right)^\mu$, of the
geometry (\ref{split})--(\ref{shift}) is assumed to be an ``almost'' Killing
vector within finite infinity regions, but to differ from the local normal
to the spatial hypersurface in general. In the absence of exact
asymptotic flatness, it is normalised according to
\beq
\ave{-\xi^\mu n_\mu}\Z{\Fi}=\ave{\ga(\tc,\bx)}\Z{\Fi}\equiv\ga\Z{\Fi}(\tc)
=1
\eeq
for all values of {\em wall time}, $\tc$. Within the virialised regions
within finite infinity it is assumed that the average geometry is
approximately a flat Minkowski geometry. The timelike direction of this
geometry coincides with the ``almost'' stationary Killing vector $\xi^\mu$.
To this extent the time variation of $\ga$ can be neglected, and we can
define static observers near mass concentrations within the virialised
regions in the usual sense as observers at fixed spatial coordinates with
4--velocity $W^\mu=\xi^\mu/(-\xi^\nu\xi_\nu)^{1/2}=\ga(\bx)^{-1}\xi^\mu$ and
4--acceleration $a_\mu=\nabla_\mu\ln\ga(\bx)$.

Outside finite infinity in freely expanding space it is assumed that
when the geometry (\ref{split})--(\ref{shift}) is averaged over regions
within voids, in which vorticity, shear and pressure can be neglected,
the shift vector then vanishes at the level of averaging,
$\be^i=0$, while for any spatial domain, $\DD\Z I$, entirely contained within
a void
$$\ga\Z{\DD\X I}(\tc)\equiv\ave{-\xi^\mu n_\mu}\Z{\DD\X I}>1.$$
The fact that the lapse function is generally greater than one is a consequence
of the positive gravitational energy associated with negative spatial
curvature. The value of $\ga\Z{\DD\X I}$ thus obtained is region--dependent.
If we consider a sequence of disjoint spatial regions of equal proper volume
within a void then we expect that
$$1= \ga\Z{\Fi}(\tc)<\ga\Z{\DD\X 1}(\tc)<\ga\Z{\DD\X 2}(\tc)<\cdots
<\ga\Z{\DD\X C}(\tc),$$
i.e., that $\ga\Z{\DD\X I}$ is progressively greater for averaging
regions closer and closer to the void centre in region $\DD\Z C$.
Effectively, with respect to finite infinity observers, a comoving particle
of rest mass, $m$, in freely expanding space possesses a quasilocal
``rest'' energy $E=\ave{\ga(\tc,\bx)}mc^2$, which increases towards a
void centre.

For cosmological scales we are interested in the combined average
over all such spatial scales contained within our present
particle horizon volume, $\cal H$. We will use an overbar to indicate this
``global'' average of the lapse function\footnote{One must be careful to
distinguish this averaged expression from the case of tilted cosmologies in
models with perfect fluids with non--zero pressure \cite{KE}. There one has a
similar non--averaged relation, $U^\mu n_\mu=-\ga$, which arises, however,
as a result of a {\em local} boost of the unit normal, $n^\mu$,
relative to the local fluid 4-velocity, $U^\mu$, which has
nontrivial 4-acceleration.}:
\beq \gb(\tc)\equiv\ave{\ga(\tc,\bx)}\Z{\cal H}=-\xi^\mu\bn_\mu
=\ave{-\xi^\mu n_\mu}\Z{\cal H}
\label{gab}\eeq
The volume--average geometry obtained by averaging
(\ref{split}), (\ref{shift}) over the horizon volume is then
\beq
\ds^2=-\gb^2(\tc)\,\dd\tc^2+\mean g_{ij}(\tc,\bx)\,\dd x^i\dd x^j\,,
\label{geom}\eeq
where the average of the spatial metric, $\mean g_{ij}$, remains to be
discussed.

We note that as is conventional the energy--momentum tensor is given by
(\ref{stress}), where the 4--velocity $\bn^\mu=\Deriv{\dd}{\ts}{x^\mu}$ is
written in terms of local proper time, $t$, on geodesics of dust
``particles'' which actually represent averaging regions which are small
enough that the variation of gravitational energy is insignificant.
The important physical distinction is that the time parameter $\tc$ of wall
observers in typical galaxies is not locally defined on a volume--average
``dust geodesic''. Equivalently, whereas the average dual normals, $\bn^\mu$,
may be considered to be tangents to the comoving average
dust geodesics with affine parameter $t$; they are geodesics with
non--affine parameter, $\tc$, if the geodesic equation is rewritten in terms
of $\Deriv{\dd}{\tc}{x^\mu}$. Since $\bn^\mu=\gb(\tc)^{-1}\xi^\mu$, there
is a sense in which $\bn^\mu$ may be considered to be the volume--averaged
extension of the vector field, $W^\mu$, associated with a static observer
within finite infinity.

In establishing a {\em non--locally synchronous gauge} set by the clocks
at finite infinity, we still have one gauge freedom remaining, as
the condition $\ave{\ga(\tc,\bx)}\Z{\Fi}=1$ merely fixes a reference
normalisation of clocks, which might equally have been chosen in
reference to observers at void centres, if that was where typical observers
were located.
To complete the identification of surfaces of average homogeneity,
we will therefore choose the gauge by the condition that when averaged over
regions, $\DD$, of freely expanding space, the quasilocally measured expansion
is a uniform function of local proper time, independent of spatial location:
\beq
\Deriv\dd t{\ell_r(t)}=v_r(t)
\eeq
where $\ell_r\equiv\Vav^{1/3}=\left[\int_{\DD\X I}\dd^3x\sqrt{\gd}\right]
^{1/3}$, for each averaging region $\DD\Z I$.
We can rephrase this in terms of a conventional Hubble parameter by choosing
the averaging regions to have equal fiducial proper volumes, even though their
spatial curvature will typically vary. With this understanding, an
equivalent statement is
\beq{1\over\ell_r(t)}\Deriv\dd t{\ell_r(t)}
=\frac13\ave\th\Z{\DD\X1}=\frac13\ave\th\Z{\DD\X2}=\cdots=\bH(t),
\label{homo}\eeq
$t$ being the local proper time in each region. It must be assumed that
each averaging region, $\DD\Z I$, is freely expanding
at its boundary. Any region, $\FI$, bounded by finite infinity must be
entirely encompassed within a $\DD\Z I$: $\FI\subset\DD\Z I$, $\forall\,I$.
No averaging region need necessarily contain a finite infinity -- in
particular void regions will correspond to regions $\DD\Z J$, with
$\th(t,\bx)>0$ throughout the region as long as the universe is expanding
as a whole -- which is assumed throughout this paper.

Eq.\ (\ref{homo}) can be understood as the physical statement that while
the locally measured proper times, $t$, and local spatial curvature may both
vary over the hypersurfaces of average homogeneity -- the {\em quasilocally
measured} average expansion is homogeneous. This provides an {\em implicit}
resolution of the Sandage--de Vaucouleurs paradox: a universe which appears
to have large voids growing more rapidly than environments around virialised
and collapsing regions, when measured by one set of clocks, can nonetheless
have an almost uniform quasilocal proper expansion.
Operationally, this is possible if we demand that
ideal comoving observers measure an isotropic CMB with no peculiar velocity
dipole, while making no demands on the synchronisation of their clocks.
This is consistent with non--uniformity in quasilocal gravitational
energy, which translates to non--uniformity in local measurements of the
mean CMB temperature and local average spatial curvature, for which we have
only one data point.

The proposed uniform proper expansion gauge arises by the assumption that on
average the differential increase in local proper volume between voids and
filaments is accompanied by a differential increase in the relative
positive gravitational energy associated with negative spatial curvature,
which feeds back on relative clock rates. We are talking about a circumstance
in which total energy is assumed to be conserved\footnote{One often sees
statements to the effect that in cosmology ``energy is not conserved'' in a
comoving volume in the presence of pressure, e.g., for a comoving photon
gas. (See ref.\ \cite{bary} for an overview.) What is true is that
{\em internal energy} is not conserved in such situations. However, since
gravitational energy cannot be localised, one should not talk about
energy conservation without a suitable quasilocal formulation.}. As the
universe decelerates much of the kinetic energy of expansion is converted to
different forms: thermal energy and localised forms of mechanical energy,
such as rotational energy, within bound systems where most of the matter is
located, and to the gravitational energy associated with negative spatial
curvature within voids. There is of course also a difference in the kinetic
energy of the expansion of space between bound systems, where this term
vanishes, and the voids where it makes a residual contribution.

To have a more direct physical understanding of the proposal, we need to
specify the actual scales relevant to our observed portion of the universe.
I will identify these scales as the bubble walls -- namely the filamentary
structures containing galaxy clusters -- and voids which the bubble
walls surround. Finite infinity domains are contained well inside bubble
walls, where the locally measured expansion\footnote{There is always
an inherent ambiguity in use of the word ``local'' until one specifies a
scale. In general relativity one strictly uses ``local'' for a measurement
at a point, and other measurements are ``quasilocal''. However, quasilocal
measurements can also be made over a great variety of scales. Since
the terminology of ``local'' Hubble flow is conventionally used, we will
assume this applies to the smallest quasilocal scales over which a linear
Hubble law can be extracted, and hope that this use of the word ``local''
does not cause confusion as compared to our earlier stricter usage.}
defines a local Hubble flow
$$
\bH\ns{w}(\tw)=\frac13\ave\th\ns{w}\equiv{1\over\aw}\Deriv\dd\tw\aw\,.
$$
The parameter $\tw$ coincides with the coordinate $\tc$ of (\ref{geom}).
We temporarily (until the end of section \S\ref{eqns}) add the subscript
``w'' to distinguish it in what follows from the time parameter $\tv$ that
results from normalising $\ga$ to be unity in dominant void centres.

In principle, the local Hubble flow could be measured by placing spacecraft
with laser--ranging devices near finite infinity. In practice, we must make
measurements from galaxies contained within zero expansion surfaces, which lie
within the finite infinity regions. Operationally, we need to infer $\bH$ from
the expansion between galaxies which have no peculiar velocity with respect to
the cosmic rest frame, within two or more disjoint finite infinity
regions for which the separation of the finite infinity boundaries is very
small. Since average galaxies such as ours do have small peculiar
velocities with respect to the cosmic rest frame, as determined by the
dipole anisotropy, in practice one would need to average over many finite
infinity regions to extract the local Hubble parameter.
\setcounter{footnote}3

Likewise, within the dominant voids the locally measured expansion defines a
local Hubble flow
$$
\bH\ns{v}(\tv)=\frac13\ave\th\ns{v}\equiv{1\over\av}\Deriv\dd\tv\av\,.
$$
Since observers are found in bound systems, not in voids, a cosmological
test to determine $\bH\ns{v}$ is a considerable challenge, as it requires
direct access to the clocks that tick at a rate, $\tv$. Since
voids appear to exist on all scales from a dominant volume fraction with
diameters $30\,h^{-1}$Mpc up to larger scales, and down to minivoids with
diameters of order $1$Mpc \cite{minivoids}, the time parameter $\tv$ must refer
to the average extreme clock rate in the centres of the {\em dominant} voids,
as measured by volume fraction. Within the filamentary bubble walls there
are many minivoids, but they contribute to the expansion of the bubble
walls, and the amount by which their clock rates vary will be smaller
as compared to the dominant void scales.

On account of the differences in quasilocal gravitational energy,
\beq \Deriv\dd\tv\tw\ne1\,,\eeq
so even though $\bH\ns{v}=\bH\ns{w}$, as soon as we try to construct
a present horizon volume average over both walls and voids, if we refer the
average to a single set of clocks, this ``global average''\footnote{It is
only global with respect to the entire present particle horizon volume,
$\cal H$, which though representing the entire presently observable universe,
is not the entire universe.} will differ from $\bH$.

To an observer sitting inside a bound galaxy, within an $\FI$ region,
within a filamentary wall, it will appear that the Hubble rate determined
from galaxies on the far side of a large local void is somewhat greater than
the Hubble rate within her wall. However, if she accounted for the fact that
due to quasilocal gravitational energy differences the comoving clocks
within the voids are ticking faster than her own clocks, the different Hubble
rates become uniform to first approximation. Void ``clocks'' are not
observed directly -- rather the photons in distant galaxies originate
within disjoint finite infinity regions. Thus different apparent Hubble
rates are determined by differing relative path--integrals along the
null geodesics.

It must be noted that a
larger apparent Hubble rate across voids as opposed to within bubble
walls, is the exact opposite of Newtonian intuition, where one thinks
of an exactly smooth Hubble flow, with local anisotropies being
a consequence of peculiar velocities induced by mass concentrations,
with larger Hubble rates expected between an observer and the
directions of largest mass concentrations. I remark that in the present
instance I am considering variations in the underlying Hubble flow,
with peculiar velocities averaged out. This is consistent with the notion
that deceleration is more rapid in regions of greater density.
The Newtonian logic, which does not
appear to give a very consistent picture of actual motion in the
local volume \cite{Whiting1,Whiting2}, needs to be revised -- a
point to which I will return in \S\ref{Newton}.

It may come as a surprise that the effect I am proposing can be large,
since when thinking about the effects of gravitational energy on clocks
we are most familiar with bound systems, where the effects of binding
energy give only very small differences between a galactic system and
spatial infinity in an asymptotically flat spacetime, similarly to
the small differences expected from (\ref{bind}). Similarly, when we
consider the integrated Sachs--Wolfe effect, we find it to be small
because we consider ourselves to be at the volume average and calculate
the effect of small perturbations about that average. The approach here
is different because rather than considering differential effects about
the volume average, I am considering the position of the observer to be
of crucial importance. Since galaxies are bound systems and are {\em not}
at the volume--average, there are potentially nontrivial
aspects of general relativity to be taken into account.

In particular, the dynamics
of expanding space is intrinsically different to the dynamics of
non--expanding space, and there is no inherent reason to expect
that when measured with respect to wall clocks that $\ga(\tc,\bx)$ should
be only infinitesimally larger than unity in the voids. There have been
10 billion years or more for the clock rates to slowly diverge. In the
absence of a dark energy component the only absolute upper bound on
the difference in clock rates is that within voids $\ga(\tc,\bx)<\frac32$,
which represents the ratio of the local expansion rate of an empty Milne
universe region to an Einstein--de Sitter one.

In short, I am proposing a physical understanding of the expansion of space
which attempts to clarify various misconceptions that have plagued
efforts to understand this notion. I suggest that within bound systems
space really does not expand at all, just as simple model calculations in
Newtonian gravity \cite{Price} and in general relativity \cite{gruyere} have
usually suggested. However, when space does expand
its effect on particles ``at rest'' should be locally indistinguishable from
equivalent motion of particles in a static space. At a foundational
level this can be understood in terms of the equivalence principle,
as I shall discuss in a separate paper \cite{equiv}. It demands that in
considering widely separated particles we must account for the quasilocal
gravitational energy variations resulting from the differential
kinetic energy of the expanding universe and variations in its spatial
curvature. While the average of each of these quantities appears in
the Friedmann equation, their {\em variation} cannot be
incorporated in the energy--momentum tensor by virtue of the equivalence
principle. Efforts to reparameterise inhomogeneous back--reaction
as a smooth dark energy field \cite{morpho} can therefore never be entirely
quantitatively successful, as they can never account for the fact
that our measurements differ systematically from those made at the volume
average.

On account of a lack of conceptual clarity and the seductive charm of the
very simple FLRW models with which we can perform successful calculations
while avoiding fundamental issues, we have come to a historical situation in
which we misidentify quasilocal cosmological
gravitational energy with ``dark energy''. As bound system observers who
perform observations on other bound systems, which are in regions of
locally non--expanding space, this circumstance is an unfortunate
consequence of an observer selection effect, and failing to account for
the fact that mass and volume averages can differ drastically. If nature
had provided us with observable freely falling clocks in the depths of
voids where space is locally expanding and negatively curved then, if my
thesis is correct, observations of such clocks could well have saved us
one or more decades of work in the progress of theoretical cosmology.

\section{A viable model for the observable universe\label{Have}}

The simplest nontrivial approximation is a two--scale model, in which
there are two homogeneous average local cosmic times, $\tw$ and $\tv$,
at finite infinity boundaries within the walls, and at dominant void
centres respectively. Since a true critical density exists the evolution
of finite infinity boundaries gives a notion of true cosmic time, $\tw$.
However, this parameter differs in general from the volume--averaged comoving
proper time parameter on surfaces of average homogeneity. A suitable
``global average'' Hubble parameter is then given by the volume weighted
average over all walls and voids within the particle horizon volume, $\cal
H$. The two--scale model was first introduced in ref.\ \cite{paper0},
where the wall and void scales were considered to evolve independently
and the only concern was to ascertain an order of magnitude for the
effect of the clock rate variations between walls and voids in determining
the overall cosmological parameters. The two--scale model will now
be generalised to include the coupling of the two scales via a
back--reaction term.

Ultimately we should reformulate an averaging scheme formally from explicit
regional averages involving quasilocal energy integrals. For the purposes of
the two--scale approximation, however, it should be possible to retain
Buchert's formalism, while recognising that the time parameter, $t$, would
represent the proper time at a volume average position in freely expanding
space. We should be able to obtain the broad features of a viable model
universe by solving Buchert's equations, but being careful to relate our
own clocks to the volume--averaged ones, and any other quantities which
result from such relations. This may be seen as analogous to using the
synchronous gauge in perturbation theory while taking care about identifying
physical observables \cite{PV}, even if calculations could be alternatively
performed in a more physical gauge, the ``uniform Hubble--constant gauge''
\cite{Bard} being the closest to our choice in \S\ref{quasiexp}.

Since our clocks, and all clocks within bound systems, are
expected to differ little from those at finite infinity, we will neglect
any small differences and take $\tw$ to represent the time on our clocks,
though ultimately this may also require some correction.

\subsection{Average expansion and kinematic back--reaction}
We construct the two scale model by assuming that on spatial hypersurface
of average homogeneity, our present particle horizon volume
can be represented as the disjoint union of regions corresponding to bubble
walls containing bound systems, and of regions corresponding to voids. In
this fashion, the entire present horizon volume is given by $\Vav=\Vav\ns i
\ab^3$, where
\beq\ab^3=\fwi\aw^3+\fvi\av^3\label{volav}\eeq
Here $\fvi$ is an initial fraction of the total volume in void regions,
and $\fwi=1-\fvi$ an initial fraction of the total volume in wall regions.
In other words the volumes of the void and wall regions are respectively
$\Vav\ns v=\Vav\ns{vi}\av^3$ and $\Vav\ns w=\Vav\ns{wi}\aw^3$, while
$\fvi=\Vav\ns{vi}/\Vav\ns i$ and $\fwi=\Vav\ns{wi}/\Vav\ns i$ .
The size of the averaging domains will be taken to be such that the
term representing the bubble walls is a union of domains containing finite
infinity regions, $\FF\Z I$, which can be assumed to be {\em spatially flat}.
The averaging regions in the void term contain no finite infinity regions,
and are assumed to have negative spatial curvature on average.

We will employ Buchert's scheme with vorticity, pressure and shear
assumed to be negligible on the scale of averaging. In accord with
(\ref{homo})
\beq\bH(t)=\gw{{1\over\aw}\Deriv\dd t\aw}=\gv{{1\over{a\ns v}}
\Deriv\dd t{a\ns v}}\,,\label{homo2}\eeq
where
\beq\gw=\Deriv\dd\tw t,\qquad\w{and}\qquad\gv=\Deriv\dd\tv t.\label{clocks1}
\eeq
The quantity $\gw\equiv\gb(\tc)$ is defined by (\ref{gab}); we have
temporarily added the subscript ``w'' to distinguish, $\gw$, the
volume--average of $\ga$ with respect to wall clocks, from $\gv$, the
volume--average of $\ga$ with respect to clocks at dominant void centres.

It is convenient to refer the expansion in the voids and in the walls
to one set of clocks. Since Buchert's scheme is given in terms of a
volume--average time parameter, $t$, over both of these scales, that is
the convenient parameter to use. Thus we define
\beq\Hw\equiv{1\over\aw}\Deriv\dd t\aw,\qquad\w{and}\qquad
\Hv\equiv{1\over\av}\Deriv\dd t\av\,.\label{homo3}\eeq
We emphasise that neither $\Hw$ nor $\Hv$ coincides with the ``local''
Hubble parameter any longer in the respective wall and void locations,
$\bH\ns{w}$ and $\bH\ns{v}$, which
are both equal to $\bH$ on account of (\ref{homo}) or (\ref{homo2}). In
fact, we will now have $\Hw<\bH<\Hv$. Since $\bH$ is the locally
measured Hubble parameter at the horizon volume average location, by
(\ref{volav})
\beq\bH=\frac13\ave\th\Z{{\cal H}}=\fw\Hw+\fv\Hv,\label{Hback}\eeq
where $\fw(t)=\fwi\aw^3/\ab^3$ is the {\em wall volume fraction},
$\fv(t)=\fvi{a\ns v}^3/\ab^3$ is the {\em void volume fraction},
and $f\ns w(t)+\fv(t)=1$.

It is convenient to define the relative
expansion rate
\beq\hr(t)\equiv{\Hw\over\Hv}<1.\eeq
Then $\bH=\gw\Hw=\gv\Hv$, where
\beq\gw=1+{(1-\hr)f_v\over\hr},\label{clocks2}\eeq
and $\gv=\hr\gw$. Also, by direct computation of the derivative of
$\fvi{a\ns v}^3/\ab^3$,
\bea \dot\fv=-\dot\fw&=&3(1-\fv)\gwf\bH\nonumber\\
&=&{3\fv(1-\fv)(1-\hr)\bH\over\hr+(1-\hr)\fv}\label{fdot}
\eea
and many equivalent versions of the same relation exist in terms of
$\fw$ or $\gv$.
Combining (\ref{Hback}) and (\ref{fdot}) with
\beq\ave{\th^2}\Z{{\cal H}}=9\left(\fw\Hw^2+\fv\Hv^2\right)\eeq
and (\ref{backr}) we find that in the absence of shear the kinematic
back--reaction is given by
\bea\QQ&=&6\fv(1-\fv)\left(\Hv-\Hw\right)^2\nonumber\\
&=&{6\fv(1-\fv)(1-\hr)^2\bH^2\over\left[\hr+(1-\hr)\fv\right]^2}\label{Q1}
\eea

R\"as\"anen \cite{Ras} has also considered a 2--scale model in Buchert's
scheme\footnote{Since R\"as\"anen \cite{Ras} considers his model to be a
toy model only, he pictures the two scales as two disjoint simply connected
regions, and is just considering the effects of averaging over the two
regions without considering their embedding in the universe as a whole.
It should be stressed that here the two scales are summed over many different
disjoint regions. The manner in which these disjoint regions are embedded
in the universe as a whole arises by the specification of finite infinity.}
in terms of a single time parameter and has written down expressions for the
average Hubble expansion (\ref{Hback}) and the average deceleration
incorporating a kinematic back--reaction term in a form equivalent to
(\ref{Q1}). Aside from the key fact that the physical interpretation
developed in \S\ref{Cop} and in eqs.\ (\ref{homo2}),
(\ref{clocks1}), (\ref{clocks2}) has not been pursued by others, another
difference in R\"as\"anen's model is that he considers
a combination of voids and collapsing regions for the two scales, modelling
the latter by the spherical collapse model, rather than by directly solving
Buchert's equations. Since the spherical collapse model only remains a good
approximation until the overdense spherical
perturbations reach their maximum size, and since this would have occurred of
order 10 Gyr ago or more for galaxies, R\"as\"anen's model would appear to be
an interesting toy model which exhibits the effects of apparent acceleration,
but which cannot be applied to the observed universe as a whole at late
epochs\footnote{R\"as\"anen claims his approximation is relevant at epochs
when the universe is 10 Gyr old \cite{Ras}, but fails to identify what the
collapsing regions are observationally. It would appear that such regions
should be necessarily larger than the comoving scale on which the local Hubble
flow is still observed at the present epoch \cite{Whiting2}. While our local
flow could be atypical of wall regions, given the actual structure of
voids and filaments that we observe, a {\em spherical dust} approximation for
large collapsing regions at late epochs seems hard to justify.}.

R\"as\"anen overlooks the possibility, we introduce here, of taking one
of the scales to represent spatially flat average regions which enclose the
collapsing regions, and which will have a close to Einstein--de Sitter
time--dependence. This
may be due to his statement that in a universe whose expansion
goes smoothly from Einstein--de Sitter behaviour ($\ab\propto t^{2/3}$)
to empty universe behaviour ($\ab\propto t$), as will be the case in the
model here, there can be {\em``no acceleration as the variance of the
expansion rate is too small''} \cite{Ras}.
This claim will be disproved in eq.\ (\ref{qm}) below. The claim is
true insofar as it applies to volume--averaged observers in freely
expanding space. However, the crucial point which to the best of my knowledge
appears to have been completely overlooked by others in the study of
back--reaction and averaging is that
in an inhomogeneous universe writing down a timelike parameter, $t$,
does not imbue it with physical meaning until one specifies how this
parameter is {\em operationally} related to our own clocks. That is the issue
I confront: it will turn out that
the back--reaction is indeed too small to register as apparent
acceleration in terms of measurements made at a volume--average
location in voids. However, when translated to measurements synchronous
with the wall time, $\tw$, which is the parameter most closely related to
our own clocks, then {\em apparent acceleration can be obtained}.

\subsection{The dynamical equations and observational interpretation%
\label{eqns}}
To implement Buchert's equations we note that (\ref{buche3}) is solved
by taking $\ave\rh\Z{{\cal H}}={\rh\Z0}\ab\Z0/\ab^3$, as in the standard
Friedmann equation, while the average curvature includes a contribution from
the void regions diluted by the larger combined volume of walls and voids:
\beq
\Rav\Z{{\cal H}}={6\kv\fv\over\av^2}+{6\kw\fw\over\aw^2}
={6\kv\fvi^{2/3}\fv^{1/3}\over\ab^2}\,,\label{kav}
\eeq
where we have assumed that the average curvatures, $\Rav\ns v=6\kv/\av^2$
and $\Rav\ns w=6\kw/\aw^2$ in the voids and walls respectively, and that
$\kw=0$ in the final step of (\ref{kav}). Furthermore, the
kinematic back--reaction (\ref{Q1}) may also be written as
\beq
\QQ={2\dot\fv^2\over3\fv(1-\fv)}\,.\label{Q2}
\eeq
The independent Buchert equations
(\ref{buche1}) and (\ref{intQ}) are then found to reduce to
\bea
{\dot\ab^2\over\ab^2}+{\dot\fv^2\over9\fv(1-\fv)}-{\al^2\fv^{1/3}
\over\ab^2}={8\pi G\over3}{\rhb\Z0}{\ab\Z0^3\over\ab^3}\,,&&\label{eqn1}\\
\ddot\fv+{\dot\fv^2(2\fv-1)\over2\fv(1-\fv)}+3{\dot\ab\over\ab}\dot\fv
-{3\al^2\fv^{1/3}(1-\fv)\over2\ab^2}&=&0\,,\label{eqn2}
\eea
if $\fv(t)\ne$ const. Here we have assumed that the spatial curvature of the
void regions is negative, and have thus defined $\al^2=-\kv\fvi^{2/3}$.

To derive cosmological parameters requires more than simply solving the
coupled system (\ref{eqn1}), (\ref{eqn2}). Firstly, we observed that
the time derivative in these equations is $\ts$, the time parameter at a
volume--averaged location, which will be in voids. Having solved the
differential equations, we must convert all quantities involving time
to expressions in terms of wall time, $\tw$, which is very close to the time
we measure. Furthermore, the average spatial curvature at the volume
average is negative, whereas the locally measured spatial curvature in
wall regions is almost zero on average. These differences between the
rods and clocks of wall observers and volume--average observers will
both contribute to the ``dressing'' of cosmological parameters.

The interpretation of the solutions to (\ref{eqn1}), (\ref{eqn2}) requires
particular care in relation to our own measurements. By the construction
of finite infinity, when the full geometry (\ref{split})--(\ref{shift}) is
averaged over an expanding region immediately containing finite infinity,
since $\ave{-\xi^\mu n_\mu}\Z{\Fi}=1$ and since the average spatial curvature
within a finite infinity region is close to zero, it follows that the
local average geometry near a finite infinity boundary is
\bea\ds^2\Z{\Fi}&=&-\dd\tw^2+\aw^2(\tw)\left[\dd\etw^2+
\etw^2\dOM\right]\nonumber\\ &=&-\dd\tw^2+{\fvf^{2/3}\ab^2\over\fwi^{2/3}}
\left[\dd\etw^2+\etw^2\dOM\right].
\label{figeom}\eea
where $\dOM=\dd\vartheta^2+\sin^2\vartheta\,\dd\varphi^2$ is the standard
metric on a 2--sphere, and the second line follows on account of (\ref{volav}).
At a void centre, the average of the local geometry
(\ref{split})--(\ref{shift}) over a small domain is similarly given by
an effective homogeneous isotropic geometry specified by the void time, $\tv$,
and the local volume scale factor $\av$,
\bea\ds^2\Z{\DD\X C}&=&-\dd\tv^2+\av^2(\tv)\left[\dd\etv^2+\sinh^2(\etv)
\dOM\right]\nonumber\\ &=&-\dd\tv^2+{\fv^{2/3}\ab^2\over\fvi^{2/3}}
\left[\dd\etv^2+\sinh^2(\etv)\dOM\right].
\label{vogeom}\eea

The volume average scale factor, $\ab$, of
the geometry (\ref{avgeom}) has been constructed as the average of
these two geometries via (\ref{volav}). If we take a common
centre coinciding with that of a finite infinity region,
the volume average geometry may be considered to have the form
\bea
\ds^2&=&-\dd\ts^2+\ab^2(\ts)\,\dd\etb^2+A(\etb,\ts)\,\dOM,\nonumber\\
&=&-\gb^2(\tc)\dd\tc^2+\ab^2(\tc)\,\dd\etb^2+A(\etb,\tc)\,\dOM,
\label{avgeom}
\eea
where we have dropped the subscript ``w'' from both $\tw$ and $\gw$.
Since we will no longer make explicit reference to the time measured in
void centres, we will assume that $\tc$ and $\gb$ now refer to wall time in
all subsequent expressions.
We also note that in (\ref{avgeom}) the area quantity, $A(\etb,\ts)$, satisfies
$\int^{\etb\X{\cal H}}_0\dd\etb\, A(\etb,\ts)=\ab^2(\ts)\Vav\ns{i}
(\etb\Z{\cal H})/(4\pi)$, $\etb\Z{\cal H}$ being the conformal distance to
the particle horizon relative to an observer at $\etb=0$, since we
have chosen the particle horizon as the scale of averaging.

It must be emphasised that (\ref{avgeom}) is not locally isometric to the
geometry in either the walls or the void centres, and thus it is
{\em not} a metric ansatz that one substitutes into the field equations
and then solves. Consequently, it does {\em not} represent a LTB model, even
though the line element formally has the same form \cite{LT,Bondi}. This LTB
form of the metric is a consequence of the fact that the effective
scale factor of the volume average does not evolve as a simple
time scaling of a space of constant Gaussian curvature. Its radial
null geodesics nonetheless have a length which is the quantity of
physical interest to a finite infinity observer making measurements
on cosmological scales, which is why it is written in a spherically
symmetric form.

Rather than substituting (\ref{avgeom}) as an ansatz in field equations,
we first solve the averaged Einstein equations, in the form of the Buchert
equations, and then reconstruct (\ref{avgeom}) as the appropriate average
geometry. In particular, we can identify radial null geodesics propagating at
fixed $(\vartheta,\varphi)$ in (\ref{avgeom}) and then
reconstruct the average spatial curvature of (\ref{avgeom}) by a suitable
average of the equations of geodesic deviation.
Comparing (\ref{figeom}) and (\ref{avgeom}), we see that the radial
null geodesics of the two geometries coincide provided that
\beq
\dd\etw= {\fwi^{1/3}\dd\etb\over\gb\fvf^{1/3}}\label{etarel}
\eeq
along the geodesics. We note that although
the volume average parameter, $\ts$, differs from wall time,
$\tc$, the radial null section of (\ref{avgeom})
coincides with that of the conformal rescaling of (\ref{figeom}),
$\gb^2(\tc)\,\ds^2\Z{\Fi}$. This follows since null geodesics are
unaffected by conformal transformations. Although it is possible to identify
the null sections in this manner, the complete conformally related
metrics will not be isometric due to differences in spatial curvature.
The conformal factor is physically postulated to encode the gravitational
energy variation between wall observers and the volume average. Likewise
the fact that the spatial sections are not fully related by a conformal
factor has a physical origin: the
volume average spatial curvature differs from the average spatially flat
curvature within walls. Consequently there is an average focussing
of light rays from the volume average in voids relative to observers such
as ourselves within walls.

Once the wall geometry (\ref{figeom}) is rewritten
in terms of the conformal time $\etb$ of the volume average geometry
(\ref{avgeom}), the coordinate $\etw$ can no longer be regarded as
constant if related to the position of a volume average comoving observer.
Rather we define $\etw$ in terms of an integral of (\ref{etarel}) on the
radial null geodesics of (\ref{avgeom}). The wall geometry (\ref{figeom})
therefore becomes
\bea \ds^2\Z{\Fi}&=&-\dd\tc^2+{\ab^2\over\gb^2}\,\dd\etb^2+
{\ab^2\fvf^{2/3}\over\fwi^{2/3}}\,\etw^2(\etb,\tc)\,\dOM\nonumber\\
&=&-\dd\tc^2+\ac^2(\tc)\left[\dd\etb^2+\rw^2(\etb,\tc)\,\dOM\right]
\label{wgeom}\eea
where $a\equiv\gb^{-1}\ab$
and $\rw\equiv\gb\fvf^{1/3}\fwi^{-1/3}\etw(\etb,\tc)$.

If any single geometry can be considered to encode relevant cosmological
parameters with respect to our own measurements as wall observers, it
is the geometry (\ref{wgeom}) written in terms of the global average conformal
time
\beq
\etb=\int_{\ts}^{\ts\X0}{\dd\ts\over\ab}=
\int_{\tc}^{\tc\X0}{\gb\,\dd\tc\over\ab}=
\int_{\tc}^{\tc\X0}{\dd\tc\over\ac}\,,
\label{conftime}
\eeq
which is constructed numerically from the solution to the equations
(\ref{eqn1}), (\ref{eqn2}). Eq.~(\ref{wgeom}) is in fact the closest
thing we have to an FLRW geometry synchronous to our own clocks: it may be
used to define the relevant conventional luminosity and angular diameter
distances with respect to our measurements. The quantity $\etb$ plays the
role of a ``comoving distance'' in the conventional sense. Since we are not
at the volume average, we do not need to construct the geometry (\ref{avgeom})
directly for simple distance measurements, although a knowledge of its
average spatial curvature would be directly relevant to the computation of
the integrated Sachs--Wolfe effect, cumulative gravitational lensing
etc.

\subsection{Volume--average and wall--average cosmological parameters}

Our identification of the clocks and rods of wall observers
relative to the volume average means that there will
be differences between the dressing of parameters in the present model
and that discussed by Buchert and Carfora
\cite{BC}. From (\ref{wgeom}) the global horizon volume average Hubble
parameter over walls and voids, as measured by the wall observer is
not $\bH={1\over\ab}\Deriv\dd\ts\ab={1\over\aw}\Deriv\dd\tc\aw$, but
\bea\Hh\equiv{1\over\ac}\Deriv\dd\tc\ac
&=&{1\over\ab}\Deriv\dd\tc\ab-{1\over\gb}\Deriv\dd\tc\gb,\nonumber\\
&=&\gb\bH-\dot\gb,\nonumber\\
&=&\gb\bH-\gb^{-1}\Dtc\gb,\label{42}
\eea
where the overdot still denotes a derivative w.r.t.\
the volume average time parameter, $\ts$.

Since $\bH$ is a measurable ``local'' Hubble parameter at any location,
eq.\ (\ref{42}) defines the relationship between this local Hubble parameter
and the conventional globally measured Hubble parameter, $\Hh$, according to
wall observers. Since both $\bH$ and $\Hh$ can be measured, this
relationship is ultimately open to observational testing, once the
scale of the ``local Hubble flow'' is empirically determined.

An interesting consequence of our definition of homogeneity is that
even though the bare Hubble parameter has no relationship to the
true critical density in Buchert's scheme for arbitrary domains of
averaging, for our particular choice of domain it does. By
definition finite infinity boundaries encompass regions of true critical
density, and insofar as $\bH=\bH\ns{w}$ is the locally measured Hubble
parameter at these boundaries, then $3\bH^2/(8\pi G)$ {\em is indeed the
true critical density}. As we shall see shortly, we nonetheless must take
care, as the volume average observer will perceive the critical
density to take a different numerical value.

Since the horizon volume average Hubble parameter as measured by wall
observers, $H$, differs from $\bH$ according to (\ref{42}), then clearly the
quantity $3H^2/(8\pi G)$ is not the critical density, contrary to our usual
na\"{\i}ve assumptions.
If we live in an epoch for which the time variation of the lapse function,
$\gb$, is small, then since $1<\gb<\frac32$ it follows from (\ref{42})
that we can over--estimate the value of the critical density by a significant
amount, typically by 40--80\% rather than the absolute upper bound of 125\%.
However, a factor 40--80\% is already enough to significantly change the
matter budget, and has the consequence that our recalibration of
cosmological parameters is going to be significant.

The bare cosmological parameters, expressed as fractions of the true
critical density, $3\bH^2/(8\pi G)$, are from (\ref{undress1}) and
(\ref{kav})--(\ref{eqn1}),
\bea
\OMM&=&{8\pi G\rhb\Z{M0}\ab\Z0^3\over3\bH^2\ab^3}\,,\label{UnM}\\
\OMk&=&{\al^2\fv^{1/3}\over\ab^2\bH^2}\,,\label{UnK}\\
\OMQ&=&{-\dot\fv^2\over9\fv(1-\fv)\bH^2}\,,\label{UnQ}
\eea
so that eq.\ (\ref{eqn0}) and eq.\ (\ref{eqn1}) coincide. These are
the parameters as measured locally by volume--average observers:
the time derivative in (\ref{UnQ}) refers to their local clocks.

The parameters (\ref{UnM})--(\ref{UnQ}) are averaged over the present
particle horizon volume, $\cal H$. This does not imply that they are
globally measured quantities outside our past light cone, as is implicitly
assumed in FLRW models. In fitting cosmological parameters, we
will work from the premise that the universe is close to critical
density overall, and that our horizon volume average {\em at the
present epoch} is below critical density. The implication is that horizon
volume averages in the past would have given values $\OMM\simeq1$, $|\OMk|
\ll1$, $|\OMQ|\ll1$. This certainly would have been the case at
the time of last scattering. To determine when the universe undergoes
a ``void dominance'' transition one must integrate the equations
(\ref{eqn1})--(\ref{eqn2}) and fit the results to all available data.

Whereas the parameters
(\ref{UnM})--(\ref{UnQ}) are the bare or ``true'' cosmological parameters,
our conventional parameters, based on the geometry (\ref{figeom}) or
(\ref{wgeom}), synchronous with our clocks, will differ. The question of
how cosmological parameters are to be dressed is an interesting one.
Indeed, if our claim in eq.\ (\ref{truecr}) is correct, then there
must be a sense in which if we restricted our considerations to the
matter purely within walls, ignoring the void contribution, then that
would give the ``true critical density'', with $\OmMw=1$ for some sort of
dressed density parameter. This suggests that the equations
(\ref{eqn1}), (\ref{eqn2}) should possess a simple first integral. Indeed,
this is the case, as we now demonstrate.

In the two--scale model, the equation for the second derivative of
the volume average scale factor (\ref{buche2}), which may also be derived
directly from a derivative of (\ref{eqn1}) in combination with (\ref{eqn2}),
is
\beq
{\ddot\ab\over\ab}={2\dot\fv^2\over9\fv(1-\fv)}-{4\pi G\over3}{\rhb\Z0}
{\ab\Z0^3\over\ab^3}\,.\label{da2}
\eeq
We may combine (\ref{eqn1}), (\ref{eqn2}) and (\ref{da2}) to obtain
\beq
6{\ddot\ab\over\ab}+3{\dot\ab^2\over\ab^2}-{2\ddot\fv\over1-\fv}
-{6\dot\ab\dot\fv\over\ab\fvf}-{\dot\fv^2\over(1-\fv)^2}=0\,.
\label{inteqn}\eeq
If we now multiply (\ref{inteqn}) by $24\pi G{\rhb\Z0}\ab\Z0^3/[\dot\fv\ab
-3\fvf\dot\ab]^3$, use (\ref{UnM}) and also note from (\ref{clocks2}) and
(\ref{fdot}) that
\beq
\gb={3\fvf\dot\ab\over3\fvf\dot\ab-\dot\fv\ab}
\label{gam2}\eeq
then the resulting equation may be recognised as
\beq
\Der\dd\ts\left(\gb^2\OMM\over1-\fv\right)=0\,.
\eeq
Its integral gives us
\beq
\OmMw\equiv{(1-\epi)\,\gb^2\OMM\over1-\fv}=1,
\label{Omtrue}\eeq
where $\epi\ll1$ is a small constant determined by initial conditions, since
at early times $\OMM\to1$, $\gb\to1$ and $\fv\to\fvi\ll1$. Using the
matter--dominated approximation back to the surface of last scattering,
at $t=t_i$, the small value of $\epi$ is given by
\beq
\epi=1-{1-\fvi\over\gbi^2\,\Omi}.
\eeq
Strictly, we should also include
radiation in considering the very early time limit, and modify
(\ref{Omtrue}) to include $\OMR$. However, provided
photon--electron decoupling occurs within the matter dominated era -- as will
occur for best--fit parameter values considered later -- then the effect of
$\OMR$ is negligible, and is omitted for the present considerations.

The fact that wall observers and volume average observers have different
clock rates, and that the frames (\ref{avgeom}) and (\ref{wgeom}) differ
by both a conformal factor and spatial curvature factor, means that there are
subtleties in the definitions of densities. Just as in special relativity,
where definition of internal energy density is frame dependent, so too is
the corresponding definition here, where there is a conformal frame issue
relating to gravitational energy and synchronisation of clocks. Systematically
differing results will be obtained when averages are referred to different
clocks. In the present case, the density $\rhb$ has been defined in
Buchert's scheme at the volume--average position in freely expanding space.
In the two scale approximation it may be considered as the sum
\beq
\rhb=\fvf\rhw+\fv\rhv
\eeq
where the mean densities of matter within walls and voids
\beq
\rhw={\rhb\Z0\ab\Z0^3\over\aw^3},\qquad
\rhv={\rhb\Z0\ab\Z0^3\over\av^3}
\label{rhsplit}\eeq
are {\em referred to the volume average frame}. This volume average observer
perceives that the expansion rate within the finite infinity regions, $\Hw$, is
slower than the locally measured expansion rate, $\bH$. Thus in fact, using
(\ref{volav}), (\ref{homo3}) and (\ref{rhsplit}) we see that
\beq
{3\Hw^2\rhw\over8\pi G}={\fwi\gb^2\OMM\over1-\fv}={\fwi\OmMw\over1-\epi}
\simeq\OmMw\,,
\eeq
consistent with (\ref{Omtrue}), given that $\fwi\simeq1$ and $\epi\ll1$.

Neither the volume average density parameter, $\OMM$, nor the ``true'' density
parameter, $\OmMw$, will take a numerical value close to those of the
``concordance'' dark--energy cosmology. It is possible to define a
{\em conventional} density parameter in terms of the
effective global average scale factor, $\ac=\gb^{-1}\ab$ of (\ref{wgeom}).
This gives a matter density parameter ``dressed'' by a
factor $\gb^3$
\beq
\OmM=\gb^3(\tc)\,\OMM\,.
\label{Omrel}\eeq
Since (\ref{wgeom}) is the closest approximation to a FLRW geometry
referred to local rulers and clocks, the value of $\OmM$ thus defined
will come the closest numerically to the conventional matter density
parameter of the standard dark--energy cosmology, even
though it is not a fundamental parameter in the way that both $\OMM$ and
$\OmMw$ are. We could similarly define other dressed parameters, by
introducing appropriate volume factors. We note that
even though the bare parameters (\ref{UnM})--(\ref{UnQ}) sum
to unity (\ref{eqn0}), this will not be true for such dressed parameters.

Eq.\ (\ref{Omtrue}) has two important consequences for the characterisation
of solutions. Firstly, only three of the parameters $\bH$, $\OMM$, $\fv$ and
$\gb$ are independent, meaning that other relations can be simplified. For
example,
\beq \OMQ={-(1-\fv)(\gb-1)^2\over\fv\gb\Y2}
={-(\gb-1)^2\OMM\over1-\gb\Y2\OMM} \,.
\label{Q3}\eeq
The global average Hubble parameter (\ref{42}) as measured by wall observers
is
\bea
\Hh&=&\gb\bH\left[2-\frac3{2\gb}+\half\OMM+2\OMQ\right]\nonumber\\
&=&\bH\left[2\gb-\frac32-{\gb(\gb-1)(3\gb-1)\OMM\over2(1-\gb\Y2\OMM)}\right]\,
.\label{hg}
\eea

The second important consequence of (\ref{Omtrue}) is that in place of
(\ref{eqn1}) and (\ref{eqn2}) we now only need to solve two {\em first
order} ODEs, namely (\ref{eqn1}) and
\beq
(2\fv-1){\dot\ab^2\over\ab^2}+{2\over3}\dot\fv{\dot\ab\over\ab}-
{\al^2\fv^{4/3}\over\ab^2}+{8\pi G\over3}{\rhb\Z0}{\ab\Z0^3\over\ab^3}
\left(1-\epi-\fv\right)=0,
\label{eqn3}\eeq
as may be found by combining (\ref{eqn1}), (\ref{gam2}) and (\ref{Omtrue}).
Numerically, this is a considerable simplification.
In fact, the combinations of $[\fv\hbox{(\ref{eqn1})}
-\hbox{(\ref{eqn3})}]$, and $[\fvf\hbox{(\ref{eqn1})}+\hbox{(\ref{eqn3})}]$
can both be directly factored, leading to the even simpler
equations\footnote{The simple form (\ref{eqn4}), (\ref{eqn5})
was only found after this paper was originally submitted. The general
analytic solution of eqns.\ (\ref{eqn4}), (\ref{eqn5}) is readily obtained,
and will be presented in a separate paper \cite{sol}.}
\bea
\fvf{\dot\ab\over\ab}-\frn13\dot\fv=
\sqrt{{8\pi G\over3}\rhb\Z0(1-\epi)\fvf{\ab\Z0^3\over\ab^3}}\,,
\label{eqn4}\\
{\dot\ab\over\ab}+{\dot\fv\over3\fv}={\al\over\fv^{1/3}\ab}
\sqrt{1+{8\pi G\over3\al^2}{\rhb\Z0\epi\ab\Z0^3\over\fv^{1/3}\ab}}\,.
\label{eqn5}
\eea

\subsection{Apparent cosmic acceleration\label{acc}}

{}From (\ref{buche2}), (\ref{Q1}), (\ref{UnM}) and (\ref{UnQ}) the bare
or volume--average deceleration parameter is
\beq
\bq\equiv{-\ddot\ab\over\bH^2\ab}=\half\OMM+2\OMQ=\half\OMM-
{2\fv(1-\fv)(1-\hr)^2\over\left[\hr+(1-\hr)\fv\right]^2}\,.
\label{Unq}\eeq
The back--reaction contribution, $2\OMQ$, to (\ref{Unq}) is negligible both
at early times when $\hr\to1$ independently of the initial small value of
$\fv$, and at very late times when $\fv\to1$. It is certainly possible for
the back--reaction term to dominate so that apparent acceleration is
obtained; it is the extent to which this is possible for reasonable
parameters which is the subject of debate. The maximum value of $|\OMQ|$
is obtained at the epoch when $\ddot\fv=0$. One may solve (\ref{eqn2}), in
combination with (\ref{eqn1}), as a quadratic equation for $\dot\fv$ at this
epoch to determine whether the local minimum of deceleration is negative for a
particular solution. One finds that whether the back--reaction is large
enough to give acceleration depends on initial conditions together with
cosmological parameters. As a simple example, if $\ddot\fv=0$ is attained
at the epoch when $\fv=4/7$ then $2\OMQ=\frac{-1}{24}\left(1-\OMM
\right)$ at the epoch in question, so that acceleration is only attained if
$\OMM<\frac1{13}$ by that epoch. This would not be reasonable if $\OMM$ were
the standard matter density fraction related to our clocks and rulers,
since $\OMM(t)$ is monotonically decreasing and the value we measure today is
believed to be larger.

The crucial point to observe, however, is that the bare deceleration
parameter (\ref{Unq}) is defined for the volume--averaged rulers and clocks
so that debates based on equations such as (\ref{Unq}) are misdirected
if they ignore the question as to how average parameters are operationally
defined in an inhomogeneous universe. To make contact with our rulers and
clocks, or indeed of other observers in bound systems, we need to use
dressed, or wall--average parameters. The wall--average global deceleration
parameter analogous to the global wall--average Hubble parameter (\ref{42}) is
\bea
\qh&\equiv&{-1\over\Hh^2{\ac}^2}{\dd^2\ac\over\dd\tc^2}\nonumber\\
&=&-\left({\dot\ab\over\ab}-{\dot\gb\over\gb}\right)^{-2}
\left[{\ddot\ab\over\ab}-{\dot\gb\dot\ab\over\gb\ab}
-{\ddot\gb\over\gb}+{\dot\gb^2\over\gb}\right]\nonumber\\
&=&\left(\gb\bH-\dot\gb\right)^{-2}\left(\gb^2\bH^2\bq+\gb\ddot\gb\right)
+\left(\gb\bH-\dot\gb\right)^{-1}\dot\gb\,,
\label{qm}\eea
Realistic solutions exist with the term proportional to $\ddot\gb$
providing a dominant negative contribution at late times.
Thus it is quite possible to obtain regimes in which
{\em the wall observers measure apparent acceleration, $q<0$,
even though void observers do not}.

For the purposes of comparison of numerical solutions to (\ref{eqn1}),
(\ref{eqn3}) with observed quantities, a number of useful relations can
be derived by combining these equations with (\ref{clocks2}) and
(\ref{fdot}). The derivative of the lapse function with respect to
void average time, $t$, is
\beq
\dot\gb=\gb^{-1}\Dtc\gb
=\gb\bH\left[\frn32\gb^{-1}-1-\half\OMM-2\OMQ\right]\,,\label{dgamma}
\eeq
where $\OMM$ is given by (\ref{UnM})
while the global average deceleration parameter (\ref{qm}) as measured
by wall observers is
\bea
\qh&=&{\gb\Y2\bH^2\over\Hh^2}\Biggl[\frac14-\frac{1}{\gb}+{2\over\fv}\gbff
+\frac32\OMM+\frac34{(16-13\fv)\over\fvf}\OMQ\nonumber\\
&&\qquad\quad-{2\over\fv}\fvf\gbff(\OMM+\OMQ)-\frac14\OMM^2-2\OMM\OMQ-4\OMQ^2
\Biggr]\nonumber\\
&=&{\gb\Y2\bH^2\over\Hh^2}\Biggl[{3\over2\gb}-{9\over4\gb\Y2}
+{2\over\fv}\fvf\gbff\Bigl(1-{\OMM\over\gb}\Bigr)-\frac{\OMM^2}4+\frac{3\OMM}2
\nonumber\\ &&\qquad\quad-{12\over\fv}\fvf\gbff^2-{2\over\fv^2}\fvf^2\gbff^3
\Bigl(1-{2\over\gb}\Bigr)\Biggr].\nonumber\\ \label{qg}
\eea
It must be recalled that in these expressions $\OMM$ is a time--varying
parameter given by (\ref{UnM}), which begins close to unity and decreases
monotonically. One of the parameters $\fv$, $\gb$ and $\OMM$ can be
further eliminated by (\ref{Omtrue}). However, the resulting expressions
are no more compact than (\ref{qg}).

\subsection{Qualitative behaviour of solutions\label{qual}}

The physical boundary conditions we apply to the problem are that at
early epochs near the surface of last scattering, the expansion rate is
essentially uniform so that $\hr=1-\ep$, $\ep\ll1$. In this limit $\dot\fv
\simeq0$. The volume fraction of voids thus grows very slowly
from some small initial value and the back--reaction is negligible.
The initial void volume fraction is a parameter input to numerical
integration; the only constraint is that initial density contrast in voids
should be consistent with observed bounds inferred from the CMB. In this
initial phase $\gb\simeq1$, $\OMM\simeq1$ and $\OMQ\simeq0$ so that by
(\ref{dgamma}) $\dot\gb\simeq0$. Thus the universe evolves essentially as
an Einstein--de Sitter universe with negligible spatial curvature, and
only very small differences between clocks in overdense and underdense
regions.

As the universe evolves density contrasts grow, bound systems form and
enter the non--linear regime where vorticity and shear cannot be
ignored. We have to average over regions larger than finite infinity
domains so that these effects can be neglected.
The voids remain in the ``linear regime'' but their volume increases
more rapidly than the wall regions where galaxies form and cluster. It is
claimed that the differences in gravitational energy grow as spatial
curvature differences between the wall and void regions increase.
This is implicit in our assumption that the true surfaces of average
homogeneity are those with a uniform quasilocally measured expansion.
Although some of the differences in quasilocal gravitational energy
between bound systems and voids may be attributed to the kinetic
energy of expansion, given the present epoch value of the Hubble
parameter, spatial curvature variations should in fact be the dominant
contribution.

From (\ref{dgamma}) we see that the rate of growth of $\gb$ is abetted by both
the decrease in $\OMM$ and the initial increase in $-\OMQ$, although there
is an eventual decrease in $-\OMQ$ as $\bH$ decreases. The rate of growth of
$\gb$ as seen by
wall clocks is amplified since $\Dtc\gb=\gb\dot\gb$. Depending upon parameter
values it is possible for wall observers to register an apparent
acceleration with the deceleration parameter of (\ref{qg}) taking values
$q<0$. Back--reaction and the rate of increase of $\gb$ are largest in an
epoch during which the universe appears to undergo a void--dominance
transition, or equivalently a transition in which spatial curvature $\OMk$
becomes significant. The reason for apparent acceleration at such an
epoch has been partly described by R\"as\"anen \cite{Ras}: in the
transition epoch the volume of the less rapidly decelerating regions increases
dramatically, giving rise to apparent acceleration in the volume average.
We must be careful to note that these statements are true, when {\em referred
to one set of clocks}, such as our own. Taking gravitational energy
differences into account the quasilocally measured expansion can nonetheless
be uniform, resolving the dilemma, stated in ref.\ \cite{IW}, of the geometry
being close to an FLRW one.

Since $\ddot\gb$ also contributes to
(\ref{qm}) apparent acceleration can be seen by observers in galaxies in a
transition from global Einstein--de Sitter--like evolution to open FLRW--like
evolution, in contradiction to the picture one would have if one were at the
volume average position in voids. In fact, we will find that the deceleration
parameter (\ref{qg}) is small and negative for typical parameter values.
One very important feature of the present model is that since we are no
longer dealing with a purely Gaussian curvature evolution, the effect of an
apparent acceleration on the luminosity distance is greater than would be
obtained for a FLRW model based on the numerical value of $q$. In particular,
even if $q$ is written in terms of wall time, cosmological evolution is no
longer simply described by any quantity derived from a single scale factor.
The deviation of $\rw$ in (\ref{wgeom}) from the conformal scale, $\etb$,
which physically amounts to the rate of change of spatial curvature
differences, will also contribute to the average increase of the luminosity
distance.

At very late times eventually $\fv\to1$ and by (\ref{Q3}) $\OMQ\to0$.
In this limit (\ref{hg}) and (\ref{qg}) yield
\beq
\Hh\simeq\bH\left[(2+\half\OMM)\gb-\frn32\right]\label{Hap}
\eeq
and
\beq
\qh\simeq{\bH^2\over\Hh^2}\left[\frn32(\gb-\frn32)+(\frn32-\frn14\OMM)\gb^2
\OMM\right].\label{Qap}
\eeq
Since we also have $\OMM\to0$, by (\ref{dgamma}) there is a stable limit
in which $\gb\to\frn32$, $\dot\gb=0$ and $q\to0$. Equivalently,
$\ab\goesas\ts$ while $\fvf\goesas\ts^{-1}$, so that within walls
$\aw\goesas\ts^{2/3}$, just as at early times. At the same time the volume
average evolution is approximately the late--time coasting phase of an open
FLRW universe. This justifies the analytic approximation of ref.\
\cite{paper0}. Thus as $t\to\infty$, $\hr\to\frn23$, the limiting
Hubble rate of the Einstein--de Sitter divided by that of the Milne universe.

Since the averaging scheme is defined
relative to a particular scale, this limit will not necessarily
ever be reached. In the present case, the relevant scale is the size
of our present particle horizon volume: the initial void volume fraction
is defined relative to this volume. At a much
later epoch we would need to re-perform the analysis, with a new volume
fraction corresponding to the largest correlated perturbation within the
past horizon volume at the epoch in question. This perturbation might be
above critical density, which would require modification to the analysis
above. We will return to this discussion in \S\ref{inflation}.

It must be noted that the late--time behaviour given by (\ref{Hap}),
(\ref{Qap}), when $\fv\to1$, is an attractor--like feature in the phase
space of the differential equations (\ref{eqn1}), (\ref{eqn3}), whereas
the early--time Einstein--de Sitter--like behaviour is not an attractor
if the equations are time--reversed. The early Einstein--de Sitter--like
phase is imposed in our case, by taking a small initial void fraction,
as a reasonable initial boundary condition consistent with the initial
uniformity of the primordial plasma as evidenced by the isotropy of the
CMB, and as consistent with the expectations of primordial inflation.

\subsection{The zero back--reaction analytic approximation\label{anap}}

It is possible to obtain an analytic approximation which demonstrates
the variation of gravitational energy, and clock rates, between the
wall and void regions by setting the back--reaction to zero and taking
$\fv=1$, $\dot\fv=0$ identically. In this case (\ref{eqn1}) is simply
the Friedmann equation for an open universe. Since the wall regions and
voids are then decoupled, a simple but effective ansatz is to assume
that the expansion rate of the spatially flat wall regions remains
close to that of an Einstein--de Sitter universe as measured by wall
clocks. As discussed above, this is justified by the fact that both
initially and at late times the local wall geometry does
expand in this manner.

This approximation was discussed in ref.\ \cite{paper0}. It should be noted
that setting $\fv=1$ and $\dot\fv=0$ identically does not give expressions
which coincide exactly with the $\fv\to1$ and $\dot\fv\to0$ limit discussed
in \S\ref{qual} above, since in the latter case we are dealing with
a system of equations with nontrivial back--reaction in which the ratio
$\dot\fv/\fvf$ remains finite as $\fv\to1$. Thus the Hubble parameter
and deceleration parameter given in ref.\ \cite{paper0} do not have
the precise analytic forms of (\ref{Hap}) and (\ref{Qap}). Nonetheless,
the Hubble parameter in particular is extremely close to (\ref{Hap}), and
the approximation of ref.\ \cite{paper0} may be sufficient for determining
the global average Hubble constant.

The approximation of ref.\ \cite{paper0} was tested against
SneIa data in ref.\ \cite{paper1}. Although it is in some ways a
crude approximation, it fits the data remarkably well with best fit $\chi^2$
values only 7\% larger than for the standard \LCDM\ model. It appears that any
apparent cosmic acceleration in the SneIa data is in fact marginal and any
model in which $q\to0$ compares favourably with observation. Remarkably,
the best-fit values of the Hubble constant derived by this analysis
\cite{paper1}, $\Hm=62.7^{+1.1}_{-1.7}\kmsMpc$ using the
2004 gold data set of Riess \etal~\cite{Riess04} or alternatively
$\Hm=60.5\pm1.5\kmsMpc$ using the 2006 gold data set \cite{Riess06},
agree with the recent measured value published by
Sandage \etal\ \cite{Sandage} to within 1--2\%.

The parameter fits of ref.\ \cite{paper1} indicate that any ratio of
non--baryonic dark matter to baryonic matter from 0:1 up to the standard
5:1 is admitted, and do not constrain non--baryonic dark matter themselves.
The match to the baryon acoustic oscillation scale does appear to constrain
non--baryonic matter, however, as will be discussed in \S\ref{cmb}.
Tests similar to that of ref.\ \cite{Allen} may also serve
to bound the amounts of non--baryonic dark matter,
but as will be emphasised in \S\ref{cluster} below, all steps in
measuring masses of galaxy clusters would need to be carefully
reconsidered.
The parameter fits of ref.\ \cite{paper1} are of course based on the
crude approximation of ref.\ \cite{paper0}. In ref.\ \cite{paper2}
they are updated to include back--reaction and its effects on cosmological
evolution.

\section{Null geodesics and observable quantities\label{obs}}

With the possible exception of high energy cosmic rays, most cosmological
information comes to us on null geodesics carried by photons. In order
to relate the average geometry to observations we have to specify how
light propagates on average. The manner in which we have related the
wall geometry at finite infinity to global average parameters on
null geodesics according to (\ref{wgeom}) in fact already allows us
to determine average quantities which depend only on average geodesic
length.

\subsection{Cosmological redshift\label{red}}

In the two scale model, the average geometry may be approximated by
(\ref{avgeom}), and it assumed that light follows radial null geodesics of
this geometry, when a common origin with a point within a finite infinity
region is chosen. Since null geodesics are unaffected by conformal
transformations, and since the radial null sections of (\ref{wgeom})
have been chosen to coincide with those of (\ref{avgeom}),
it is simplest to use the wall geometry (\ref{wgeom}) in order
to determine the redshift we measure, since wall time, $\tc$,
is assumed to differ little from that of our own clocks.
As discussed in ref.\ \cite{paper0} the cosmological
redshift can be derived in the standard fashion, but there is a
difference in values of redshifts measured between our location and
the volume average location in voids.

The cosmological redshift, $z$, determined by wall observers is related
to that determined by volume average comoving observers, $\bz$, by
\beq
1+z={\an\over\ac}={\ab\Z0\gb\over\ab\gc}={\gb\over\gc}(1+\bz)\,.
\label{redshift}
\eeq
This is a direct consequence of the quasilocal energy difference,
and variation of clock rates, between the two locations. Since
freely--falling matter in voids is so diffuse as to be unobservable, the
quantity $\zb$, which is larger than $z$, is not directly relevant to
emitters or absorbers in any cosmological test yet conceived.

While eq.\ (\ref{redshift}) seems little more than a reparameterisation
which relates observed redshifts to numerical solutions of the
volume--average Buchert equations (\ref{eqn1}), (\ref{eqn3}), there are more
profound ramifications. In particular, the redshift of the CMB as measured
at a comoving volume average position is greater than in bound systems.
Since the volume average CMB temperature is used to calibrate several
parameters associated with the primordial plasma, we need to recalibrate
all quantities associated with the early universe. Local physics at that
epoch is unchanged but our interpretation is systematically changed.
It is these recalibrations, which account for quasilocal energy variations,
which will allow us to obtain a viable model of the universe without
dark energy. These recalibrations are performed in \S\ref{cmb}.

\subsection{Luminosity and angular diameter distances}

The luminosity and angular diameter distances can also be determined
in the standard fashion from (\ref{wgeom}), since for all known cosmological
observations apart from the CMB both the emitters and observers are in
bound systems within finite infinity regions, which are close to spatially
flat. Furthermore, the surface of last scattering is located at an epoch
when $\gb\simeq1$, making the distinction between the two possible
classes of observer irrelevant for measurements involving the CMB.

While there are changes in both photon energies and the
average focussing of geodesics, as null congruences exit and enter voids and
finite infinity regions, the average effect of this depends only on the
average comoving distance, $\etb$, to a source, as determined from
the solution to (\ref{eqn1}), (\ref{eqn3}). Consequently, the relative
difference in the absolute luminosity of a source in a wall region
and the observed flux in another wall region is accounted for by
taking the standard luminosity distance derived from (\ref{wgeom}),
\beq
\dL=\an(1+z)\,\rw={\gc}^{-1}\ab\Z0(1+z)\,\rw\label{dL}
\eeq
where
\bea
\rw(\tc)&=&\gb\fvf^{1/3}\int^{\tc\X0}_{\tc}{\dd\tc\over\gb\fvf^{1/3}\ac}
\nonumber\\
&=&\gb\fvf^{1/3}\int^{\ts\X0}_{\ts}{\dd\ts\over\gb\fvf^{1/3}\ab}
\label{rwo}
\eea
for a source emitting a photon at wall time, $\tc$, or volume average
time, $t$. Note that $\rw$ coincides with the conventional conformal
scale $\etb=\int^{\ts\X0}_\ts\dd\ts/\ab$ at early times when
$\fv\to0$ and $\gb\to1$.

If we multiply (\ref{dL}) by the bare Hubble constant, it follows that
\beq
\Hb\dL={\gc}^{-1}\ab\Z0\Hb(1+z)\,\rw\,.
\eeq
Equivalently using (\ref{42}) and (\ref{UnK}) also,
\beq
\Hh\Z0\dL=(1-{\gc}^{-1}\gc')\al f\ns{v0}^{1/6}\OM\Z{k0}^{-1/2}(1+z)\,\rw\,,
\eeq
in terms of the wall--measured average Hubble constant, $\Hh\Z0$,
where $\gc'\equiv{\Hb}^{-1}\left.\Deriv\dd t\gb\right|_{t_0}$.

The effective angular diameter distance to sources observed by wall
observers may now be defined in the standard fashion
\beq d\Z A={\dL\over (1+z)^2},\label{dA}\eeq
where $\dL$ is given by (\ref{dL}).

\subsection{Numerical example\label{nex}}

As a proof of principle, we will give the results of one particular
numerical example in this section. It is has not been determined to
be the ``best--fit'' example, but is one case which fits all known
tests well, including the broad features of the CMB as will be demonstrated
in \S\ref{cmb}. We adopt the terminology {\em Fractal Bubble} (FB) universe
to describe our new cosmological model, for ease of reference in what follows.

We numerically evolve the equations (\ref{eqn1}), (\ref{eqn2})
forward in volume--average time, in dimensionless units of $\Hb\ts$,
from suitable initial conditions. We simultaneously integrate eqs.\
(\ref{conftime}) and (\ref{rwo}), and also wall time with respect to
volume--average time $\tc=\int^t\X0\dd\ts/\gb$, to determine the
value of the lapse function at the present epoch. This is then used to
recalibrate the solutions in terms of wall--measured times and redshifts.

The initial time is taken to coincide with the epoch of last scattering,
when the universe is smooth and we deem a suitable initial condition to be one
which is consistent with the amplitude of the observed primordial density
perturbations by this epoch. In particular, whereas the density contrast of
photons and the baryons coupled to them is of order $\de\rh/\rh\goesas10^{-5}
$, the density contrast in non--baryonic dark matter can be of order
$\de\rh/\rh\goesas10^{-3}$, and this may also be taken as representative
of a typical void perturbation.

By our assumptions, if we take our present horizon volume, $\cal H$,
then at last scattering the greatest proportion of its volume, $\fwi$,
will be in perturbations whose mean is exactly the true
critical density. However, in addition to these perturbations there will
be one underdense dust perturbation, of initial density contrast
$\left(\de\rh\over\rh\right)\ns{vi}<0$ occupying a small fraction, $\fvi$,
of $\cal H$ so that the overall density contrast of $\cal H$ at that epoch is
\beq
\left(\de\rh\over\rh\right)_{\!\cal H\hbox{\sevenrm i}}=\fvi\left(\de\rh\over
\rh\right)\ns{vi}\goesas-10^{-6}\w{to}-10^{-5}.\label{initc}\eeq
This might be achieved in various ways, by having $\fvi\goesas10^{-3}$
and $(\de\rh/\rh)\ns{vi}\goesas-10^{-3}$, or $\fvi\goesas-10^{-2}$ and
$(\de\rh/\rh)\ns{vi}\goesas-10^{-4}$, or $\fvi\goesas3\times 10^{-3}$
and $(\de\rh/\rh)\ns{vi}\goesas-3\times 10^{-3}$ etc. The important
point to note is that $\fvi$ does not refer to a single smooth
underdense region, but a region with other perturbations embedded in it,
which is significantly larger than the particle horizon at last scattering,
and only becomes correlated within our past light cone on time scales which
are a significant fraction of the present age of the universe, as will be
further discussed in section \ref{inflation}.

Detailed cosmological parameter fits will be presented elsewhere
\cite{paper2}. Here and in \S\ref{cmb}, we will present one numerical
example, as a demonstration that the new model fits the major current
observations which support the \LCDM\ paradigm, while in addition resolving
some anomalies at odds with the \LCDM\ paradigm.
Decoupling occurs at a scale $\zdec\simeq1100$ with respect to wall observers
to within approximately 2\%. We cannot assume a value of $\zdec$ to the
accuracy quoted for WMAP \cite{wmap}, since the value quoted
includes statistical uncertainties specific to the \LCDM\ model. The
recalibration of the detailed peak fitting algorithms for the CMB anisotropy
spectrum in the present model requires an immense undertaking and is left for
future work. We will offer a proof of principle, by demonstrating agreement
with observations at the few percent level.

The example we have chosen has an initial void fraction of $\fvi=5.5\times
10^{-4}$ at $\zdec$, which with $\left(\de\rh\over\rh\right)\ns{vi}\goesas
-2\times10^{-3}$ gives
$\left(\de\rh\over\rh\right)_{\!\cal H\hbox{\sevenrm i}}\goesas-10^{-6}$.
Integrating the equations we find that by the present epoch, the void
fraction has risen to $f\ns{v0}=0.759$, with a density parameter
$\OMMn=0.127$ with respect to the volume average. This translates to
a conventional density parameter (\ref{Omrel}) of $\OmMn=0.33$, as
being the value relative to the wall geometry (\ref{wgeom}) normalised
to global average parameters. The difference in the rates of
the volume average clocks with respect to our clocks at the present
epoch is $\gc=1.38$. The global average Hubble constant takes the value
$\Hm=62.0\kmsMpc$, while the bare Hubble constant which would represent
the value we should determine from measurements restricted to lie
{\em solely within local filamentary walls}, is $\Hb=48.4\kmsMpc$.

\begin{figure}[htb]
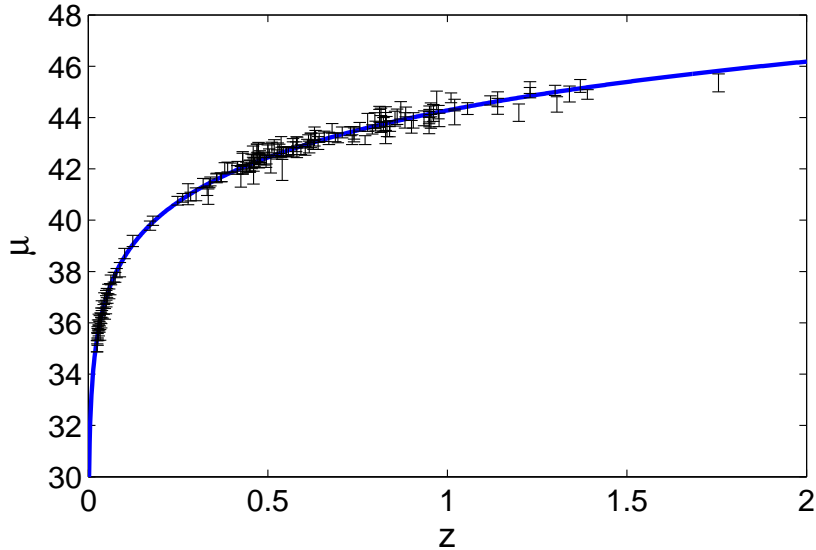

\vbox{\figld
\caption{\label{fig_ld}%
{\sl Distance modulus, $\mu\equiv m-M = 5\log_{10}(d\Z{L}) + 25$, versus
redshift, $z$, with $\dL$ in units Mpc. The theoretical curve for a FB model
with $\Hm=62.0\kmsMpc$, $\gc=1.38$, $f\ns{v0}=0.759$ is compared to the 182
SneIa, excluding the ``Hubble bubble'' points at $z\le0.023$
of the Riess \etal\ Gold06 data set \cite{Riess06}. For these
parameter values $\chi^2=163.2$, or $0.9$ per degree of freedom.}}}
\end{figure}

The distance modulus is plotted in Fig.\ \ref{fig_ld}, in comparison to
the Riess Gold06 data set \cite{Riess06} of 182 supernovae. We find
a $\chi^2$ of $0.9$ per degree of freedom, making the goodness of fit
essentially indistinguishable from the \LCDM\ model, since on a statistical
basis $\chi^2$ of order 1 per degree of freedom is to be expected.
We follow Riess \etal\ \cite{Riess06} in excluding supernovae at extremely
low redshifts within the ``Hubble bubble'' from the analysis. Whereas the
reasons for doing so are not theoretically well--motivated in the \LCDM\
model, in the present case there are imperative reasons for doing so,
as is further discussed in \S\ref{bubble}.

\begin{figure}[htb]
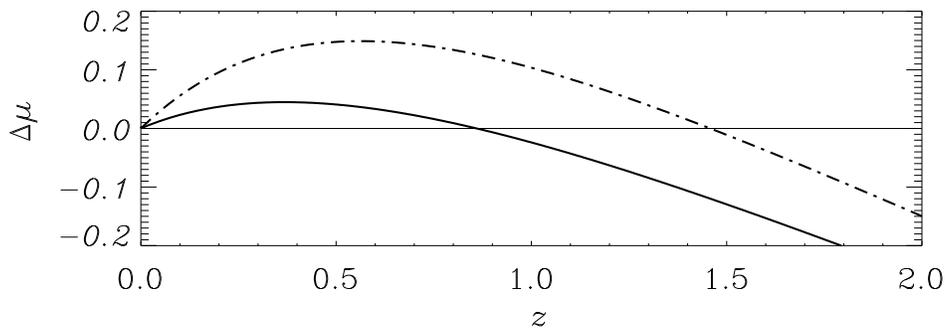

\vbox{\figdelmu\vskip-\baselineskip
\caption{\label{fig_delmu}%
{\sl The difference, $\DE\mu=\mu\Ns{FB}-\mu\ns{empty}$, of the distance
modulus of the FB model with $\Hm=62.0\kmsMpc$, $\gc=1.38$, $f\ns{v0}=0.759$
and the distance model that of an empty coasting (Milne) universe with the
same Hubble constant, versus redshift (solid line). Positive values of
$\DE\mu$ correspond approximately to apparent acceleration. As a comparison
we plot the corresponding difference of distance moduli,
$\DE\mu=\mu\Ns{$\scriptstyle\Lambda$CDM}-\mu\ns{empty}$
for a flat \LCDM\ model with $\Om=0.268$, $\Omega_\Lambda=0.732$, and the same
value of the Hubble constant (dot--dashed line).}}}
\end{figure}

In Fig.~\ref{fig_delmu} we compare the difference in distance moduli
of the FB model example and a \LCDM\ model example with the distance
modulus of the empty Milne universe, for the same value of the Hubble
constant in all cases. We note that apparent acceleration starts later
for the FB model, and is closer to the coasting Milne universe at late epochs.

\begin{figure}[htb]
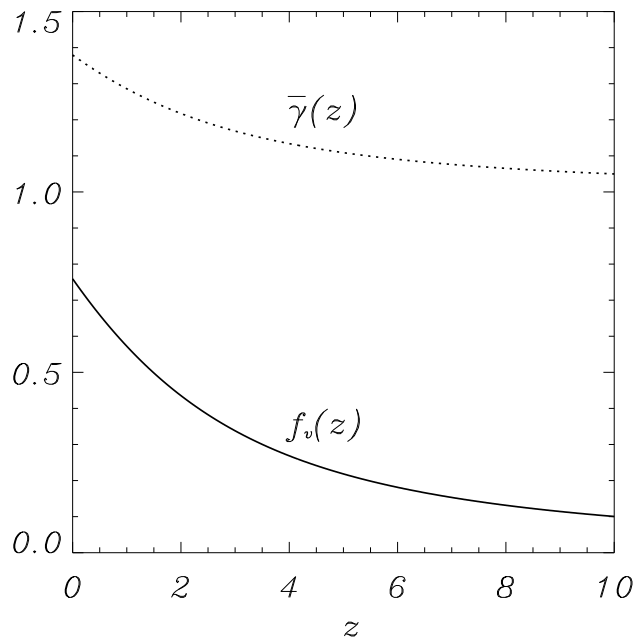

\vbox{\figfvgb
\caption{\label{fig_fvgb}%
{\sl The void fraction, $\fv$, (solid line), and
clock rate of volume--average observers with respect to wall observers
in a galaxy, $\gb$, (dotted line) as a function of redshift for the FB model
with $\Hm=62.0\kmsMpc$, $\gc=1.38$, $f\ns{v0}=0.759$.}}}
\end{figure}
The deceleration parameter as measured by a wall observer (\ref{qg})
is $\qh=-0.0428$, whereas a volume--average observer measures a deceleration
parameter (\ref{Unq}) given by $\bq=0.0153$. This gives one concrete example
of our finding in \S\ref{acc} that cosmic ``acceleration'' is an apparent
effect, depending crucially on the position of the observer and local clocks.
Both observers register a deceleration parameter close to zero, a general
feature of a universe which undergoes a void--dominance transition. According
to a wall observer in a galaxy, apparent acceleration begins at an epoch
$z=0.909$ for the present parameters, when the universe is 7.07 Gyr old,
a little under half its current age. The void fraction at this epoch is
$\fv=0.587$. The void fraction, $\fv$, and the mean lapse function or
``gravitational energy parameter'', $\gb$, are shown in Fig.~\ref{fig_fvgb}.

Since we are no longer dealing with a pure Gaussian curvature evolution,
as discussed above the evolution of the luminosity distance cannot be
characterised by one single parameter such as the deceleration parameter.
In addition to the change of sign of the deceleration parameter as
measured by observers in galaxies, another important cosmological milestone
is the epoch at which the void fraction reaches $\fv=0.5$, since the
coefficient of $\bH^2$ in (\ref{eqn3}) changes sign at this epoch. For the
present parameters, this event happens somewhat earlier than the
apparent acceleration transition, at a redshift of $z=1.49$ when the
universe is 5.12 Gyr old as measured by a wall observer in a galaxy.

\begin{figure}[htb]
\vbox{\figage
\caption{\label{fig_age}%
{\sl The expansion age as seen by wall observers in galaxies for a FB
model with $\Hm=62.0\kmsMpc$, $\gc=1.38$, $f\ns{v0}=0.759$ (solid line), is
compared to the expansion age as seen by volume--average observers
for the same parameters (dashed line). The expansion age for standard \LCDM\
cosmology with the ``concordance'' parameters $\Hm=71.0\kmsMpc$, $\OmMn=0.268$,
$\Omega\Z{\Lambda0}=0.732$ (dot--dashed line) is also shown for
comparison.}}}
\end{figure}

The expansion age history of the universe, as shown in Fig.~\ref{fig_age},
is also particularly interesting in comparison with the ``concordance''
\LCDM\ model. As measured by wall observers in a galaxy, the universe is
14.7 Gyr old at present, 1 billion years older than the concordance
model\footnote{We are using best--fit parameters for the WMAP1 data with
$h=0.71$, which lie within the 1$\si$ errors of the WMAP3 data, so as not to
overstate our case. WMAP3 raises the best--fit Hubble parameter for \LCDM\
to $h=0.73$ and drops the matter density $\OmMn=0.238$ for a spatially flat
model, which combine to have little net effect on the expansion age. However,
if WMAP3 is combined with other datasets a small amount of positive spatial
curvature appears possible. Inclusion of a small positive spatial curvature,
$\Omega\Z{k0}=-0.01$, would decrease the age of the universe
slightly and increase the contrast with the FB model.}.
The difference in ages is mainly due to the smaller Hubble constant in
the FB model. The \LCDM\ model would have an even greater age for the lower
Hubble constant consistent with the measurement of Sandage \etal\
\cite{Sandage}; only that value is not the best fit to WMAP in the
\LCDM\ case. As we shall see in \S\ref{cmb}, by contrast, the lower Hubble
constant provides a good fit very well to those features of the CMB that
we have been able to determine for the FB model.

What is perhaps most significant about Fig.\ \ref{fig_age} is the earlier
history since the fractional difference in the expansion age of the FB model
example from the concordance \LCDM\ model is much greater at higher redshifts,
where complex structures are observed. At $z=1$ with an age of 6.67 Gyr the
FB model is 12\% older than the \LCDM\ model, at $z=2$ with an age of 4.02
Gyr it is 20\% older, and at $z=6$ with an age of 1.24 Gyr it is 30\% older.
These values are for one particular choice of parameters. Other examples with
an age of the universe greater than 15 Gyr are also in regions of parameter
space which accord well with the supernovae data \cite{paper2}.

\subsection{Averaging the optical equations\label{nullav}}

As noted above, local spatial curvature as related to volume average
parameters via (\ref{wgeom}) is all that is required to determine quantities
such as luminosity and angular diameter distances (\ref{dL}), (\ref{dA})
relevant to our own measurements. The determination of the volume average
spatial curvature may therefore seem an academic exercise. However, it could
be relevant for any calculations which may rely on the average differences in
spatial curvature, including the integrated Sachs--Wolfe effect, cumulative
gravitational lensing, and possibly even weak gravitational lensing
measurements. The new interpretational framework presented in this paper
requires a systematic re--examination of such effects from first principles.
We leave such issues for future work. Here we will merely try to introduce
a possible framework for calculating the volume average spatial curvature.
As such, the results derived elsewhere in this paper are independent of
the calculations in this section.

In the case of the standard FLRW models,
given the comoving scale at which a null geodesic originated,
it is trivial to compute the average focussing of the spatial geometry to that
scale, as the geometry is completely homogeneous. Furthermore, we have found
that since the wall geometry is spatially flat on average, we are still able
to construct average luminosity and angular diameter distances (\ref{dL}),
(\ref{dA}), because we still have an effective homogeneous reference system
at the positions of both source and observer. In an arbitrary position in an
inhomogeneous universe, however, the problem of determining the local
spatial curvature is very nontrivial. In the present case the full volume
average geometry (\ref{avgeom}) is not even conformally isometric to local
wall geometry. Thus to determine the volume average spatial curvature
we are left with the general problem of employing the equation of geodesic
deviation for null geodesics in the form of the Sachs optical equation
\cite{Peebles},
\beq
{1\over\sqrt{\cal A}}{\dd^2\sqrt{\cal A}\over\dd\la^2}=-\half
\left({\cal R}_{\mu\nu}Y^\mu Y^\nu+w^{\mu\nu}w_{\mu\nu}\right)
\label{sachs}\eeq
where $\cal A$ is the cross--sectional area and $w_{\mu\nu}$ the shear tensor
of a null congruence, with tangent $Y^\mu=\Deriv\dd\la{x^\mu}$,
$\la$ being an affine parameter. The shear and expansion of the null
congruence are, of course, distinct from those of the spacelike hypersurfaces
used in Buchert's averaging scheme.

The aim here is to determine a quantity equivalent to the angular diameter
distance in FLRW models at the volume average position. Thus rather than
dealing with individual congruences we must deal with an average over the
whole sky for congruences of some initial fiducial proper size, which
originated at a given redshift of the average geometry. We will neglect the
shear: such a term will have some small effect through cumulative
gravitational lensing. However, based on the evidence of studies
in FLRW models \cite{lensing1}, it is expected to be much less than
the dominant effect from the average geometry.

We will relate the expansion of the average null congruence to the volume
expansion of the spacelike hypersurfaces in Buchert's scheme, by performing
an average in the two--scale approximation of \S\ref{Have}, whereby
we assume that the geometry is approximately locally homogeneous and
isotropic separately within the wall and void regions. Thus
\beq
{\cal R}_{\mu\nu} Y^\mu Y^\nu=-2\left(\Deriv\dd\la t\right)^2\left(
{\ddot\av\over\av}- {\dot\av^2\over\av^2}-{\kv\over\av^2}\right)
\eeq
within a connected void region. Within such a region (\ref{sachs}) then
becomes
\beq
\left(\Deriv\dd\la t\right)^2\left({\ddot\dA\over\dA}-\frac13\th_{\rm v}
{\dot\dA\over\dA}\right)=-\frac{2}3\left(\Deriv\dd\la t\right)^2
\left(\dot\th_{\rm v}-{\cal R}_{\rm v}\right)\,
\label{sachv}\eeq
where $\dA=\sqrt{\cal A}$, and a similar expression applies to a wall region.

To perform a Buchert average of the optical equation requires care
since the proper area, $\cal A$, of the null congruence of a source of some
fiducial proper size is not a volume scalar defined on spacelike
hypersurfaces of the sort appearing in the commutation rule (\ref{comm}).
Since Buchert's scheme does not average the null geodesic equations
themselves, we have assumed that the appropriate null geodesics are
those of the average geometry (\ref{avgeom}) in the case of the two--scale
model. Essentially, rather than taking an ensemble of null geodesics, we
are looking for an idealised geodesic in the average spatial geometry
that is the best fit to a homogeneous isotropic one at any epoch
in the cosmological evolution.

Likewise, in the case of average null geodesic deviation, we will
assume that the average area of a fiducial congruence of such geodesics
responds dynamically to the spatial Buchert average of the r.h.s.\ of
(\ref{sachs}). Formally, for the purposes of the two--scale model we must
assume two conditions in combining (\ref{sachv}) with a similar expression
for the wall regions:
\begin{itemize}
\item[(i)] The affine parameter ratio $\left(\Deriv\dd\la t\right)^2$
may be cancelled in numerator and denominator when performing the averages.
In doing so we assume that just as the affine parameter, $t$, is related
to timelike geodesics in the average geometry, $\la$ is
related to the null geodesics of the same geometry.
\item[(ii)] The average angular diameter scale is defined such that
$\ave{\th\,\ln(\dA)}=\ave{\th}\ave{\ln(\dA)}$, or equivalently
that $\ave{\Der\dd\ts\ln(\dA)}=\Der\dd\ts\ave{\ln(\dA)}$, where $\dA$ is
given in units $c/\Hb$. We note that $\dot\dA/\dA$ is the time derivative
of an angular scale.
\end{itemize}
The two--scale average of (\ref{sachs}) then becomes
\beq
{\dd^2\ave{\dA}\over\dd\ts^2}-{\dot\ab\over\ab}{\dd\ave{\dA}\over\dd\ts}
=\Biggl({\ddot\ab\over\ab}-{\dot\ab^2\over\ab^2}-\half\QQ-\frn16\Rav\Biggr)
\ave{\dA}\,,
\label{sacha}\eeq
where we emphasise that the commutation rule, $\ave{\dot\th}=\Der\dd\ts
\ave{\th}-\frn32\QQ$, has been applied only in the limit that the
background shear averages to zero: $\langle\si\rangle=0$.

It is convenient to define an {\em effective volume--average comoving
lensing scale} $\rb(\ts)$ by
\beq
\ave{\dA}=\ab(\ts)\,\rb(\ts)\,\de\,
\label{optco}\eeq
for a fiducial source which subtends an angle, $\de$. We substitute
(\ref{optco}) in (\ref{sacha}) and use (\ref{kav}) and (\ref{Q2}) to finally
obtain
\beq
\ddot\rb+{\dot\ab\over\ab}\dot\rb+\Bigl({\dot\fv^2\over3\fv(1-\fv)}-
{\al^2\fv^{1/3}\over\ab^2}\Bigr)\rb=0\,.
\label{optical}\eeq
The important quantity to note in (\ref{optical}) is the kinematic
back--reaction term, which reduces the effect of the spatial curvature
term. In a smooth FLRW geometry obtained by integrating
the Friedmann equation there is no need to go through an analysis
similar to that presented above, since once one knows the comoving
scale at which a null geodesic originated, one can read off the associated
proper length scale from (\ref{optco}). Equivalently, when $\dot\fv=0$ and
$\fv=$ const, (\ref{optical}) reduces to the FLRW null geodesic equation.

If we assume that $\rb=0$ is taken to denote the position of a volume average
observer, then the relationship between the solution to (\ref{optical}) and
the average spatial curvature is obtained by noting that the volume average
area function, $A(\etb,\ts)$, of (\ref{avgeom}) may be taken to satisfy
$A\propto\ab^2(\ts)\rb^2(\etb)$, where $\rb$ is the solution to (\ref{optical})
for radial null geodesics of (\ref{avgeom}) with volume average
conformal time, $\etb$, given by (\ref{conftime}). We note that $\rb(\etb)$
lies in the range $\etb\le\rb(\etb)\le\sinh(\etb)$. The lower bound on $\rb$
coincides with the purely spatially flat case, in which $\fv=0$,
$\dot\fv=0$, so that we have an Einstein--de Sitter model. For a given
value of $\OM\Z{M0}$, the upper bound on $\rb$ coincides with the
open FLRW model value for that choice of $\OM\Z{M0}$. For different void
fractions and back--reaction, the actual value of $\rb$ will lie
in between these bounds, giving an average spatial curvature between
both extremes, dependent on cosmological parameters.

We should be careful to note that if the solution to (\ref{optical})
were to be actually used to derive a luminosity distance or angular
diameter distance, then the result could only apply to a situation in which
both observer and source are at volume average locations in freely expanding
space. Additional calculations would be needed if the source was in a bound
system. However, since actual observers and sources are both in bound
systems, the solution to (\ref{optical}) does not apply in this fashion.

\section{The CMB and resolution of observational anomalies\label{cmb}}

Detailed data analysis of the new cosmological model, and a best--fit
parameter determination will be presented elsewhere \cite{paper2}.
In this section we will illustrate how the recalibration of cosmological
parameters, particularly those relating to the early universe as determined
from the CMB, can resolve a number of observational anomalies.

An important consequence of the variation in clock rates between ideal
comoving observers in galaxies and voids, is that the temperature of
the CMB radiation will be lower when measured at the volume average
in voids, taking a value
\beq \bT=\gb^{-1}T \label{vtemp}\eeq
at any epoch, where $T$ is the temperature
of the CMB as seen by wall observers. At the present epoch for the numerical
example of \S\ref{nex}, with a measured value of $T\Z0=2.725$ K locally, we
have $\bT\Z0=1.975$ K, for example, at the volume average location.
Essentially, there is an extra factor of $\gc$ in the redshift of CMB photons
as seen by void observers on account of gravitational energy differences.
Since observed matter sources and absorbers of photons are located in
galaxies and gas clouds within the filamentary walls at late epochs, this
does not have any obvious direct implications for local physical processes.

Whereas the underlying physics at the epochs of primordial nucleosynthesis
and recombination, when $\gb\simeq1$, is no different than usual, the fact
that volume average observers should measure a mean CMB temperature of
$2.725{\gc}^{-1}$ at the present epoch will affect all the usual
calibrations of radiation--dominated era parameters inferred relative to
present epoch comoving observers. We are faced with the task of systematically
rederiving all the standard textbook calculations \cite{Peebles,KT}
associated with the hot big bang, and making recalibrations where
necessary.

This section proceeds chronologically through the history of the universe.
We explicitly resolve two anomalies in \S\ref{nuc} and \S\ref{curve}:
those associated with the primordial lithium abundance, and with average
spatial curvature. In \S\ref{pole} the anomalies associated with large
angle multipoles in the CMB anisotropy spectrum are discussed.
In \S\ref{age} we outline qualitatively how the anomalies associated with
the expansion age, early formation of galaxies and emptiness of voids in
N--body simulations should be naturally resolved. In \S\ref{bubble} we
explain the Hubble bubble feature. Further observational
questions concerning void and galaxy cluster scales will be dealt with in
\S\ref{Newton}.

\subsection{Primordial nucleosynthesis bounds and the baryon fraction
\label{nuc}}

The first important parameter which is directly altered, given that a
volume average observer would measure a lower CMB temperature to us,
is the baryon--to--photon ratio inferred from primordial nucleosynthesis
bounds, which will lead to a recalibration of the present epoch baryon
fraction.

The number density of CMBR photons as measured by wall observers,
\beq
n_\ga={2\ze\over\pi^2}\left(\kB T\over\hbar c\right)^3\,,
\label{n_photons}\eeq
evaluated at the present epoch with $T\Z0=2.725$ K, yields the
conventional value $n_{\ga0}=4.105\times10^8\w{m}^{-3}$.
However, the corresponding number density measured at the volume average
is
\beq
\bn_\ga={2\ze\over\pi^2}\left(\kB\bT\over\hbar c\right)^3={n_\ga\over
\gb^3}\,,
\label{nv_photons}\eeq
yielding $\bn_{\ga0}=4.105\,{\gc}^{-3}\times10^8\w{m}^{-3}$ at the present
epoch.
The number density of baryons measured by a volume average observer is
given by
\beq
\bnB={3\Hb^2\bOmBn\over8\pi G\,m_p}
={11.2f\Z B\OMMn h^2\over(\gc-\gc')^2}\,\w{m}^{-3},
\label{nv_baryon}\eeq
where prime denotes the dimensionless derivative, $\gc'\equiv\frac1{\bH\X0}
\left.\Deriv\dd t\gb\right|_{t_0}$,
$m_p$ is the proton mass, $f\Z B\equiv\bOmBn/\OMMn$ is the volume--average
ratio of the baryonic matter density to the total clumped matter density
(including any non--baryonic dark matter), and we have related the bare
Hubble constant to our observed global average Hubble constant using
(\ref{42}). It follows that the volume--average baryon--to--photon ratio is
\beq
\etBg={\bnB\over\bn_\ga}={2.736\times10^{-8}f\Z B\OMMn\gc^3 h^2
\over(\gc-\gc')^2}\,,
\label{Bg_ratio}\eeq
as compared to $\et\ws{FLRW}=2.736\times10^{-8}f\Z B\Om h^2$ in the
standard FLRW model.

Early time calculations of nuclear processes are not affected
in the new scenario, as they occur at epochs when spatial curvature is
unimportant, and the effects of inhomogeneities are no more significant
than in standard cosmological scenarios. The principal departure is
firstly that (\ref{Bg_ratio}) differs somewhat from the conventional value,
and secondly, by (\ref{Omtrue}), that a small ``bare'' value of $\bOmBn$ as
determined by volume--average observers translates to a much larger equivalent
value, $\OmBw\simeq\gc^2\bOmBn/(1-f\ns{v0})$, as locally measured by observers
within finite infinity regions, which have average critical density.
This value of $\OmBw$ can be translated to an equivalent number density of
baryons within walls, $n\Z B$. One finds that the ratio $n\Z B/n_\ga$, is
a factor $\gc^{-1}(1-f\ns{v0})^{-1}$ larger than the volume--average ratio
(\ref{Bg_ratio}). However, it is the volume--average ratio (\ref{Bg_ratio})
which corresponds directly to that
conventionally used in the standard FLRW models. Using conventional big bang
nucleosynthesis bounds, we will find that the ratio of baryonic to
non--baryonic dark matter will generally be found to be increased, and it is
this difference which is the key to resolving the lithium abundance anomaly.

Prior to the
the detailed measurements of the Doppler peaks in the CMBR, the values
quoted for $\etBg$ tended to be somewhat lower than the WMAP best--fit value.
For example, Olive, Steigman and Walker \cite{bbn1} quoted two possible
ranges at the 95\% confidence level: $\etBg=1.2$--$2.8\times10^{-10}$ or
$\etBg=4.2$--$6.3\times10^{-10}$, depending on whether one accepted higher
or lower values of the primordial D/H abundance. At a similar time, Tytler
\etal\ \cite{bbn2}, accepting the lower D/H abundances, quoted a range
$\etBg=4.6$--$5.6\times10^{-10}$ at the 95\% confidence level.
The WMAP parameter estimates moved the best--fit
range of the baryon to photon ratio to $\etBg=6.1^{+0.3}_{-0.2}\times10^{-10}
$, at the very edge \cite{bbn1}, or beyond \cite{bbn2}, the earlier
95\% confidence limits. The underlying physical reason for this parameter
shift is that by increasing the fraction of baryons relative to
non--baryonic cold dark matter one increases baryon drag in the
primordial plasma, which
suppresses the height of the second Doppler peak relative to the first.
Such a suppression was required in order to fit the WMAP data
\cite{wmap}, even though it led to results at odds with primordial
lithium abundance measurements, and pushed agreement with $^4$He abundances
to the previous $2\si$ confidence limit.

To demonstrate the resolution of the lithium abundance anomaly, let us
assume a value of $\etBg=4.6$--$5.6\times10^{-10}$ in the range of
Tytler \etal\ \cite{bbn2}, the lower end of which accords with lithium
abundance measurements, but which lies outside the best--fit range for WMAP
using the standard \LCDM\ model \cite{bbn3}. There is an intrinsic tension
in the data between lithium and deuterium abundance measurements \cite{bbn4}
at the opposite ends of the range we adopt. Using the parameters of the
numerical example of \S\ref{nex}, with $h=0.62$, we find that the
bare volume--average baryon density parameter is
$\bOmBn\simeq0.027$--$0.033$.
Since the total clumped matter density is $\OMMn=0.127$,
the relative ratio of non--baryonic dark matter to baryonic
matter is roughly 3:1 in this particular example. Therefore it
is likely that one can admit enough baryon drag to fit the ratio of
the heights of the first two Doppler peaks, while simultaneously
having a baryon--to--photon ratio favoured by big bang nucleosynthesis,
with the bounds accepted prior to WMAP \cite{bbn1,bbn2}. In particular,
concordance with the lithium abundance observations can be obtained.

The example we have given is one simple illustration. Taking supernovae
data alone, one finds similarly to the case of the zero back--reaction
approximation \cite{paper1}, that any ratio of non--baryonic to baryonic
matter from close to 0:1 up to the standard \LCDM\ value of 5:1 is possible.
Other data sets -- in particular, the CMB data and baryon acoustic
oscillation signatures in galaxy clustering statistics -- place restrictions
on this range. In general the range of non-baryonic to baryonic matter
ratios is restricted from 2:1 to 5:1 by these other tests \cite{paper2}. Thus
while the present model leads to a recalibration of cosmological parameters,
it appears that non--baryonic dark matter must be retained.

In a sense, the small baryon density parameters of order $\bOmBn\simeq0.03$
favoured prior to WMAP are still correct for a volume--average
void observer. It is simply the case that the mean CMB temperature and
photon density as seen by the void observer is also lower, and thus a
recalibration with respect to our measurements in galaxies is required,
resulting in much larger possible ratios of baryons to non--baryonic dark
matter, as locally measured in galaxies.

\subsection{Average spatial curvature\label{curve}}

Since the first measurement of the angular position of the first Doppler
peak in the CMB anisotropy spectrum by the Boomerang experiment in 2000
\cite{boom}, it has been assumed that the average spatial curvature of the
universe is close to zero, and that models such as the present one, which has
a sizeable negative spatial curvature in the observed portion of the universe
at the present epoch, would be ruled out. In fact, the derivation of the
average spatial curvature has assumed a FLRW geometry which evolves
according to the Friedmann equation with the conventional identification
of comoving clocks. In the present model the entire
analysis of the CMB needs to be redone, and conclusions which pertain to
FLRW evolution cannot be simply carried over.

The essential point, which has been overlooked, is that the angular positions
of the Doppler peaks really are only a measure of {\em local}
average spatial curvature, and an indirect one at that. It is only if
one {\em assumes} that the locally measured spatial curvature is identical to
the volume average that one can claim to ``measure'' the average spatial
curvature of the universe. That average spatial curvature and local
spatial curvature can be different was already borne out by the
toy--model investigation of Einstein and Straus, who concluded
that: ``the field in the neighbourhood of an individual star
is not affected by the expansion and curvature of space'' \cite{gruyere}.
In the present model, all bound systems lie
within finite infinity regions, where the immediate local expansion is
given by the geometry (\ref{figeom}), which is spatially flat. There is
a significant difference in spatial curvature between the voids and
wall regions at the present epoch, which must be manifest in a relative
focussing between the volume average position and galaxies within finite
infinity regions.

If there is a sizeable negative average spatial curvature at
the present epoch, then there must be ways of detecting it other than
via the measurement of the angular scale of the Doppler peaks.
Such effects do arise when one considers more subtle measurements associated
with average geodesic deviation of null geodesics.
Indeed one prediction is that there should be nontrivial ellipticity
in the CMB anisotropies on account of greater geodesic mixing.
This effect is in fact observed, and is an important anomaly for the
standard \LCDM\ paradigm, which has been overlooked by the majority of the
community to date. The effect is most significant in the WMAP3 data
\cite{elliptic2}, although it has been seen in all earlier data sets going
back to the measurements by COBE \cite{elliptic1}. Since this measurement
of spectral ellipticity does not rely
on a calibration of cosmological parameters which assume a standard
FLRW evolution, it is arguably a much less model--dependent measurement
of average spatial curvature than the angular position of the first
Doppler peak is. This {\em better} determination of spatial curvature clearly
supports the new paradigm, and conceivably can be further developed
to constrain the model presented in this paper.

Since the present model implicitly solves the anomaly of refs.\
\cite{elliptic2,elliptic1}, it remains for us to show that in the new
paradigm, average negative spatial curvature at the present epoch
is in fact {\em consistent} with the angular scale of the first Doppler
peak, and the corresponding baryon acoustic oscillation scale in
galaxy clustering statistics. We shall now proceed to do so.

The underlying physical effects which determine the angular size of
the first Doppler peak in the angular power spectrum of CMB anisotropies
boil down to chiefly two quantities:
\begin{itemize}
\item[(i)] the proper length associated
with the sound horizon, $D_s$, at the epoch of recombination;
\item[(ii)] the average
integrated lensing of that scale by the average geometry in the intervening
15 Gyr; including the difference between our location and the volume
average at the present epoch.
\end{itemize}
For a wall observer in a galaxy, the second quantity must be computed from
the effective angular diameter distance (\ref{dA}), $d\Z A=D_s/\de=\ac(\ts)
\,\rw(\ts)$, using a solution to (\ref{rwo}). However, whereas
in the FLRW model the average spatial curvature only affects the angular
diameter distance, in the present model we must also recalibrate the sound
horizon. We will now present the required steps.

\subsubsection{Recalibration of the early universe: recombination and
decoupling\br}\br
In the standard FLRW models volume--average comoving observers are used to
calibrate quantities associated with the CMB. Since the Buchert equations
are also written with respect to these observers, in recalibrating the
standard calculations associated with the radiation--dominated era, it is
easiest to again use the volume average. However, the changed relationship
between the volume--averaged quantities and our observations at the present
epoch must be accounted for.

One direct route to the required recalibration comes from the observation
that at early times $\gb\to1$, an approximation which is valid not only
in the radiation--dominated era but also for much of the early
matter--dominated ``dark ages'' of the universe before galaxies form.
From eq.\ (\ref{redshift}) we see directly that our observed redshifts, $z$,
are then related to the equivalent volume--averaged quantity, $\bz$, by
\beq
1+z\simeq{1\over\gc}{\ab\Z0\over\ab}={1\over\gc}(1+\bz)\,
\label{redwall}
\eeq
for processes at early epochs.

The first quantities of interest are those associated with recombination
and matter--photon decoupling. To discuss such effects we must add the energy
density of radiation
\beq
\rhb\Z R={\rhb\Z{R0}\ab\Z0^4\over\ab^4}={\pi^2g_*\over30}
{(\kB\bT)^4\over\hbar^3 c^5},
\label{rhv_rad}\eeq
to the r.h.s.\ of eq.\ (\ref{eqn1}). Here the degeneracy factor
relevant for the standard model of particle physics, $g_*=3.36$, is
assumed. The bare cosmological parameters (\ref{UnM})--(\ref{UnQ}) are
then supplemented by the bare radiation density parameter
\beq
\OMR={8\pi G\rhb\Z{R0}\ab\Z0^4\over3\bH^2\ab^4}\,.\label{UnR}\\
\eeq
Similarly to (\ref{Omrel}) the bare parameter (\ref{UnR}) is related to
a conventional dressed parameter
\beq
\OmR=\gb^4(\tc)\,\OMR\,,
\label{Omrrel}\eeq
relative to wall observers if referred to the frame (\ref{wgeom}) synchronous
with our clocks at any epoch.

In addition to adding (\ref{rhv_rad}) to the energy density in the r.h.s.\ of
eq.\ (\ref{eqn1}), we also note that with our chosen boundary conditions
the universe is homogeneous and isotropic with close to zero
spatial curvature initially, $\fv\simeq0$ and $\dot\fv\simeq0$, so that
in the very early epochs near last scattering (\ref{eqn1}) simply becomes
the Friedmann equation
\beq
{\dot\ab^2\over\ab^2}={8\pi G\over3}\left({\rhb\Z{M0}\ab\Z0^3\over\ab^3}
+{\rhb\Z{R0}\ab\Z0^4\over\ab^4}\right)\,,
\label{eqn_early}
\eeq
pertinent to a spatially flat universe containing matter and radiation.
Thus all the standard early time calculations apply, with the crucial
proviso that {\em eq}.\ (\ref{redwall}) {\em must be used to relate
$\ab\Z0/\ab$ to the redshifts we determine as wall observers at the present
epoch}. The epoch of matter--radiation equality occurs when
\beq
{\OMMn\over\OM\Z{R0}}=\gc(1+z\ns{eq})={\gc\Omega\Z{M0}\over\Omega\Z{R0}}
\eeq
so that $1+z\ns{eq}=\Omega\Z{M0}/\Omega\Z{R0}$ as is standard in terms of the
``conventional'' normalisation of parameters (\ref{Omrel}), (\ref{Omrrel})
defined with respect to (\ref{wgeom}).

The conditions which define the epochs of decoupling and recombination
do not differ at all from the standard case in so far as volume--averaged
quantities are used. In particular, recombination may be defined as the
epoch when the ionisation fraction decreases to $X_e=0.1$, where $X_e$
is given by the Saha equation for the equilibrium ionisation fraction,
\beq
{1-X_e\over{X_e}^2}={4\sqrt{2}\ze\etBg\over\sqrt{\pi}}\left(\kB\bT
\over m_ec^2\right)^{3/2}\exp\left(B\over\kB\bT\right),
\label{saha}\eeq
where $m_e$ is the mass of the electron, $B$ the binding energy of
hydrogen and $\etBg$ is given by (\ref{Bg_ratio}). Decoupling occurs roughly
at the epoch when
\beq
\bH\simeq{1\over\bn_e\si\Ns{T}}={1\over\etBg\bn_\ga X_e\si\Ns{T}}\,,
\eeq
where $\bn_\ga$ is given by (\ref{nv_photons}), $\bn_e$ is the (bare) number
density of electrons, $X_e$, is given by solving (\ref{saha}) with $\bT/\bT\Z0
=\ab\Z0/\ab$, and $\si\Ns{T}$ is the Thomson cross-section.

While the locally measured energy scales of recombination and
decoupling are unchanged
-- modulo any small differences from small changes to $\etBg$ -- since
these are physical scales determined by quantities such as the binding
energy of hydrogen, what is important for cosmological calibrations is that
decoupling occurs at a parameter value
\bea
{\ab\Z0\over\ab\ns{dec}}&=&{\bT\ns{dec}\over\bT\Z0}={\gc T\ns{dec}\over T\Z0}
={\gc\an\over a\ns{dec}}=\gc(1+z\ns{dec})\nonumber\\
&\simeq&1100\gc\,.
\eea
In the numerical example of \S\ref{nex}, we see that a volume--average
void observer therefore estimates last--scattering to occur at a redshift,
$\bz\ns{dec}+1\simeq1518$, as compared to the wall value
$z\ns{dec}+1\simeq1100$.

The sound horizon can be calculated in the standard fashion from
(\ref{eqn_early}), using the fact that the speed of sound in the
plasma prior to decoupling is
\beq
c_s={c\over\sqrt{3(1+0.75\,\rhb\Z{B}/\rhb_\ga)}}\,.
\eeq
We therefore find that the proper distance to the comoving scale of the
sound horizon at any epoch is
\beq
\bD_s={\ab(t)\over\ab\Z0}{c\over\sqrt{3}\,\Hb}\int_0^{\mx\ns{dec}}
{\dd\mx\over\sqrt{(1+0.75\,\OM\Z{B0}\mx/\OM\Z{\ga0})(\OMMn\mx+
\OM\Z{R0})}}\,,
\eeq
where $\mx=\ab/\ab\Z0$, so that $\mx\ns{dec}=\gc^{-1}(1+z\ns{dec})^{-1}$.
There are small differences in each of the factors in the integrand as
compared to the corresponding expression for the FLRW model, as well as
the factor of ${\gc}^{-1}$ in the upper limit of integration. Furthermore,
the overall Hubble distance is the bare one, referring to the volume average.
Relative to the average Hubble constant measured by wall observers we
have
\beq
{c\over\Hb}={c(\gc-\gc')\over \Hm}\,.
\eeq

For the numerical example of \S\ref{nex}, $c/\Hb=6200$ Mpc, while
the proper length scale of the sound horizon is $\bD_s(t\ns{dec})\simeq
0.152\pm0.002$ Mpc at last scattering, or $\bD_s(t\Z0)\simeq230\pm2$ Mpc at the
present epoch. Since this second scale refers to the volume--average
comoving rulers, in conversion to the conventional frame (\ref{wgeom})
synchronous with our clocks, we have $D_s(\tc\Z0)={\gc}^{-1}\bD_s(t\Z0)
=166.8\pm1.5$ Mpc. We note that the proper length scale at last--scattering
is the same whether determined by the wall observer or the void observer,
since $\gb\simeq1$ at that epoch; i.e., it is a standard ruler set by
local physics.

Since the wall measure $D_s(\tc\Z0)\simeq167$ Mpc is the one that most
directly compares to the corresponding scale in the FLRW model, it is the
one which should be compared with the WMAP best--fit value of
$D\Z{s\lcdm}(t\Z0)\simeq147$ Mpc \cite{wmap}.
Since these values are computed with values of the observed Hubble
constant which differ by 14\%, it actually makes more sense to compare
them as $D_s(\tc\Z0)=103.4\pm0.8\,h^{-1}$ Mpc for our case (with $h=0.62$),
and $D\Z{s\lcdm}(t\Z0)=104\,h^{-1}$ Mpc for the concordance \LCDM\ model
(with $h=0.71$). Allowing for the different Hubble
parameters, we see that the comoving scales in fact agree.

The comoving baryon oscillation scale of
$\goesas100\,h^{-1}$ Mpc is now independently verified in the concordance
\LCDM\ model from its statistical signature in galaxy clustering statistics
\cite{bao}. This determination is of course related to angular
diameter distances in the standard FLRW geometry, and the details of the
analysis should be thoroughly checked in the present model. Since the geometry
(\ref{wgeom}) is by our arguments the closest thing to a FLRW geometry in
terms of luminosity distance and angular diameter distance measurements,
the fact that the scale determined by Eisenstein \etal\ and Cole \etal\
is reproduced in
the present model should not be a surprise if it is
correct. The numerical example of \S\ref{nex} was chosen as one case of
parameters which fit both the supernovae data and the first Doppler peak
while resolving the lithium abundance anomaly, rather than by its match to
the baryon acoustic oscillation scale in galaxy clustering statistics.
Other parameters exist which
allow one to match the angular scale of the sound horizon but not the
comoving baryon acoustic oscillation scale, but these also have a poor
fit to the supernovae \cite{paper2}. Since galaxy clustering statistics are
an important model discriminator \cite{Blanchard}, the fact that our
new concordance parameters appear to match the comoving baryon oscillation
scale is encouraging. The details of galaxy clustering statistics in the
new cosmological model still need to be thoroughly investigated from first
principles.

\subsubsection{Fitting the first Doppler peak in the CMB anisotropy
spectrum\br}\br
The angle subtended by the sound horizon provides the rough measure of the
angular position of the first Doppler peak that we will adopt. Actual
computation of the angular power spectrum would unfortunately require
that we systematically rewrite the existing numerical codes which are
calibrated to the Friedmann equation. Nonetheless, the proposed measure
is close physically, and as a proof of principle we simply need to check
whether the model gives the corresponding angular scale determined for the
concordance \LCDM\ model, $\de\Z{s\lcdm}=0.01004\pm0.00004$ rad, where the
error is a statistical one from parameter fits \cite{wmap}.

For the numerical example
of \S\ref{nex} we find that the quantity $\rw$ which solves (\ref{rwo})
for the comoving distance to the
last scattering surface at $\ab\Z0/\ab\ns{dec}=1518$ is $r\ns{w\;dec}=3.73
c/(\ab\Z0\Hb)$.
Thus
\beq
\dAdec={\ab\Z0r\ns{w\;dec}\over\gc(1+z\ns{dec})}=\ab\ns{dec}r\ns{w\;dec}
=15.2\w{Mpc},
\label{dAs}\eeq
and since $\bD_s(t\ns{dec})\simeq0.152\pm0.002$ Mpc, the observed angular
scale $\de_s=\bD_s/\dAdec=0.01$ rad, is reproduced, for values of the
baryon--to--photon ratio which concord with lithium abundance measurements.
We should note that the angular scale of the sound horizon does not coincide
exactly with the angular position of the first peak in the \LCDM\ model and
cannot be expected to do so in the present model either. Although the entire
detailed analysis of the
WMAP data needs to be redone, since we are assuming entirely standard physics
for the photon--matter plasma and since the effect of dark energy is
negligible at the time of last scattering in the case of \LCDM, we should
expect that the small difference between the exact angular position of the
first peak and the angular scale of the sound horizon should be similar in
the present model. However, until the detailed calculations have been
performed, in looking at the broader parameter space for the new cosmology
we should only expect to match the angular scale of the
sound horizon of the concordance \LCDM\ model at the level of a few percent,
rather than as precisely as the numerical example considered here.

We have therefore established that the model is viable in terms of the
angular position of the first Doppler peak. In future, physicists and
astronomers must take care in saying that this angular scale is a
``measure'' of average spatial curvature. It is not -- it might perhaps
be said to be a measure of
local spatial curvature; in an inhomogeneous universe this need not
coincide with the spatial curvature measured at the volume average.
In changing the na\"{\i}ve assumption that our measurements coincide
with the volume average, we have shown that a systematic analysis
accounting for both gravitational energy and spatial curvature variations
can agree quantitatively with observation.
In the present model the angular position of the
first Doppler peak might be said to be a ``measure'' of
average negative spatial curvature at the present epoch, since zero average
spatial curvature now gives a much smaller value of $\rw$, and
consequently an angle too large by a factor of two. However,
I believe it would be safer to reserve statements about ``measures''
of spatial curvature to tests \cite{elliptic2,elliptic1}
which do not make model--dependent assumptions about global averages.

\subsection{Anomalies in large angle CMB multipoles\label{pole}}

There are other anomalies which can be understood in the general framework
of inhomogeneous cosmologies, whose resolution does necessarily rely on
the particular recalibration of cosmological parameters presented here,
but which may be naturally understood in the present framework. One
particular anomaly in this class is the so--called ``axis--of--evil''
\cite{axis}.
It has been known for some time that realistic large matter inhomogeneities
nearby can generate a significant contribution to the CMBR dipole via the
Rees--Sciama effect \cite{RS}. Foreground inhomogeneities cannot generate
the entire dipole since generically the fractional temperature anisotropy in
the quadrupole generated by the Rees--Sciama effect is of the same order
as that due to the dipole \cite{MM}. However, it is
very possible that realistic foreground inhomogeneities can produce
fractional temperature anisotropies of order $|\Delta T|/T\goesas10^{-5}$
in both the dipole and quadrupole \cite{RRS}.

It is therefore very interesting to note that a statistical
study of several possible systematic errors in the WMAP data by Freeman
\etal\ \cite{Freeman} indicates that, of the several effects they studied,
a 1--2\% increase in the magnitude of the peculiar velocity attributed
to the CMB dipole was the only one which may potentially resolve anomalies
associated with large angle multipoles. This is precisely the order of
magnitude of effect we would expect from a $|\Delta T|/T\goesas10^{-5}$
contribution from a Rees--Sciama dipole. Effectively, our current estimate
of the magnitude of the peculiar velocity would include a 1--2\% systematic
error due to a small anomalous boost. Disentangling the small Rees--Sciama
dipole from the dominant contribution of our own peculiar velocity with
respect to the cosmic rest frame would require an enormous computational
effort, and the sky maps would have to be redrawn. However, in the interests
of our fundamental understanding of the universe, these steps should be taken.

\subsection{Expansion age and structure formation\label{age}}

It is a general feature of the present model that the expansion age
is larger at any given epoch than that of the currently favoured
\LCDM\ model, allowing more time for structure formation. Since
old structures are observed at epochs which often seem somewhat too
early in terms of the standard paradigm, this is particularly pleasing.
Since the age of the universe is position--dependent in the new
paradigm, there are in fact different aspects to the solution presented
here.

First of all, in terms of wall time, which is relevant to all bound
systems, the age of the universe is generally about 1 to 1.5 billion
years older than the best--fit age to the WMAP data \cite{wmap} using
the standard \LCDM\ model. Partly, this can be understood in terms of
the value of the Hubble constant, which is about 14\% lower in our case, in
agreement with the recent value of Sandage \etal\ \cite{Sandage}.

The second important issue surrounding the expansion age is that it is
considerably larger for volume average observers than for observers
in galaxies. Whereas wall time is the appropriate time parameter for
actual measurements in bound systems, including the oldest globular clusters
with nucleochronology dates of order $15\pm1$ Gyr, the question arises as
to which parameter is the one that is closest to the parameter assumed in
the $N$--body simulations in Newtonian gravity by which structure formation
is modelled at present. Since such simulations refer to perturbations
about a cosmic mean, it would appear logically that volume--average
time is the relevant parameter. If this is the case, then the fact
that too much structure is obtained within simulated voids, as compared to
actual observation \cite{P_void,P_void2}, could have a very simple
explanation: {\em the integration time assumed is too short}.

For the numerical example of \S\ref{nex} the relevant $N$--body simulations
should be run for 18.6Gyr of volume--average time. Since the relative
proportions of baryonic and non--baryonic dark matter are also recalibrated,
the choice of boundary conditions in the simulations would also be changed.

One of the ultimate lessons of the present paper is that Newtonian concepts
are inadequate to fully understand the large scale structure of the universe.
We ultimately need a better post--Newtonian approximation scheme which
accounts for the existence of a finite infinity scale, and gravitational
energy gradients. Nonetheless, by simple recalibration of various parameters,
accounting for the clock effects, it is quite possible that some of the
puzzles of $N$--body simulations could be resolved. This issue should be
fully investigated.

\subsection{The Hubble bubble and scale of homogeneity\label{bubble}}

Recent analysis of SneIa data by Jha, Riess and Kirshner \cite{JRK} confirms
an effect which has been known about for some time, and has been interpreted
as our living in a local void \cite{Tom1}, the ``Hubble bubble''. By the
latest estimates, if one excludes SneIa within the Hubble bubble at redshifts
$z\lsim0.025$, then the value of the Hubble constant obtained is lower by
$6.5\pm1.8$\% \cite{JRK}. It is for this reason that SneIa at $z\le0.023$ have
been excluded from the Riess \etal\ Gold06 dataset \cite{Riess06}, whereas
they were included in the Gold04 dataset \cite{Riess04}. The Hubble bubble
is particularly problematic for the standard \LCDM\ paradigm, since there is
no natural grounds for choosing a locally measured Hubble constant,
or alternatively excluding the Hubble bubble. Jha, Riess and Kirshner show
that the difference between these two choices using their MLCSk2 data--fitting
techniques and a simulated sample for the ESSENCE survey, can lead to
differences of up to order 20\% in the parameter, $w$, deduced for the
dark energy equation of state $P=w\rh$ \cite{JRK}. Analysis of the first
results of the ESSENCE survey \cite{essence} confirms that this effect could
contribute as much as $0.065$ in the systematic error budget for the $w$
parameter in actual data obtained to date.

The existence of the Hubble bubble, if interpreted conventionally as a large
local void expanding into a more slowly expanding surrounding space, gets
to the heart of the Sandage--de Vaucouleurs paradox. If this is typical, rather
than a statistical fluke, how can the overall expansion rate still be uniform?
The answer in the present model is definitive. The expansion is actually
uniform if differences in spatial curvature and clock rates are taken into
account. If we sample standard candles below the scale of homogeneity, then
the na\"{\i}ve assumption that all clock rates are equal will lead us to
see a sizeable variance in the Hubble flow in spite of its statistical
quietness.

If we make observations within a filamentary bubble wall, such as towards
the Virgo cluster, we should infer a lower Hubble constant than the global
average, $\Hm$. Ideally this would approach the ``bare'' or ``true'' Hubble
constant, $\Hb$, which for the example of \S\ref{nex} is 22\% lower than
$\Hm$. The effect of minivoids will increase the
inferred value. Similarly, if we make measurements to galaxies
on the other side of the voids on the dominant scale, $\goesas30h^{-1}$ Mpc,
then we should infer a Hubble constant which is greater than the global
mean, $\Hm$, by a commensurate amount. This occurs since the path integral
of the photon geodesics through voids includes regions of high relative
positive gravitational energy where clock rates are faster.

Since voids occupy a greater volume of space than the bubble walls, as
long as we measure the Hubble constant by averaging isotropically over all
directions on distance scales less than the scale of homogeneity we will
infer a higher value for the Hubble constant than the global average, $\Hm$.
As we sample on larger and larger scales that approach the scale of homogeneity
then the average Hubble constant will decrease until it levels out at
the global average $\Hm$.

The scale of homogeneity is only defined statistically, since structure
formation is a non--linear process. However, there is a clear operational
definition of this scale: {\em it is the baryon acoustic oscillation scale}
\cite{bao}, which effectively demarcates the regime of linear
perturbations from non--linear ones, on account of the causal evolution
of initial density fluctuations. For the numerical example of \S\ref{nex},
this scale is at $\goesas167$ Mpc in terms of the fiducial geometry
(\ref{wgeom}) synchronous with our clocks. This corresponds to a redshift
$z=0.033$, and thus the Hubble bubble feature presently observed is
sampled at about 2/3 of the scale of homogeneity, consistent with our
interpretation.

The Hubble bubble is not a statistical fluke which marks our location
out as anything special. In the new paradigm it is a feature we can
expect on average, consistent with the Copernican Principle.

\section{Towards dynamical models on intermediate scales\label{Newton}}

\subsection{Variance in the Hubble constant}

The model proposed in \S\ref{Have} should not only be tested against
all observational data in terms of standard tests of types that have
already been considered, but also in terms of its unique predictions.
One prediction is provided by eq.\ (\ref{42}), which relates
the local Hubble parameter, $\bH$, at any location to the
horizon volume average parameter, $H$.

It is difficult to empirically define a locally measured $\bH$, as will
be discussed below. However, I wish to stress that it is a feature of
this model that it is not only the mean value of the Hubble parameter
which is important but also its variance, as is given by the two
extremes in (\ref{42}).
In particular, we see that resolving the Sandage--de Vaucouleurs
paradox also provides new fundamental insights into another dispute
associated historically with those two names, namely what is the value of
the Hubble constant? Sandage and collaborators for many decades favoured
lower values, and de Vaucouleurs higher values.

While it is beyond question that most of the history of the great Hubble
debate has been associated with the resolution of tricky systematic issues
to do with statistical biases and the astrophysics of the standard objects
in the cosmic distance ladder \cite{Tamm}, it is also evident that in the
present model we should measure some underlying variance in the Hubble
constant, depending on the scale over which it is determined, as
discussed in \S\ref{bubble}. Once all astronomers agree on the scale of
observations, as well as on the cosmic distance ladder, then values of
the Hubble constant will converge to $\Hm$. However, ``local'' measurements
within an ideal bubble wall would give a lower value, $\Hb$.
The determination of both of the parameters, $\Hm$ and $\Hb$
is of crucial importance to the new paradigm, as it should ultimately be
related to cosmic variance in the density perturbation spectrum, as will
be discussed in \S\ref{inflation}.

\subsection{Local Hubble flow\label{local}}

A crucial observational question, given the extent to which Newtonian gravity
is employed by astronomers on large scales, and by those engaged in $N$--body
structure formation simulations, is: on what scale may Newtonian gravity
be applied in cosmology? There are three elements to the Newtonian limit:
(i) the weak field limit of general relativity is assumed, $g_{\mu\nu}=
\eta_{\mu\nu}+h_{\mu\nu}$, $\left|h_{\mu\nu}\right|\ll1$; (ii) all
characteristic velocities are smaller than that of light, $v\ll c$; and
(iii) pressures are negligible in comparison to energy densities,
$P\ll\rh c^2$. One point that is often overlooked in applying Newtonian gravity
at cosmological scales is that even when the second and third approximations
hold, since space is expanding assumption (i), the weak field limit,
never strictly holds in cosmology and therefore application of Newtonian
mechanics is completely {\em ad hoc} unless the validity of the weak
field limit is considered within the context of an expanding space.

To identify the scales of averaging over which the weak field limit can be
applied, one first needs to answer the question as to what is the largest
scale on which the equivalence principle can be applied \cite{equiv}?
Having established such a scale, the
Newtonian limit can be applied when in addition to matter satisfying
properties (ii) and (iii) the expansion rate of space is small,
or equivalently Newtonian dynamics may apply on scales over which
quasilocal gravitational energy differences can be neglected.
However, to determine this scale empirically is far from trivial.
One might apply Newtonian gravity successfully to systems of tightly bound
clusters of galaxies which have virialised, but it is clear that at
some level Newtonian dynamics must be modified, and the determination
of large--scale streaming motions is a point in issue.

I envisage that once the effects of quasilocal gravitational energy are
better quantified then a new post--Newtonian approximation scheme relevant
to expanding cosmological backgrounds might be developed. Within the context
of such a scheme it may be possible to quantify the maximum quasilocal
energy variations in minivoids of various sizes by application of the
Traschen integral constraints \cite{Traschen1} to the evolution of initial
underdense perturbations.

The hope is that an effective post--Newtonian approximation for expanding
space may yield much better fits to observation than is currently achieved.
Techniques such as POTENT do not appear to be particularly successful, and
detailed studies of the best observational environment within which a
linear Hubble flow is observed, namely the 10 Mpc$^3$ ``local volume'',
yield a picture which is full of puzzles \cite{Whiting1,Whiting2}. These
puzzles may just be due to poor statistics; however, in the view of the
work of the present paper, it is possible that model assumptions must
be revised already at this local scale.

One key question is the extent of anisotropy in the Hubble law.
There have been a number of attempts to determine anisotropy within the
local volume, culminating in the work of ref.\ \cite{Whiting2}, where the
principal eigenvalues of the local Hubble tensor are determined to be
$\{86\pm15\;(31),53\pm8\;(14),40\pm20\; (34)\}\kmsMpc$. Here Whiting's
$1\sigma$ confidence limits are quoted with the 90\% confidence limits in
brackets. At the 90\% level there is no evidence for anisotropy, and only
tentative evidence at the $1\sigma$ level. Furthermore, the directions of the
eigenvectors can vary by $90^\circ$ within the large uncertainties, and
so are effectively arbitrary. It is quite possible that the 10 Mpc$^3$
volume studied in refs.\ \cite{Whiting1,Whiting2} is too small, and that
one needs to study the ``local'' Hubble flow galaxies on the far side of
the closest void of a $30h^{-1}$ Mpc diameter, relative to filament
directions, to see an effect.

On very large redshift scales, there is tentative evidence of anisotropy
in SneIa data \cite{Bochner}, which is limited by present statistics.
It is vitally important to the present proposal that such studies are
extended with new data, and any relevant scales determined. If any
systematic anisotropies are observed it is possible and consistent
with the present proposal that they could be due to large voids
in the foreground, rather than the background.

At this stage it is too early to quantify what should be expected in the
local volume. The greatest empirical difficulty lies in disentangling the
peculiar velocities of bound systems within finite infinity regions from
underlying expansion between such regions. A dense cluster such as
Virgo should clearly be expected to be within a finite infinity region,
and exhibit the increased peculiar velocity dispersion and tidal torques
that one expects from Newtonian gravity. Those aspects of Newtonian
intuition will remain empirically valid. It is only when significant
regions of freely expanding space are incorporated in the scale under
consideration that we should exercise caution: here a revised
post--Newtonian approximation scheme is required.

Since regions of expanding space beyond finite infinity possess positive
quasilocal gravitational energy, effectively the ``energy required to escape''
beyond finite infinity is larger than we would estimate from the dynamics
of a bound system alone. In terms of the Newtonian Kepler problem finite
infinity effectively sets the scale at which ``$E_\infty=0$''. However,
regions beyond finite infinity have $E>0$ and the energy required to reach
them is larger. A Newtonian physicist might think of this effect as
``deepening'' the gravitational wells of galaxy clusters. However, since
space is not static and particles which escape beyond finite infinity do
not return to the potential well from which they originated,
this is a conceptually flawed interpretation which I do
not wish to encourage. The actual circumstance is best compared with the
case of a standard FLRW cosmology: one may solve the geodesic equations
that follow from (\ref{FLRW}) to show that a particle with an initial peculiar
velocity, $v_i$, as measured locally by an isotropic observer at an epoch
when $\ab=\ab_i$ has a peculiar velocity (in units $c=1$)
\beq v(t)={v_i\ab_i\over\sqrt{(1-v_i^{\ 2})\ab^2(t)+v_i^{\ 2}\ab_i^{\ 2}}}\,,
\label{decay}\eeq
at later epochs as measured locally by isotropic observers. This well--known
result may be interpreted as saying that peculiar velocities decay.

In the case of inhomogeneous cosmologies (\ref{decay}) must be replaced by a
suitable average quantity. The question of averaging the geodesic equations
becomes an interesting problem over and above the averaging in Buchert's
scheme, which just considers the field equations. At the largest cosmological
scales, we have assumed that light follows null geodesics of the average
geometry (\ref{avgeom}), and the same could be expected to be true of
timelike geodesics. However, below the scale of homogeneity -- identified
with the baryon acoustic oscillation scale -- this average of geodesic
motion would have to be refined.

For individual voids, where an observer will detect a differential rate of
expansion from zero at the zero expansion surface to a maximum at the void
centre, it would appear from quasilocal energy considerations that we should
have sharper localised decays of peculiar velocities than for a homogeneous
model. This needs to be confirmed by detailed modelling; the LTB models would
be appropriate for studying individual voids. This question is left for future
work; but has clear direct implications for the statistical quietness of
the average Hubble flow, and the Sandage--de Vaucouleurs paradox.

\subsection{Dynamics of galaxy clusters\label{cluster}}

Even though a revised post--Newtonian approximation scheme remains to
be established, one clear physical difference that the present model
will make is in any situation in which $3H^2/(8\pi G)$ is taken to
be a measure of the closure density which demarcates bound from
unbound systems. One situation in which this regularly occurs in
observational cosmology is through the Navarro--Frenk--White (NFW) model
\cite{NFW} for galaxy clusters, which are estimated to have a density profile
\beq
\rh(r) = \rhcr{\de\Z C \over\left(r/r\Z S\right)
\left(1+ r/r\Z S\right)^2}\,,\label{nfw}
\eeq
where
$\de\Z C = 200C^3/[3\ln(1+C) -C /(1+C)]$,
$r$ is a radial distance within the cluster, and $r\Z S$ and $C = r\Z{200}/r$
are two empirical parameters, respectively the ``scale radius'',
and ``concentration parameter''. When the NFW model is used
in cosmological tests -- for example, in order to determine the ratio of
baryons to dark matter as a function of redshift \cite{Allen}, the critical
density is calibrated with redshift as $\rhcr(z)=3H^2(z)/(8\pi G)$.

Physically, as long as $\rhcr$ represents a closure density, then
it must correspond to the true critical density, $3\bH^2/(8\pi G)$, and
as indicated earlier this may be typically 40--80\% of the value of
the critical density estimated from the global average Hubble constant
at late epochs.
Unfortunately, the NFW model is essentially an empirical fit to the
results of N--body CDM simulations in Newtonian gravity. Thus one cannot
make any simple qualitative statement about how its use might change
if the closure density is to be recalibrated. If similar empirical
fits apply then the only obvious deduction we can make is
that if $\rhcr$ is effectively overestimated at late epochs, then the
density contrast $\de\Z C$ is effectively underestimated.
Consequently, in the present model as long as eq.\ (\ref{nfw}) remains
empirically valid we would expect the density contrast in galaxy
clusters to be higher than is usually assumed.

If there are any testable consequences that follow from such considerations,
then they may possibly apply to highly dynamical non--equilibrium
circumstances, such as that of the collision of the galaxy clusters observed
in the ``bullet cluster'' 1E0657-56 \cite{bullet1}. The high velocity of the
gas shock front trailing the smaller sub--cluster in 1E0657-56 appears
anomalously high \cite{bullet2} as compared to expectations from the
masses of the sub--clusters inferred by using weak gravitational lensing
and with the NFW or King \cite{King} profiles. Others have argued
that such high velocities may not be all that rare statistically
speaking \cite{Hayashi}. Since it is quite possible that the dynamics
of such systems may change in the present model, without invoking new
forces of nature \cite{bullet2}, this issue deserves further investigation.

It is clear that any model--dependent assumptions related to weak
gravitational lensing need to be carefully re--examined. Furthermore,
in the strong lensing regime time delays
in gravitational lensing events, which exhibit some puzzles
\cite{Kochanek}, are a clear circumstance in which the present model
may give differences from the standard \LCDM\ model. Much detailed
modelling remains to be done. However, it is clear that the new
paradigm is likely to give a number of predictions which differ from the
standard \LCDM\ model, which can be tested.

\subsection{Corrections within finite infinity domains\label{fid}}

If we demand that a revised post--Newtonian approximation scheme is required
beyond finite infinity, a natural question to ask is: to what extent
will such effects persist within the finite infinity domains, given the
fact that there must be a nontrivial distance between the zero expansion
surface and finite infinity? Could the zero expansion surface lie so
close to galaxies that the (presumably small) differences of spatial
expansion could influence their rotational dynamics?

I will not answer this question here, since if it extends to near galactic
levels then we are dealing with scales where vorticity cannot be neglected,
as we have in our averaging approximations. It is quite possible that given
the non--linearities introduced by vorticity, the first step in the
weak--field limit of assuming Minkowski space plus a small perturbation is
inappropriate at the galactic scale, even with only a pressureless dust
energy--momentum tensor. Arguments of this sort have been presented by
Cooperstock and Tieu \cite{CT1,CT2}. While such arguments may have promise,
unfortunately the analysis of refs.\ \cite{CT1,CT2} contains potential
flaws which have been much debated\footnote{In particular, quite apart from
any criticisms which have been raised concerning the physical nature of
singularities in the $z=0$ plane of the galactic matter distribution,
Cooperstock and Tieu have given their results in terms of a
radial coordinate which does not correspond to the physical proper radius.
Using the metric (1) of ref.\ \cite{CT2} one sees that the invariant
proper circumference of their axial Killing vector vanishes at $r=\e^{w(r,z)}
N(r,z)$, and that the roles of $\pt/\pt\phi$ and $\pt/\pt t$ as spacelike
and timelike vectors would be reversed for $r<\e^w N$. In fact, since the
invariant norm of the Killing vector $\pt/\pt t$ diverges at $r=\e^w N$,
there is a singularity there and the region $r<\e^w N$
is not physical. The true centre of the galaxy when expressed in terms
of proper distances is at the coordinate surface $r=\e^w N$. Since the
locally measured rotational velocity tends to the speed of light as one
approaches this ``galactic centre'' the approximations used break down there.
This is the explicit manifestation of the criticisms by others \cite{dust},
which have been commented on in refs.\ \cite{CT1,CT2},
that comoving coordinates cannot be chosen globally in the presence of
vorticity. Cooperstock and Tieu point out that their choice of boundary
conditions removes the potential problems at $r=\e^w N$. With different
boundary conditions, Balasin and Grumiller \cite{BG}
avoid any potentially singular source in
the $z=0$ plane, but have unfortunately made the error of failing to
identify the true centre of the galaxy, so as to excise the unphysical region.
It is possible that the main results of ref.\ \cite{BG} may survive the
correction of this error.}.
The variant of the Cooperstock--Tieu model proposed by Balasin and Grumiller
\cite{BG} is particularly interesting, since these authors find a significant
reduction in dark matter, but not its complete elimination -- which accords
well with our cosmological parameter estimates based on the baryon acoustic
oscillation scale \cite{paper2}.
Whatever results from a careful analysis of the proposals of
refs.\ \cite{CT1,BG}, it has only an indirect bearing on our discussion,
since the present proposal relates to scales of averaging very much
larger than galactic scales, and the problem of galactic dynamics
could involve more ingredients than those considered by these authors.

The only direct consequence of the present proposal vis--\`a--vis dark
matter, is that the recalibration of cosmological parameters will
potentially result in substantial changes to the matter budget, both
in the overall fraction of matter in our observed portion of the universe
relative to the critical density, and in
the ratio of non--baryonic to baryonic dark matter. The numerical example
of \S\ref{nex}, which fits observations well, has a 3:1 ratio of
non--baryonic dark matter to baryonic matter. As indicated in \S\ref{cluster},
the exact value of this ratio cannot be fully resolved until
gravitational lensing and the question of the estimate of masses of
clusters of galaxies has been re--examined. This may involve
some subtle recalibrations which are likely to change the overall
dark matter budget in a fashion which is difficult to foresee.

If any further changes are to be made at the level of galactic dynamics,
such changes should be made by solving geodesic equations within the context
of general relativity, for appropriate non--asymptotically flat backgrounds.
If any modification of Newtonian dynamics is encountered at the level of
galactic bound orbits, then it may primarily be due to non--linear
couplings between dust and spatial vorticity. Whether the present
proposal has anything further to add at galactic scales can only be decided
once we construct a dynamics for systems embedded in finite infinity regions.
The scale at which the negative curvature of the voids becomes manifest
may be an important boundary condition.

It is my view that we should think deeper and harder about general
relativity, rather than resorting to ad hoc Newtonian
forces or adding terms to the gravitational action which are not justified
from any other observations or any fundamental physical principles.
When non--baryonic dark matter candidates are postulated, then those which
naturally have almost no standard model interactions \cite{AKG} would seem
easiest to reconcile with the rest of cosmology, rather than those that
result from fundamental modifications to gravity. While
I do not see any obvious role for MOND \cite{mond}, we should be open to the
possibility of phenomenological changes to Newtonian gravity, and
actively investigate models in which the assumptions of asymptotic flatness
are changed.

Direct changes to Newtonian expectations could conceivably arise for particles
in orbits unbound to galactic structures, as discussed at the end of
\S\ref{local}. For example, if we were to send a spacecraft on a hyperbolic
orbit with a large enough escape velocity to leave the galaxy, and ideally
the local group, then we might well expect to register clock anomalies
as compared to Newtonian predictions. It should be noted that the Pioneer
spacecraft do {\it not} qualify as their escape velocities from the solar
system should leave them still bound to the galaxy \cite{Slava}. The
present proposal relies on space in bound systems being non--expanding,
and if it is correct attempts to explain the Pioneer anomaly \cite{Pioneer}
in terms of the expansion of the universe must fail. Indeed, careful attempts
to consider the Pioneer anomaly in this fashion give an effective Newtonian
acceleration in the {\em opposite} direction to the observed anomaly
\cite{CG}. In my view, it is far more likely that the Pioneer anomaly -- if
it is anything other than instrumentational -- is more likely to be explained
in terms of a careful consideration of quasilocal gravitational binding
energy differences between the solar system and the galactic frames.

A question of principle does remain. In the present context, gravitational
binding energy differences may be larger than is assumed when dealing with
ideal asymptotically flat geometries. We
have assumed that differences in total gravitational energy between
our location and finite infinity can be neglected for cosmological averaging.
It is conceivable that the question of ``where is infinity?'' which is
central to my proposal may also be related to issues associated to
gravitational binding energy. The Pioneer and other flyby anomalies all
appear to involve a potential misunderstanding of energy transfer processes
\cite{transfer} as spacecraft gain energy to move into orbits which are
hyperbolic with respect to either the solar system frame, or planetary frame
as relevant. Thus the understanding
of gravitational energy in relation to the true notion of asymptotic
infinity may reveal subtleties not yet considered.
To date discussions of gravitational binding energy generally appeal to
asymptotic flatness \cite{KLB}, and it is clear that at some level this
must be corrected. The notion of finite infinity may be relevant to
the framework within which such issues should ultimately be debated.

One experiment that would settle this issue would be a space probe mission
to fly a cosmic microwave background imager on a trajectory similar
to those of the Pioneer spacecraft, to measure the mean temperature of
the CMB to the accuracy required if it is assumed that the observed
anomaly is a clock effect related to gravitational energy. A very small
deviation in the mean CMB temperature, showing a consistent decrease at
very large distances, would be expected over the course
of the mission. It is quite possible that measurements of the required
accuracy are not yet technically feasible. Nonetheless, such a mission
may become feasible over the course of the next century. A commensurate
decrease in the angular scale of the CMB anisotropies would also be
expected in principle, due to variations in spatial curvature, but at a
level so tiny that it is likely to be beyond technical feasibility for
even longer.
\setcounter{footnote}3

Collisions between galaxies could conceivably provide an astrophysical arena
in which individual stars are ejected at high enough escape velocities for
them to escape from galaxy clusters, and make the local clock effects I am
concerned with readily measurable\footnote{I thank Alex Nielsen for this
suggestion.}. In practice, such events are at such distances that it is
unlikely that we could resolve individual stars ejected far from the
galaxies. In the case of galaxy or galaxy cluster mergers, what we principally
observe are the bound systems within finite infinity domains, for which there
will be no direct clock effects. The only indirect tests might involve mass
determinations, as in the case of the bullet cluster mentioned above.
Since we observe such mergers at one epoch in the progress of the collision,
and cannot play the movie forwards or backwards, the issue of statistics
\cite{Hayashi} is likely to make claims of a definitive test on these
lines hard to justify, however.

\subsection{Traversal of voids}

The immediate question that is asked by many of those who are presented with
this model is: surely we must observe clocks in voids that would rule
out these effects? The answer, unfortunately, is that voids by definition
have large negative density contrasts, and thus we do not actually observe
matter within them. It is possible to find samples of galaxies in very
thin filamentary structures within voids. However, a galaxy by definition
is a bound system, and is within a finite infinity domain. The proper
distance to finite infinity is much less for a void galaxy than a cluster
galaxy; however, it is still within a region where clock differences can be
expected to be negligible when measured at cosmological distances.
Even the gas cloud precursors to galaxies, which are clumped with
a sufficient density to be detectable via absorption signals, are contained
within finite infinity domains.

It is only if the effects of quasilocal gravitational energy
persist within finite infinity regions that we might expect to see
differences, for example, in rotation curves of void galaxies as opposed
to cluster galaxies. However, as I indicated in \S\ref{fid}, without
detailed modelling such a possibility remains a pure speculation, which
does not follow in an obvious fashion from considerations about
the quasilocal energy of the expansion of space or spatial curvature
variations.

Apart from photons, the only particles that regularly traverse voids are
high--energy protons and other cosmic rays. One might be concerned that the
GZK bound \cite{gzk} would be altered since the CMB temperature measured
by isotropic observers is lower in voids in the present model, allowing
higher charged particle energies to achieve the equivalent
centre--of--momentum energy associated with the standard calculation of
the GZK bound. However, since high--energy charged particles originate
from sources in in galaxies within the filamentary walls, there will be a
compensating decrease in their energies as measured by a void observer. This
is equivalent to saying that the decay of peculiar velocities of charged
particles traversing voids is expected to be greater. Thus the GZK bound
should not be grossly affected. There may be some changes due to the fact
that the energy of massive and massless particles do not scale in the
same way with redshift. However, detailed calculations, with gravitational
energy changes expected from individual voids need to be considered. It is
quite conceivable that there could be alterations to the statistics of
charged particles travelling cosmological distances in relation to the GZK
bound, due to something akin to an integrated Sachs-Wolfe effect for massive
particles. To determine whether this is the case would require detailed
calculations taking account of whatever relations replace (\ref{decay}).

\section{Primordial inflation and the origin of inertia
\label{inflation}}

I will now outline why the model described in the previous sections, with
its two cosmic times, is the natural outcome of primordial inflation, via
cosmic variance in the spectrum of primordial density perturbations.

We detect temperature fluctuations in the mean CMB temperature of order
$\left|\DE T\right|/T\goesas10^{-5}$, which themselves arise via the
Sachs--Wolfe effect from primordial density perturbations believed to be of
order $\de\rh/\rh\goesas10^{-3}$ in non--baryonic dark matter at the time of
decoupling. The precise value of $\de\rh/\rh$ responsible for the observed
temperature anisotropies via the Sachs--Wolfe effect may actually require
recalibration since it turns out that the calibration makes use of
present day values of the Hubble constant and matter density for a
smooth FLRW model. However, it is unlikely that it should change much.

The new cosmological model requires systematic recalibration of all observed
quantities, but in a manner which should ultimately be consistent with the
inflationary spectrum and the early evolution of density perturbations
according to established theory. Since the universe is homogeneous
and isotropic at early epochs, there is no change to the underlying physics
-- it is merely the calibration which changes, wherever quantities have been
determined with present epoch parameters of a smooth FLRW model. In practice,
this occurs at very many steps.

It is a consequence of primordial inflation that the density fluctuations
from which all structures ultimately grow are close to scale invariant.
Thus provided our particle horizon volume is smaller than the scale of the
largest perturbation at any epoch then at some scale of averaging
we will always be sampling the non--linear structure that arose from
perturbations nested inside some larger single perturbation which is still
itself effectively in the linear regime, giving an effective almost--FLRW
geometry on the largest averaging scale. We just have to be very careful
about relating our observational parameters to that geometry.

One problem with the standard linearised analysis of cosmological
perturbations, quite separate from the thorny issues associated with
gauge choices, is that
its methodology introduces language which becomes totally inappropriate
for dealing with averaging at late epochs. Once large volumes of non--linear
structures exist, it is more appropriate to think in terms of real space,
rather than Fourier space. As an example, much of the debate that followed
a suggestion of Kolb \etal\ \cite{kolb1}, which was corrected in their
later work \cite{kolb2}, concerned the issue of ``super--horizon sized''
as opposed to ``sub--horizon sized'' modes.

Of course, only matter
within our past lightcone can have a causal influence on the geometry
of the universe we observe today. However, in reality all density
perturbations are finite regions of space, nested within each other
like Russian dolls.
As with all other observers in bound systems we are sitting on overdense
galaxy perturbations embedded in underdense regions embedded in overdense
galaxy cluster perturbations etc. If we follow this hierarchy up to the
largest observable scale, then we can imagine that we are sitting inside an
underdense perturbation that is commensurate with the size of our present
horizon volume. If this is the case, when did its effects become apparent to
us? Those versed in linearised analysis at early epochs speak about
perturbations starting to have an influence when they ``cross the
horizon''. However,
for perturbations which are relatively large in comparison to our Hubble
volume, the perturbation ``crossing time'' is significant. Does a
perturbation begin to have a noticeable effect when 50\% of its spatial
volume is within one's past light cone, or 75\%, or what? Since general
relativity is causal it is certainly not true that a density perturbation
has an instantaneous impact on spacetime geometry the moment that
the boundary of a perturbation pops through the past light cone at some event.

For the reasons above, I observe that while present geometry must result
from Einstein's equations applied to sub--horizon--sized volumes, arguments
about whether the perturbations responsible for back--reaction at late
epochs are a few times smaller than our present horizon volume, or slightly
larger than our present horizon volume, are not productive. Two scales are
involved:
\begin{itemize}
\item[(i)] The finite infinity region scales are very much smaller
than the particle horizon volume. These surround bound systems,
are truly in the ``non--linear regime'' and define the reference point of
our clocks.
\item[(ii)] The large scales comparable to the horizon volume are
characterised by perturbations still in the ``linear regime'' and define
large--scale geometry. They define the clocks of observers at volume average
positions in freely expanding space which differ from those in bound systems.
\end{itemize}
In dealing with the large scales we are talking not about smooth
uniform perturbations with no substructure but about statistical
correlations between disjoint regions that emerge over billions of years
after galaxies have started to form. That
is why an averaging scheme, which deals with a void volume fraction as
in this paper, is preferred. Once a best--fit to cosmological parameters
is found, one can determine the size of the primordial perturbations
responsible.

The effects of individual perturbations will continue to have an influence on
the geometry of spacetime over time scales commensurate with the size of each
perturbation. As long as the volume of the observed universe is smaller
than the scale of the largest perturbation, cosmic variance will ensure that
back--reaction is important. If this notion is not familiar to the
reader, I will now explain the idea from first principles.

\subsection{Particle horizon volume selection bias and cosmic variance}

Take an initial spatial
hypersurface corresponding to the end of the inflationary epoch. Onto this
hypersurface randomly scatter small blobs of all proper sizes, allowing
blobs to fall inside other blobs, from some
smallest scale, up to some largest scale \BB. Larger blobs
will naturally contain all smaller blobs, according to a distribution
which could in principle be calculated. The fact that there exists a
largest scale, \BB, of perturbations is simply due to the fact that
inflation ends, so there must be a cut--off to the spectrum of
perturbations at some upper bound. Scales larger than \BB\
will be called the {\em bulk}, and the density averaged in the bulk
will be extremely close to the true critical density.

The density of the blobs is assumed to be Gaussian--distributed about
the true critical density, and the distribution
is assumed to be scale--invariant in the following sense. Let us consider
a volume, \VV, of the spatial hypersurface very much larger than the cut--off
\BB. Then for density perturbations of any given proper volume scale the
fraction of the proper volume of \VV\ contained in the blobs of any
particular scale will be roughly equal, with the same mean density as
the bulk. Tiny blobs, \TT, each have a small
volume but are much more numerous. Furthermore, when sampled
on the bulk scale, a collection of perturbations on any scale
will have the same mean density as the bulk, whether it is the tiniest
scales, \TT, or any intermediate scale, \SS.

Now let us consider the intermediate scale, \SS, which we will suppose
is the scale of the perturbation which will be the dominant one
determining the present epoch geometry of our observed portion of the
universe, which is nearly FLRW. We will assume that \SS\ is commensurate with
our present horizon volume, possibly somewhat smaller or larger.
Since we are dealing with a single perturbation \SS, rather than a
statistical ensemble of them, its mean density can be different from
the bulk critical density, and since our observed universe is
void--dominated at present, we assume it to be underdense.

We also immediately run into a further sampling issue.
While the tiny scales, \TT, sampled within \SS\ will still have
a mean density distributed about the bulk critical density, just as
if they had been sampled on the bulk scale \VV, as soon as we approach
scales, \LL, which are smaller than but close to the scale \SS, then a
discrepancy will arise. Each density perturbation, being
a macroscopic region of space with a mean density has to be wholly contained
in some other density perturbation. If two perturbations were to ``overlap''
then the overlap region is just a third perturbation with a mean density
of the mean of the first two perturbations. As the scales of perturbations
become comparable there are fewer ways to fit one perturbation inside
another. For scales
\LL\ close to that of \SS, the number of perturbations in the sample
will be very small indeed, and on average we expect the mean density of the
limited sample to differ from the sample of the same scales within the
bulk, through a $\sqrt{N}$ statistic.
I will call these effects -- combined with the fact that the overall single
perturbation represented by the volume \SS\ has an average density
different from the mean of the whole distribution -- the {\em
particle horizon volume selection bias}.

Density perturbations at last scattering will, of course, give rise
to temperature fluctuations in the CMBR through the dominant Sachs--Wolfe
effect, and what I have just defined as the particle horizon volume selection
bias is the reason for {\em cosmic variance} in the two-point angular
correlation function of CMBR temperature anisotropies \cite{LL} being much
larger at low angular multipoles, as is well--understood. As the phrase
cosmic variance is often used solely for the distribution of
observed CMB temperature anisotropies, I have introduced an alternative
terminology. However, the particle horizon volume selection bias is just
one consequence of the variance in the underlying spectrum of nearly
scale invariant density perturbations. It is that underlying variance
in the perturbations, which I am referring to as cosmic variance in a
broader sense.

\subsection{Cosmic perturbations and cosmic evolution\label{structure}}

Given the picture above of the macroscopic volume of space at last scattering
from which our observable universe formed, the story of its evolution to
the present day can be understood. If we consider the spatial extent of
our present horizon volume at last scattering, then it will contain an
initial underdense dust void volume fraction, $\fvi$, consistent with
(\ref{initc}) correlated on scales very large compared to the particle
horizon at last scattering. Different values of $\fvi$ consistent with
the bound (\ref{initc}), and values of $(\de\rh/\rh)\ns{vi}$ consistent
with the primordial spectrum, give different initial void volume fractions
for numerical modelling.

An observer starting from an initial point which goes on to form a galaxy
will be inside an overdense region, which itself will be a perturbation
contained within the large fraction of \SS\ which averages to critical.
Such perturbations can be treated as constrained perturbations in an
Einstein--de Sitter background. An observer in such a perturbation
will at first perceive that she lives in a closed universe while the
perturbation remains within the linear regime, but after the perturbation
turns around and collapses, it is other larger perturbations which remain
in the linear regime and which define the dominating perturbation and
back--reaction at any particular epoch. An observer starting at one of
the initial parcels of fluid which develops to be a void at the present
epoch will have quite a different interpretation of cosmic history.

The problem of explaining the existence of voids in structure formation
simulations is now readily understood to be a consequence of the fact that
the wrong mean density has been assumed. It is the true critical density
which must be assumed in structure formation simulations: the average
density we measure on the scale of \SS\ today has not evolved smoothly
by the Friedmann equation from the true critical density. By assuming
that it does we come to choose random perturbations about the wrong mean.
Once a suitable initial void volume fraction, $\fv$, is determined by fitting
the overall cosmological evolution to Type Ia supernovae and gamma--ray
burster luminosity distances, the CMB, and the acoustic oscillation scale
in galaxy clustering statistics, then suitable boundary conditions for
structure formation simulations can be specified.

\subsection{Mach's principle}

It is interesting to note that since primordial inflation ensures that
the expansion rate of the universe is uniform at the time of last
scattering, and since this gives a true universal critical density which
defines the true surfaces of homogeneity, and therefore a reference
point for inertial frames, primordial inflation effectively gives us a
version of Mach's principle, even if not the one Mach would have originally
envisaged. Because there is a true critical density there is a real sense
in which it is all the matter in the Universe which defines the reference
point of inertial frames. This
just relies on the general properties of inflation, rather than any
specific inflationary model. Specific versions of Mach's principle
which relate to rotating frames can be understood in terms of the
causal evolution of angular momentum perturbations \cite{Mach1,Mach2},
with an initial distribution as expected from inflation. The arguments in
the present paper amount to the statement that those aspects of
the definition of inertial frames that relate to the normalisation of
clocks can similarly be understood in terms of the causal evolution
of density perturbations, with an initial distribution as expected from
inflation. Further refinements of the formalism and predictions should be made
by exploiting the Traschen integral constraints \cite{Traschen1,Traschen2}.

\section{Discussion\label{dis}}

In this paper I have proposed a dramatically new interpretation of averaged
cosmological quantities, in the context of standard general relativity.
The Einstein equations are retained, as is the geodesic postulate
(or theorem for those who regard it as such) that a test particle follows
a timelike geodesic, and that it measures a proper time defined by the
metric. What is abandoned is a simplifying assumption about
averaging a cosmological spacetime geometry.

Given that space within bound systems is not expanding the pertinent question
is not: ``What effect does expanding space have on bound systems?''
\cite{gruyere}; but ``How does expanding space affect local determinations
of cosmological parameters within expanding regions in a way which would
distinguish them from local determinations of cosmological parameters made
within bound systems?'' The conventional assumptions
that our clocks tick at a rate equal to that of a volume--averaged
comoving observer, and that the average spatial curvature we measure in galaxy
clusters also coincides with the volume average, are required neither by
theory, principle nor observation.

The first key step in understanding this is that {\em total gravitational
energy} -- including the quasilocal energy of the expansion of space and
spatial curvature variations, as well as gravitational potential differences
in bound systems -- can play a role in gravitational time dilation and this
can be globally manifest in a universe whose spatial matter distribution is
as inhomogeneous as the one we live in. Different classes of ideal isotropic
observers can each measure a uniform CMB with no dipole anisotropy, while
measuring a different mean temperature and different angular anisotropy
scales.
The second key understanding is that since the universe did initially expand
at a uniform rate, there is a true critical density despite present day
inhomogeneity, which does define a universal energy scale that demarcates
bound systems from unbound ones. With the identification of finite infinity
this allows us to identify a global mean expansion rate -- we just
need to define this expansion rate operationally, as it will generally
differ from the average expansion when measured over our entire past horizon
volume today according to a single clock. The
third key understanding is that apart from the CMB all our cosmological
observations are on bound systems, leading to an observer selection effect:
even if our clock rates and spatial curvature measurements differ negligibly
from those in distant galaxies, the same measurements can differ
systematically at the volume average of our
observable universe in freely expanding space. These understandings lead to a
definition of average homogeneity by the obvious requirement that
the expansion {\em rate} is homogeneous, once quasilocal expansion of space
is operationally defined with respect to local measurements.

The new picture of the universe is both familiar and strange to twentieth
century physicists. There is an average homogeneous isotropic geometry, but
one which does not evolve according to the Friedmann equation. The fact that
our clocks and measurements inferred from them do not coincide with those
at a volume average position means that the universe has an age which is
position--dependent, being as old as 21 billion years in the centres of voids.
Yet all observers in bound systems will nonetheless agree on it being of
order 15 billion years old. The ultimate fate of the universe -- whether it
will expand forever or not -- is {\em undecidable}, at the present epoch
at least. We must wait until the scale of the largest primordial
perturbation is correlated within our past light cone before we can know
whether the universe will ultimately collapse.

The new proposal renders the terminology of ``open'' and ``closed''
universes somewhat obsolete. I therefore propose to call this model
the {\em Fractal Bubble Universe}. I use the word ``fractal'' here
with some trepidation, as it has come to be popularly associated with
those who wish to impose a mathematical structure on the universe as an
extra principle of nature\footnote{It should be noted that the Copernican
Principle advanced in \S\ref{Cop} is distinct from Mandelbrot's Conditional
Cosmological Principle \cite{Mandel}, which is further discussed by Mittal
and Lohiya \cite{ML}. Mandelbrot assumes galaxies are fractally distributed
and the universe appears the same at any galaxy, but not in voids. I argue
that there is a notion of homogeneity whether viewed from galaxies or voids;
one just needs to be careful about relating this to local measurements.}
\cite{Mandel}. In my view, that is taking one
notion of mathematical beauty too far. Physics proceeds from principles to
do with basic observable quantities such as energy and momentum; my proposal
aims to refine the understanding of those quantities by better characterising
quasilocal gravitational energy in cosmology. This relates to the operational
understanding associated with averaging the left hand side of Einstein's
equations, rather than imposing a fractal structure on the energy--momentum
tensor \cite{ML}. While fractals are
ubiquitous in nature, such structures arise from physical processes and
are always limited to a range of scales.
It is certainly true that a void--dominated universe at the present epoch
would be consistent with a fractal distribution of galaxies at some scales,
and on those scales the observed universe may appear hierarchical in the sense
that de Vaucouleurs argued for \cite{deVau}. I suggest the word ``fractal''
insofar as it aptly encompasses those observations. Whether the distribution
of galaxies is strictly fractal or not in some precise mathematical sense
is a point which has been much debated \cite{paradox1,fractals}, and will
no doubt continue to be debated, but is not in any way directly relevant to the
arguments I have presented.

This new proposal, if it is correct, means that present epoch ``dark energy''
is an historical accident resulting from a misidentification of gravitational
energy, which is not local and cannot be fully described by the
internal energy of a fluid--like quantity.
Furthermore, this new understanding should provide
a much richer and deeper framework for theoretical cosmology and observational
modelling. The changes are not small, as every average cosmological parameter
must be systematically recalibrated. However, once these changes are made
many avenues of new research directions will open up. Qualitatively,
cosmic variance becomes as important a feature as cosmic averages in the new
paradigm. While the historical argument among astronomers over the value of
the Hubble constant is for the most part due to systematic issues, there
is an intrinsic variance related to the choice of averaging scale which may
have contributed to these contentions. This variance, as quantified by
eq.\ (\ref{42}), should ultimately be related to cosmic variance in the
primordial density perturbation spectrum.

The coincidence problem in the standard \LCDM\ model, as to why the fraction
of dark energy should be commensurate with the fraction of clumped
matter at the present epoch, is eliminated since present epoch dark energy
is eliminated. Furthermore, the other coincidence as to why cosmic
``acceleration'' should occur at the same epoch when the largest structures
form is naturally solved. Voids are associated with negative spatial
curvature, and negative spatial curvature is associated with the positive
gravitational energy which is largely responsible for the gravitational
energy gradient between bound systems and the volume average. Since
gravitational energy directly affects relative clock rates, it is at the
epoch when the gravitational energy gradient changes significantly that
apparent cosmic acceleration is seen.

Some potential directions for future work and a number of potentially
distinctive cosmological tests have been suggested in \S\ref{cmb},
\S\ref{Newton} and \S\ref{structure}. Clearly, variations in quasilocal energy
should be modelled from a more suitable formalism, e.g., a generalisation of
\cite{KLB,quasi1,quasi2}, accounting for the integral constraints
\cite{Traschen1,Mach1,Traschen2}.
Any better understanding of quasilocal gravitational energy in one dynamical
situation can only help improve our understanding in other situations,
such as the choice of average slicing for numerical studies of black hole
encounters. The mean lapse or ``gravitational energy parameter'', $\gb$,
which is in a sense related to the non--affinity of wall time, $\tc$, on
volume--average dust geodesics, appears to play a role analogous to the
surface gravity in black hole models, although the latter is
defined for null foliations. In the cosmological context, the first integral
(\ref{Omtrue}), which may also be written as $\fw\simeq\gb^2\OMM$, is perhaps
the single most important equation that follows from the definition of finite
infinity.

As theoretical physicists we are altogether too much inclined to
add all sorts of terms to the gravitational action, even if they potentially
violate basic principles such as causality as in the case of ``phantom
energy'', rather than thinking deeply about the basic operational issues of
our subject. I believe we should guard the principles that have worked until
they can be proved to fail. It is my own view that Einstein
was correct about general relativity, and what I have presented here
follows logically from his theory when combined with initial conditions
given by primordial inflation. As a consequence even our understanding of
Einstein's most famous equation for the rest energy of a test particle,
$E=mc^2$, must be further refined in cosmology since in truly expanding regions
two widely separated inertial observers cannot be at rest relative to each
other. However, the refinement of the notion of ``rest'' energy merely
clarifies an aspect of the theory, quasilocal gravitational energy, which
Einstein knew to be incomplete.

Nature is of course the final arbiter of all ideas; so it is pleasing
that in addition to achieving concordance with supernovae data and the baryon
acoustic oscillation scale, the present model may resolve a number of
observational anomalies.
In addition to the fact that the expansion age is generally larger in the
present model, allowing more time for structure formation, the specific
observation of ellipticity in the CMB anisotropies \cite{elliptic2,elliptic1}
is clearly consistent with the present paradigm. The problem of lithium
abundances \cite{lithium} can likewise be understood, since the baryon
fraction obtained from standard nucleosynthesis bounds for a given
volume--average baryon--to--photon ratio, $\etBg$, can be higher.
Although we have given generic arguments and explicit calculations to show
that the generic features of the CMB anisotropy spectrum such as the angular
scale of the first Doppler peak and the ratio of the heights of the first
and second peaks will fit the recalibrated cosmological parameters, much
work needs to be done to write detailed numerical codes to perform
fits to the WMAP data in as much detail as has been done for the standard
\LCDM\ model \cite{wmap}.

Whatever the outcome of observational tests on the new cosmological model, the
questions I have raised are, I believe, so fundamental to the foundations
of cosmology that they must be seriously considered. If I am wrong, then the
following questions remain: What is the energy content of an expanding space
of varying curvature and what is its influence on particle clocks? Can
average spatial curvature vary to the extent that we infer different
angular scales in the CMB anisotropy spectrum to those at the volume average?
What is the largest scale on which the equivalence principle can be applied?
Where is the effective spatial infinity vis--\`a--vis a bound system?
How do we resolve the Sandage--de Vaucouleurs paradox? How do we explain the
existence of voids? Why does the observed universe appear to undergo
``accelerated'' expansion just at the epoch when structure forms? How do we
define average surfaces of homogeneity in a very lumpy universe? A working
framework which seeks to
quantitatively answer all of these issues without changing the principles of
our best theory of gravitation stands a better chance of success, I believe,
than alternative hypotheses which ignore these foundational questions.

\ack

This work was supported by the Mardsen Fund of the Royal Society of New
Zealand. I am particularly grateful to Jorma Louko and Jenni Adams for their
detailed comments on initial drafts of this manuscript. I also wish to thank
many people for discussions, correspondence, questions and criticisms
which have helped me to clarify my ideas and presentation, including:
Roy Kerr, Matt Visser, Paul Davies, Peter Szekeres, Thomas Buchert,
Syksy R\"as\"anen, Alan Whiting, Steve Maddox, Hui Yao, Benedict Carter,
Alex Nielsen, Cindy Ng, Ben Leith, Tam Nguyen Phan, Marni Sheppeard,
Ishwaree Neupane, Dharamvir Ahluwalia-Khalilova, Harvey Brown, Ewan Stewart,
Brett Bochner, Tony Fairall, Richard Easther, and Don Page.
I also thank Prof.\ Y.M.~Cho for hospitality at Seoul National University.

\section*{References}

\end{document}